\documentclass[a4paper,11pt]{article}
\pdfoutput=1
\usepackage{jheppub}
\usepackage[T1]{fontenc}
\usepackage{amsmath}
\usepackage{amssymb}
\usepackage{epsfig}
\usepackage{subfigure}
\usepackage{float}
\usepackage{verbatim} 
\usepackage{pstricks}
\usepackage{color}
\usepackage{slashed}
\usepackage{multirow}
\usepackage{pstricks}
\usepackage{color}
\usepackage{booktabs}
\usepackage{subfigure}
\usepackage{epstopdf}
 \usepackage{braket} 
\allowdisplaybreaks[3]

\newcommand{\sw}{s_{W}}
\newcommand{\cw}{c_{W}}

\newcommand{\mc} {\mathcal}
\newcommand{\mb} {\mathbb}
\newcommand{\norm}[1]{\left\lVert#1\right\rVert}

\begin{document}

\rightline{\footnotesize USTC-ICTS/PCFT-20-29}
\title{Elastic positivity vs extremal positivity bounds in SMEFT: a case study in transversal electroweak
gauge-boson scatterings}
\author[a]{Kimiko Yamashita,}
\author[a,b,c]{Cen Zhang,}
\author[d,e]{Shuang-Yong Zhou}

\abstract{
The positivity bounds, derived from the axiomatic principles of
quantum field theory (QFT), constrain the signs of Wilson coefficients and their
linear combinations in the Standard Model Effective Field Theory (SMEFT).  
The precise determination
of these bounds, however, can become increasingly difficult as more and more SM
modes and operators are taken into account.  
We study two approaches that aim
at obtaining the full set of bounds for a given set of SM fields: 
1) the traditional elastic positivity approach, which exploits the elastic scattering
amplitudes of states with arbitrarily superposed helicities as well as other
quantum numbers, and 
2) the newly proposed extremal positivity approach, which
constructs the allowed coefficient space directly by using the extremal
representation of convex cones. 
Considering the
electroweak gauge-bosons as an example, we demonstrate how the best analytical
and numerical positivity bounds can be obtained in several ways.  
We further compare the
constraining power and the efficiency of various approaches, as well as their
applicability to more complex problems. While the new extremal approach is more
constraining by construction, we also find that it is analytically easier to
use, numerically much faster than the elastic approach, 
and much more applicable when more SM particle states and operators are taken
into account. As a byproduct, we provide the best positivity bounds on the transversal
quartic-gauge-boson couplings, required by the axiomatic principles of QFT,
and show that they exclude $\approx 99.3\%$ of the parameter space currently
being searched at the LHC.
}

\affiliation[a]{Institute of High Energy Physics, Chinese Academy of Sciences, Beijing 100049, China}
\affiliation[b]{School of Physical Sciences, University of Chinese Academy of Sciences, Beijing 100049, China}
\affiliation[c]{Center for High Energy Physics, Peking University, Beijing 100871, China}
\affiliation[d]{Interdisciplinary Center for Theoretical Study, University of Science and Technology of China, Hefei, Anhui 230026, China}
\affiliation[e]{Peng Huanwu Center for Fundamental Theory, Hefei, Anhui 230026, China}

\emailAdd{kimiko@ihep.ac.cn}
\emailAdd{cenzhang@ihep.ac.cn}
\emailAdd{zhoushy@ustc.edu.cn}

\maketitle

\section{Introduction}

The dimension-8 (dim-8) Wilson coefficients
\cite{Henning:2015alf,Murphy:2020rsh,Li:2020gnx} that give rise to $s^2$
dependence of a two-to-two scattering amplitude in the Standard Model Effective
Field Theory (SMEFT) are not allowed to take arbitrary values
\cite{Zhang:2018shp,
Bi:2019phv, Bellazzini:2018paj, Remmen:2019cyz, Remmen:2020vts, Zhang:2020jyn,eepaper}.  
By assuming that the SMEFT admits a UV completion that satisfies the
fundamental principles of quantum field theory (QFT), including analyticity,
unitarity, crossing symmetry, locality and Lorentz invariance,
the so-called positivity bounds can be
derived \cite{Adams:2006sv, deRham:2017avq, deRham:2017zjm, Pham:1985cr,
	Pennington:1994kc, Ananthanarayan:1994hf, Comellas:1995hq,
	Manohar:2008tc, Low:2009di, Sanz-Cillero:2013ipa, Bellazzini:2016xrt,
	nimahuanghuang, deRham:2018qqo, deRham:2017imi, Baumann:2015nta,
Bellazzini:2015cra, Cheung:2016yqr, Bonifacio:2016wcb, Du:2016tgp, Bellazzini:2017fep,
Hinterbichler:2017qyt, Bellazzini:2017bkb,  Bonifacio:2018vzv, Bellazzini:2019xts,
Melville:2019wyy, Melville:2019tdc, deRham:2019ctd, Alberte:2019xfh,
Alberte:2019zhd, Ye:2019oxx, Wang:2020jxr, Alberte:2020jsk, Huang:2020nqy,
Tokuda:2020mlf}, determining the signs of certain linear
combinations of dim-8 coefficients (plus possible dim-6 coefficients at the
squared level).
Since the ultimate goal of the SMEFT is to
determine its UV completion, one should restrict the search for operators
only within these bounds, and optimize the search strategy accordingly.

Alternatively, one might also use these bounds to experimentally
test the fundamental principles of QFT \cite{Distler:2006if,eepaper}.
An even more important application of positivity is to
infer or to exclude possible UV states using precision measurements,
analyzed at the dim-8 level,
in a completely model-independent way \cite{eepaper},
which potentially provides an answer to the ``inverse problem''
\cite{ArkaniHamed:2005px,Dawson:2020oco,Gu:2020thj}.
In particular, if experimental observation continues to agree with the SM,
this allows to set exclusion limits on all UV states up to certain scales,
which cannot be lifted by cancellations among various UV particles \cite{eepaper}.
Dim-8 operators and their positivity nature are thus crucial for SM tests.  In
either case, as the LHC has started to probe the dim-8 SMEFT operators in many
occasions
\cite{Ellis:2018cos,Alioli:2020kez,Sirunyan:2019der,CMS:2020meo,Sirunyan:2020tlu},
and more opportunities can be foreseen at the future lepton colliders
\cite{Ellis:2019zex,Ellis:2020ljj,eepaper},
it has become increasingly important to understand the positivity bounds on
their coefficients.

In SMEFT, there are at least two approaches to derive positivity bounds, which
we dub ``elastic positivity bounds'' and ``extremal positivity bounds'' in this
paper.  The first is the conventional approach which makes use of the {\it
elastic} 2-to-2 forward scattering amplitude. Using analyticity of the
amplitude, the Froissart bound and the optical theorem, one can show that its
second order derivative is positive:
\begin{flalign}
	M^{ij}\equiv \frac{d^2}{ds^2}\bar M(ij\to ij)(s,t=0)\ge0
	\label{eq:def1}
\end{flalign}
where $\bar M(ij\to ij)$ is the elastic scattering amplitude between states
$\ket{i}$ and $\ket{j}$, with poles subtracted (to be explained later) and
$s,t$ are the standard Mandelstam variables.  The $i,j$ indices label the SM modes.
Since $M^{ij}$ involves only low-energy physics, it can be computed within  the
SMEFT.  This leads to, at the tree level, a set of linear homogeneous
inequalities for dim-8 coefficients $C_\alpha$
\cite{Zhang:2018shp, Bi:2019phv, Remmen:2019cyz}:
\begin{flalign}
	\sum_\alpha C_\alpha p_\alpha^{ij}\ge0
\end{flalign}
where $p_\alpha^{ij}$ only involve SM parameters. These are exactly a set of
positivity bounds on SMEFT Wilson coefficients. 
 While the squared contribution of the dim-6 coefficients may also enter the l.h.s.,
in this work we will mainly focus on dim-8 coefficients, for reasons that will be
discussed later. We will, however, investigate the impact of the dim-6 coefficients
in Section~\ref{sec:dim6}.

The above results depend on the choice of basis for particle states.
While physics is independent of the basis for states, the notion of elasticity
is not. The elastic amplitude in one basis (e.g.~the mass-eigenstate basis)
may involve non-elastic amplitude components when transformed to a different
basis (e.g.~the gauge-eigenstate basis). To maximize the constraining power,
one should consider all basis, or equivalently,
consider the elastic scattering of arbitrarily superposed states, e.g.~the
scattering of $u^i\ket{i}$ and $v^j\ket{j}$ states, with $u,v$ arbitrary complex
vectors. This leads to the following infinite set of bounds:
\begin{flalign}
	\sum_\alpha C_\alpha p_\alpha(u,v)\ge0
	\label{eq:13}
\end{flalign}
where $u,v$ are arbitrary complex vectors, and $p_\alpha(u,v)$ are quartic
polynomials of $u,v,u^*,v^*$.  $C_\alpha$ needs to satisfy a number of inequalities for
the above to hold for all possible values of $u,v$, and it is this set of
inequalities that we call elastic positivity bounds.  Since all superpositions
are explored, the full set of elastic positivity bounds is basis-independent,
and so one can start from an arbitrary particle basis.  While this approach is
convenient when the number of particle states (or the dimension of $u,v$) is
small, identifying the full set of bounds can quickly become difficult as the
number of states increases. One way to see this is that Eq.~(\ref{eq:13}) is
equivalent to the determination of the positive semi-definiteness of a
quartic polynomial of $u,v$, which is a NP-hard problem.

The second approach that we will discuss in this work is the extremal
positivity approach that has been recently proposed by some of the
authors~\cite{Zhang:2020jyn}.  This approach has the advantage that one is
guaranteed to obtain the best bounds allowed by the fundamental QFT principles.
Indeed, bounds tighter than the elastic positivity Eq.~(\ref{eq:13}) can
potentially be obtained, and an explicit example has been presented in
\cite{Zhang:2020jyn}.  In this approach, instead of using elastic channels to
probe the bounds, which are the boundary/facets of the allowed parameter space,
one first constructs the edges, or the {\it extremal rays} (ERs), of the
allowed space. The convex hull of these rays determines the bounds.  This is
efficient because the ERs can be directly written down via group theoretical
considerations.  The approach essentially describes the allowed parameter space
as a convex cone via the {\it extremal representation} of cones, and thus its
name.  When the cones are polyhedral, positivity bounds are the facets of the
cones, and can be identified through a vertex enumeration algorithm.

The systematic application of both approaches to the SMEFT dim-8 operators is
currently very limited. Positivity bounds in most literature come from the
elastic positivity approach, exploiting no or only limited superpositions of
states.  A particular interesting topic is the positivity bounds for anomalous
quartic gauge-boson couplings (aQGC), as these couplings are currently being
used as a theory framework to interpret the vector boson scattering (VBS)
and the tri-boson production measurements at the LHC \cite{Rauch:2016pai}.  In
Ref.~\cite{Zhang:2018shp,Bi:2019phv}, following the elastic
positivity approach, we worked in the mass eigenstate basis (i.e.~in the broken
phase of the SM electroweak symmetry),
but only considered the superposition of various polarization states. The
mixing between different gauge components and between $W$ and $B$ were not
considered.  This already significantly constrains the physical space of
the aQGCs,
with only about 2\% of the total space satisfying the bounds.\footnote{This
	2\% is computed without including the operators $O_{T,10}$ and $O_{T,11}$,
	which have been missing in the conventional aQGC parameterization,
	see Section~\ref{sec:smeft}.}  In
addition, the authors of \cite{Remmen:2019cyz} worked in the unbroken phase,
and considered the superposition of Higgs and Goldstone bosons but not
that of the gauge modes. This gives a set of complementary constraints. 
On the other hand, the extremal representation approach has been so far only
considered in Ref.~\cite{Zhang:2020jyn}, where simple examples have already
shown that it is efficient and powerful at least for cases in which low energy
modes lie within the same irreducible representation (irrep) (such as for the
Higgs doublet or the $W$-boson triplet). More general applications of this
approach are yet to be explored.

In this work, we apply both approaches---the elastic positivity from all
superposed states, and the extremal representation approach---to the set of the
dim-8 operators that parameterize the transversal aQGCs, i.e.~the couplings of
four $W$-bosons, of four $B$-bosons, and of two $W$- and two $B$-bosons.  Our goal
is three-fold:
\begin{enumerate}
	\item We will use the aQGC operators as a realistic case, to establish
		the methodology for both approaches.  We focus on the potential
		difficulties that arise when one aims to extract the best
		positivity constraints that involve a large number of particle
		states. For the elastic approach, the main difficulty is that
		all possible superpositions of the following modes need to be
		explored:
\begin{flalign}
W_x^1,W_y^1,W_x^2,W_y^2,W_x^3,W_y^3,B_x,B_y
\end{flalign}
where the subscripts $x,y$ for $W,B$ are the transversally polarized modes, 
and the superscripts of $W$ are the $SU(2)$ index in the $\mathbf{3}$
representation.  The $u,v$ vectors are thus elements of $\mb C^{8}$, and
together they have 32 real degrees of freedom. One then needs to exhaust all
possible directions in a 32 dimensional space, to
check if a given set of coefficients can be excluded by some $u,v$ vectors.
Alternatively, for the extremal approach, the main difficulty
is that when both $W$ and $B$ are involved, degenerate irreps show up, and so
there are multiple ERs that are continuously parameterized by some real
parameters. One needs to find a way to identify the boundary spanned by these
``continuous rays''. In both cases, we will discuss how to deal with the
difficulties and to solve the problem both analytically and numerically. The
latter allows us to quantify the accuracy of any approximations we will use in
the analytical approach.

\item We will compare several aspects of both approaches, to understand their
	advantages and disadvantages. In terms of the final
	results, we know that the extremal approach gives tighter constraints,
	and we will quantify the actual improvement in this realistic case.
	Another important aspect to be compared, is to what extend these
	approaches can be implemented algorithmically, and thus made applicable
	to more general and complicated problems (e.g.~full set of bounds for
	the entire SMEFT).  Finally, the speed and accuracy of the numerical
	approaches are also important.

\item We will obtain the best positivity bounds on the transversal aQGCs, which
	alone are a very important physics result.  Searching for possible
	beyond the SM physics in the form of aQGC is one of the main goals of
	the current electroweak program at the LHC. These couplings can be
	measured in the VBS or the tri-boson production channels. Knowing their
	bounds from positivity will undoubtedly provide guidance for future
	theoretical and experimental studies.  Existing results
	\cite{Zhang:2018shp,Bi:2019phv,Remmen:2019cyz} followed the elastic
	approach and only explored limited superpositions. In this work we will
	unify and supersede these results, by exploring the elastic channels of
	all possible superposed states.  On the other hand, the extremal
	positivity approach has only been considered for 4-$W$-boson operators,
	but it already gives better bounds than the conventional elastic
	approach \cite{Zhang:2020jyn}. We will show that this continues to hold
	when both $W$ and $B$-bosons are taken into account.
\end{enumerate}

The paper is organized as follows. 
In Section~\ref{sec:smeft}, we will list the full set of effective operators that
are relevant in this study.
In Section~\ref{sec:positivity}, we will present the theoretical basis of this work -- the two approaches to positivity bounds.
In particular, in Section~\ref{sec:geometry} we will present a geometric interpretation
for the positivity problem, based on which we will explain the elastic positivity approach in
Section~\ref{sec:elasticapproach} and the extremal positivity approach in
Section~\ref{sec:extremalRep}, using 4-Higgs and 4-$W$ operators as toy examples.  
In Section~\ref{sec:elastic} and Section~\ref{sec:extremal},
we will tackle the problem using the elastic approach and the extremal
approach, respectively.  In both cases we will present analytical and numerical
results. 
The quadratic dim-6 contributions will be discussed in Section~\ref{sec:dim6}.
Finally, we will discuss all the results and compare the two approaches
in Section~\ref{sec:discussion}, and summarize in Section~\ref{sec:summary}.
For readers who are mainly interested in the physics results, i.e.~positivity bounds
on aQGC coefficients, our best analytical results are given in
Eqs.~\eqref{eq:erbound1}-\eqref{eq:erbound2}.

\section{Effective operators}
\label{sec:smeft}

The aQGCs already appear at dim-6 in the SMEFT. They are however not independent
of the triple gauge-boson couplings at dim-6, which are often assumed
to be {severely} constrained through other channels such as the di-boson production. 
Dim-8 Wilson coefficients are thus
used to parameterize the truly independent aQGCs, which are measured in the
VBS and tri-boson processes. They are also used to cover all possible helicity combinations
involved, and to account for cases in which the dominant effects are beyond
dim-6, possibly due to loop-suppression at dim-6\footnote{ There are three
independent degrees of freedom at dim-6, two of which can only be generated at
the loop level, in a weakly coupled UV completion.} or the helicity selection
rule \cite{Azatov:2016sqh}. Experimental studies have been extensively
performed, by both the ATLAS and the CMS collaborations at the LHC, to set
limits on the sizes of aQGCs, and we refer to
\cite{Sirunyan:2019der,CMS:2020meo,Sirunyan:2020tlu} for some recent
progresses.  A compilation of current limits on aQGC can be found in
\cite{cmstwiki}. The HL-LHC projection for the dim-8 aQGC
operator sensitivities can reach the TeV level and beyond \cite{Azzi:2019yne}.

The dim-8 aQGC operators are often categorized in three different
types: the $S$-type, $M$-type, and $T$-type
operators~\cite{Eboli:2006wa,Degrande:2013rea,Eboli:2016kko}.  The $S$-type
operators have a schematic form of $(D\Phi)^4$, where $D\Phi$ is the covariant
derivative of the Higgs doublet.  The $M$-type operators involve two covariant
derivatives of the Higgs doublet and two field strengths, schematically forming
$(D\Phi)^2F^2$.  The $T$-type operators are constructed by four field strength
tensors, $F^4$. 

In this paper, we focus on the $T$-type operators, which
involve four transversal gauge modes. There are 10 such operators.
Using the convention in \cite{Degrande:2013rea}, we define
\begin{equation}
\hat{W}^{\mu\nu} \equiv i g \frac{\sigma^{I}}{2} W^{{I}, \mu\nu},
\, \, \, \hat{B}^{\mu\nu} \equiv i g' \frac{1}{2} B^{\mu\nu}
\label{eq:WandBhat}
\end{equation}
and
\begin{equation}
\tilde{W}_{\mu\nu} \equiv i g \frac{\sigma^{I}}{2} \left(\frac{1}{2} \epsilon_{\mu\nu\rho\sigma} W^{{I}, \rho\sigma}\right),
\tilde{B}_{\mu\nu} \equiv i g' \frac{1}{2}\left(\frac{1}{2} \epsilon_{\mu\nu\rho\sigma} B^{\rho\sigma}\right),
\label{eq:dual}
\end{equation}
where $ \epsilon_{\mu\nu\rho\sigma}$ is the Levi-Civita tensor,
\begin{equation}
W^{I}_{\mu\nu} = \partial_\mu W^{I}_\nu - \partial_\nu W^{I}_\mu
+ g\epsilon_{{IJK}}W^{J}_{\mu}W^{K}_\nu,
\, \, \, B_{\mu\nu}= \partial_\mu B_\nu - \partial_\nu B_\mu,
\label{eq:WandB}
\end{equation}
and $g$ and $g'$ are $SU(2)_L$ and $U(1)_Y$ gauge couplings, respectively.
With these notations, the $T$-type dim-8 aQGC operators are listed in
Table~\ref{tab:f4_operators}.
Note that two operators, $O_{T,10}$ and $O_{T,11}$, are included in addition to
the conventional $T$-type aQGC operators.
These two operators are conventionally missed in the set of aQGC operators, and
this has been pointed out recently by Ref.~\cite{Remmen:2019cyz}. They are also
included in Refs.~\cite{Li:2020gnx,Murphy:2020rsh}.  

We do not consider parity-odd operators.  The effective Lagrangian for the
aQGC interactions is:
\begin{align}
	\mathcal{L}_{aQGC} =  \sum_i \frac{F_{T,i}}{\Lambda^4}\mathcal{O}_{T,i} \label{eq:lag},
\end{align}
where $F_{T,i}$ is the Wilson coefficient for the corresponding operator
$\mathcal{O}_{T,i}$, and $\Lambda$ is the characteristic scale of new physics.

The scattering amplitude of two transversal gauge bosons could also receive
contributions at the $s^2/\Lambda^4$ level from the dim-6 operators. These
contributions are proportional to the squares of the dim-6 Wilson coefficients.
They come from diagrams in which triple-gauge-boson couplings (or potentially
the $HVV$ couplings) can be inserted twice. In the Warsaw basis
\cite{Grzadkowski:2010es}, such a contribution only comes from the $O_W$
operator:
\begin{flalign}
	O_W=\varepsilon^{IJK}W_\mu^{I\nu}W_\nu^{J\rho}W_\rho^{K\mu}  .
\end{flalign}
We define $\bar a_W\equiv a_W/g^2$ for convenience, where $a_W$ is the
coefficient of this operator.

Finally, all scattering amplitudes used in this paper are computed at the tree
level using standard
tools~\cite{Christensen:2008py,Alloul:2013bka,Hahn:2000kx,Hahn:1998yk}.

\begin{table}[t]
\center
\begin{tabular}{ll}
$O_{T,0} = \mathrm{Tr}[\hat{W}_{\mu\nu}\hat{W}^{\mu\nu}] \mathrm{Tr}[\hat{W}_{\alpha\beta}\hat{W}^{\alpha\beta}]$
&$O_{T,1} = \mathrm{Tr}[\hat{W}_{\alpha\nu}\hat{W}^{\mu\beta}] \mathrm{Tr}[\hat{W}_{\mu\beta}\hat{W}^{\alpha\nu}]$\\
$O_{T,2} = \mathrm{Tr}[\hat{W}_{\alpha\mu}\hat{W}^{\mu\beta}] \mathrm{Tr}[\hat{W}_{\beta\nu}\hat{W}^{\nu\alpha}]$
& $O_{T,10} = \mathrm{Tr}[\hat{W}_{\mu\nu}\tilde{W}^{\mu\nu}] \mathrm{Tr}[\hat{W}_{\alpha\beta}\tilde{W}^{\alpha\beta}]$\\
\\%
$O_{T,5} = \mathrm{Tr}[\hat{W}_{\mu\nu}\hat{W}^{\mu\nu}]\hat{B}_{\alpha\beta}\hat{B}^{\alpha\beta}$
& $O_{T,6} = \mathrm{Tr}[\hat{W}_{\alpha\nu}\hat{W}^{\mu\beta}]\hat{B}_{\mu\beta}\hat{B}^{\alpha\nu}$\\
$O_{T,7} = \mathrm{Tr}[\hat{W}_{\alpha\mu}\hat{W}^{\mu\beta}]\hat{B}_{\beta\nu}\hat{B}^{\nu\alpha}$
&$O_{T,11} =  \mathrm{Tr}[\hat{W}_{\mu\nu}\tilde{W}^{\mu\nu}]\hat{B}_{\alpha\beta}\tilde{B}^{\alpha\beta}$\\
\\%
$O_{T,8} = \hat{B}_{\mu\nu}\hat{B}^{\mu\nu}\hat{B}_{\alpha\beta}\hat{B}^{\alpha\beta}$
& $O_{T,9} = \hat{B}_{\alpha\mu}\hat{B}^{\mu\beta}\hat{B}_{\beta\nu}\hat{B}^{\nu\alpha}$
\end{tabular}
\caption{$T$-type aQGC operators}
\label{tab:f4_operators}
\end{table}

\section{Theoretical framework}
\label{sec:positivity}

The positivity bounds derived from forward scattering amplitudes have been well established
and widely used, for example, in Refs.~\cite{Adams:2006sv,Bellazzini:2015cra}. A
generalization to the non-forward case has been discussed in
Refs.~\cite{deRham:2018qqo,deRham:2017zjm,Manohar:2008tc,Bellazzini:2016xrt,nimahuanghuang}.
{However, in the SMEFT, non-forward scatterings at the leading order} do not
lead to new bounds on the $s^2$ dependence of the amplitudes. In this work we
thus focus on the forward case.

In this section, we will first derive a dispersion relation for general
non-elastic forward scatterings. We will then present a geometric
interpretation for the parameter space allowed by this dispersion relation, and
show that positivity bounds arise from the latter.  We will show that both the elastic
approach and the extremal approach can be easily understood from this
geometric picture.  Toy examples, including four-Higgs operators and four-$W$ operators,
will be discussed to demonstrate
how the full set of bounds can be extracted following both approaches.
Finally, for the elastic approach, we will also show that the superposition of low
energy modes only needs to take real values. This is an important simplification for
the elastic approach.

\subsection{Dispersion relation}
\label{sec:dispersion}

Positivity bounds arise from the dispersion relation,
which has been extensively discussed in the literature (see
e.g.~Refs.~\cite{Bi:2019phv, Zhang:2020jyn}).
Here we only give a brief outline, skipping unnecessary details. The main difference
is that here we consider inelastic scattering amplitudes.

Let us denote $\tilde M^{ijkl}$ as the second-order $s$ derivative of the
scattering amplitude $ij\to kl$ with low-energy poles subtracted
\begin{equation}
\tilde M^{ijkl} =  \frac{1}{2} \frac{{d}^{2}}{{d} s^{2}} {M}({i j \rightarrow k l})\left(s=\frac{M^{2}}2,t\right)+c . c . 
\end{equation}
where $c.c.$ stands for the complex conjugate of the previous term. $i,j,k,l$ indices
run through either a subset or the full set of SM modes.  Using the
analyticity of the amplitude and the Froissart-Martin bound, a contour integral
around a low energy point $M^2/2$ can be deformed to go around the branch cuts
and the infinity, which leads to
\begin{flalign}
	\tilde M^{ijkl}=&\int_{M_\mathrm{th}^2}^{\infty}\frac{ds}{2i\pi}
	\frac{\mathrm{Disc} M(ij\to kl)(s,t)}{(s-\frac{1}{2}M^2)^3}+
	\int^{-(M^u_{\rm th})^2}_{-\infty}\frac{ds}{2i\pi}
	\frac{\mathrm{Disc} M(ij\to kl)(s,t)}{(s-\frac{1}{2}M^2)^3}
+c.c.
\end{flalign}
where $M_\mathrm{th}$ and $M^u_{\rm th}$ are the $s$ and $u$ channel threshold scale for process $ij\to kl$, and $M^2\equiv m_i^2+m_j^2+m_k^2+m_l^2$.
We can calculate the amplitude to a desired accuracy within an EFT up to
energy scale $\epsilon\Lambda$ (with $\epsilon\lesssim 1$), so we can subtract
the dispersive integral up to $\epsilon\Lambda$, and get
\begin{flalign}
	 M^{ijkl}=&\int_{(\epsilon\Lambda)^2}^{\infty}\frac{ds}{2i\pi}
	\frac{\mathrm{Disc} M(ij\to kl)(s,t)}{(s-\frac{1}{2}M^2)^3}+
	\int^{-(\epsilon\Lambda)^2}_{-\infty}\frac{ds}{2i\pi}
	\frac{\mathrm{Disc} M(ij\to kl)(s,t)}{(s-\frac{1}{2}M^2)^3}
+c.c.
\end{flalign}

For inelastic scattering between particles with different masses, the forward
limit $\theta=0$ is generally not the same as the limit $t=0$. However, since
the integral starts from $\epsilon\Lambda\gg m_i$, we can take the
approximation $s\gg m_i^2$, in which the forward limit $\theta=0$ becomes the
same as $t=0$ and we can make use of the simple crossing relation ${\rm Disc}
M(ij\to kl)(s,0)\simeq -{\rm Disc} M(il\to kj)(s,0)$. Also, by the same token,
we approximate the denominator $s-M^2/2
\simeq s$. (Note that $M^2$ depends on the indices $i,j,k,l$.) 
These approximations are consistent with $M^2\ll \Lambda^2$, which is the condition
for a SMEFT expansion to be valid. Any deviations from them are a higher order
effect in the SMEFT.
We then have
\begin{flalign}
\label{disrel2}
	M^{ijkl}=&\int_{(\epsilon\Lambda)^2}^{\infty}\frac{ds}{2i\pi}
	\frac{\mathrm{Disc}M(ij\to kl)(s)}{s^3}+(j\leftrightarrow l)+c.c.
\end{flalign}
Thanks to Hermitian analyticity $M(kl\to ij)^*(s+i\varepsilon) =
M(ij\to kl)(s-i\varepsilon)$ and the (generalized) optical theorem $M(ij\to kl)
- M(kl\to ij)^*= i\sum'_X M(ij\to X) M(kl\to X)^*$, where
$\sum'_X$ denotes the sum of all possible states $X$ together with their
phase space integration, we can re-write the dispersion relation as
\begin{flalign}
	M^{ijkl}=  \int_{(\epsilon\Lambda)^2}^{\infty} \sum_{X}\!{}'
	\!\sum_{K=R,I}  \!\frac{d \mu \, {m_K}^{ij}_X {m_K}^{kl}_X }{\pi
		\mu^3} 	  
 +(j\leftrightarrow l) 
 \label{eq:1}
\end{flalign}
where we have defined $M(ij\to X)\equiv m_{R_X}^{ij}+im_{I_X}^{ij}$,
with $m_{R_X}$ and $m_{I_X}$ both real matrices.
Eq.~\eqref{eq:1} is the master equation on which all positivity bounds discussed
in this work are based.

In a model-independent EFT, we shall make no assumption on $m_{K_X}^{ij}$, and
so a priori they can be arbitrary matrix functions of $s$ and the phase
space of $X$. However, since all other factors in the integrand are
non-negative, the dispersion relation implies that $M^{ijkl}$ is not
allowed to take arbitrary values, i.e.~there are bounds on
$M^{ijkl}$.
The easiest way to see this is to consider an elastic scattering amplitude.
When $i=k$ and $j=l$, the integrand of the r.h.s.~is a sum of squares and hence
positive semi-definite (PSD). We have
\begin{flalign}
	M^{ijij}\ge0
\end{flalign}
i.e.~the 2nd $s$ derivative of the (subtracted) elastic scattering amplitude is
non-negative.

More generally, consider two arbitrary vectors $u,v\in \mb C^n$, where $n$ is
the number of low energy modes being considered. Contracting Eq.~(\ref{eq:1}) with
$u^iv^ju^{*k}v^{*l}$, we find
\begin{flalign}
	u^iv^ju^{*k}v^{*l}M^{ijkl}=  \int_{(\epsilon\Lambda)^2}^{\infty} \sum_{X}\!{}' 
	\!\sum_{K=R,I}  \!\frac{d \mu }{\pi \mu^3} 	  
		\left[
			\left| u\cdot m_{K_X}\cdot v \right|^2
		       +\left| u\cdot m_{K_X}\cdot v^* \right|^2
		\right]\ge0
	\label{eq:QQ}
\end{flalign}
where the summation over repeated $i,j,k,l$ indices is omitted.
Defining a quartic polynomial
\begin{flalign}
	P(u,v)\equiv u^iv^ju^{*k}v^{*l}M^{ijkl},
	\label{eq:elasticpositivity}
\end{flalign}
we find that the dispersion relation implies $P(u,v)\ge0\ \forall u,v\in \mb C^n$.
In other words, $P(u,v)$ is a quartic PSD polynomial on $\mb C^{2n}$.  

The above statement can be thought of as coming from the 2nd order derivative
of the elastic channel $M(ab\to ab)$, where the states $a,b$ are superposed states
characterized by the $u,v$ vectors:
\begin{flalign}
	\ket a=u^i\ket i,\quad \ket b=v^i\ket i
\end{flalign}
We have essentially shown that positivity bounds can be derived
from the elastic scattering of any superposed states that mix 
possibly different helicities and other quantum numbers, as the fact that
$u^iv^ju^{*k}v^{*l}M^{ijkl}\ge0$ only requires Eq.~(\ref{eq:1}). 

In the SMEFT, $M^{ijkl}$ can be computed in terms of higher-dimensional
operators. At the tree level, since $M^{ijkl}$ is defined as the 2nd order $s$
derivative, its leading contribution comes from either a subset of the dim-8
Wilson coefficients (which give rise to $s^2$ dependence in the amplitude, see
\cite{Remmen:2019cyz}) at the linear level, or, potentially, also from
dim-6 coefficients but at the squared level.  Neglecting the latter, we can write
\begin{flalign}
M^{ijkl}=\Lambda^{-4} C_\alpha M^{ijkl}_\alpha
	\label{eq:tree}
\end{flalign}
where $C_\alpha$'s are the dim-8 coefficients and the summation over $\alpha$ is implicit. Defining the 4-th order polynomials
\begin{flalign}
	p_\alpha(u,v)\equiv M^{ijkl}_\alpha u^iv^ju^{*k}v^{*l}
\end{flalign}
the elastic positivity then simply requires $C_\alpha p_\alpha(u,v)\ge 0$.
This condition constrains the possible values of $C_\alpha$.  
While plugging any given $u,v$ vectors into this relation will lead to a
valid bound, our goal, however, is to derive the full set of necessary and sufficient
conditions for:
\begin{flalign}
	C_\alpha p_\alpha(u,v)\ge 0,\quad \forall\ u,v\in \mb C^n
	\label{eq:psdpoly}
\end{flalign}
which is independent of the basis of particle states.  This problem is equivalent to
the determination of a PSD quartic polynomial, which is a NP-hard problem.  Its
difficulty grows with $n$, the number of modes. In
Section~\ref{sec:elasticapproach}, we will show that one in fact only needs to
consider $C_\alpha p_\alpha(u,v)\ge 0$ for real $u,v$. This is an extremely
useful simplification, because it halves the number of variables to be
considered.

Eq.~\eqref{eq:psdpoly} is exactly how we extract the elastic positivity bounds.
As for the extremal positivity approach, we postpone the discussion
to Section~\ref{sec:extremalRep}, which will be more intuitive based
on the geometric picture that we will introduce in the next section.

Before proceeding, let us briefly comment on other contributions to $M^{ijkl}$
that might arise in the SMEFT.  First, we have so far neglected the quadratic
contribution from the dim-6 coefficients. In practice, this is a useful
simplification as it leads to compact results on the dim-8 aQGC operators.
Furthermore, these results are still valid in general: as we have shown in
Ref.~\cite{Bi:2019phv}, the removal of dim-6 contributions only makes these
bounds more conservative, at least in the elastic positivity approach.  On the
other hand, including these contributions is also straightforward with all the
methods we will demonstrate in this work. For simplicity, we will postpone a
discussion about the dim-6 contributions to Section~\ref{sec:dim6}, while for
the rest of the paper we will not consider such contributions, and aim at
deriving the full set of bounds for the dim-8 coefficients.  On a different ground,
in Section~\ref{sec:dim6}, we will demonstrate that, even in the extremal
positivity approach, neglecting dim-6 contributions is a safe approximation, as
it only makes the bounds more conservative.

In addition, beyond the tree level, Eq.~(\ref{eq:tree}) needs to be augmented
with loop corrections,  from the SM and even dim-6 operators. Some of these
effects are discussed in Ref.~\cite{Bi:2019phv}. For example, the SM
corrections are shown to be not important given the current experimental
sensitivity, but they might play a role in the future. In any case, positivity
bounds should be viewed as bounds on the amplitude $M^{ijkl}$: the latter can
be mapped to the SMEFT coefficients to any desired order by an explicit SMEFT
calculation, but such a calculation beyond the tree level is by itself a
separate and nontrivial problem.  Since the goal of this work is to study the
methodology for extracting bounds, in the rest of the paper we will only use
tree level expressions for $M^{ijkl}$, namely Eq.~(\ref{eq:tree}). If loop-level
amplitudes are available in the SMEFT, our approaches can be easily improved by
a loop-level matching from $M^{ijkl}$ to the Wilson coefficients.

\subsection{A geometric interpretation of positivity}
\label{sec:geometry}

A convex geometric perspective can provide useful guidance for the extraction
of bounds. Let us first introduce a few basic concepts and facts about convex
geometry. 
\begin{itemize}
\item A {\it convex cone} (or simply a {\it cone}) is a subset of some vector space,
  closed under additions and positive scalar multiplications.
  A {\it salient} cone is a cone that does not contain any straight line. In other
  words, if $\mc C$ is salient, then having $x\in \mc C$ and $-x\in \mc C$ implies
  $x=0$.
 \item  An {\it extremal ray} (ER) of a cone $\mc C_0$ is an element
  $x \in\mc C_0$ that cannot be a sum of two other elements in
  $\mc C_0$. If we write an extremal ray as $x=y_1+y_2$ with $y_1,y_2\in
  \mc C_0$, then we must have $x=\lambda y_1$ or $x=\lambda y_2$,
  with $\lambda$ some a real constant. The extremal rays of a
  polyhedral cone are its edges.
 \item The {\it convex hull} of a given set $\mc X$ is the set of all
  convex combinations of points in $\mc X$, where a convex combination is
  defined as a linear combination of points, where all the combination
  coefficients are non-negative and sum up to 1.
 \item The {\it conical hull} of a given set $\mc X$ is the set of all
  positive linear combinations of elements in $\mc X$, denoted by cone($\mc X$).
  Obviously, cone($\mc X$) is a convex cone, and its extremal rays 
  are a subset of $\mc X$.
  \item The {\it dual cone} $\mc C_0^*$ of the cone $\mc C_0$ is the set
	$\mc C_0^*\equiv\left\{y\ | \ y\cdot x\ge0,\ \forall x\in \mc C_0\right\}$,
	where $\cdot$ represents inner product.  We have $(\mc C_0^*)^*=\mc C_0$, and if
	$\mc C_1\subset \mc C_2$, then $\mc C_1^*\supset \mc C_2^*$.
\end{itemize}

With these in mind, we observe from Eq.~(\ref{eq:1}) that the dispersion
relation implies that $M^{ijkl}$, viewed as a vector, live in a convex
cone. To see this, note that the integration is a limit of summation
 and all other factors in the integrand are positive, so $M^{ijkl}$ is a
 positively weighted sum of $m^{ij}m^{kl}+m^{il}m^{kj}$, where $m^{ij}$ is a
 $n\times n$ matrix, with $n$ the number of low energy particles in the EFT.
 Since we do not assume any specific UV completion, $m^{ij}$ is arbitrary. 
Therefore all possible values of $M^{ijkl}$ must live in the following
set $\mc C$:
\begin{flalign}
	\mc C\equiv \mbox{cone}\left(\left\{m^{ij}m^{kl}+m^{il}m^{kj}\right\}\right)
	\label{eq:cone}
\end{flalign}

An immediate important fact is that $\mc C$ is a salient cone, i.e.~it does not
contain any straight line. This can be seen by contracting its elements with
$\delta^{ik}\delta^{jl}$, which always gives a positive summation in the
integrand of the r.h.s., $m^{ij}m^{ij}>0$, unless the element itself vanishes.
So if $x$ is in $\mc C$, $-x$ is not. Thus, intuitively, the $\mc C$ cone
constrains
$M^{ijkl}$ to be in certain (positive) directions in the total parameter space,
and this constraint applies to all directions.
In other words, the $\mc C$ cone represents the ``directional information'' of
the amplitude that can be extracted out of the dispersion relation.
In this view, the goal of finding all possible
positivity bounds can be achieved by finding the boundary of $\mc C$.
In the SMEFT, all particles are charged under the SM gauge symmetries, so
$m^{ij}$ is not totally arbitrary but subject to constraints from symmetries.
Nevertheless, Eq.~\eqref{eq:cone} represents all requirements from the
axiomatic principles.

The convex cone $\mc C$ defined by Eq.~(\ref{eq:cone}) can be characterized in
two ways. Positivity bounds are directly related to the {\it inequality}
representation, which follows from the Hahn-Banach separation theorem. The
theorem states that $\mc C$ is an intersection of half-spaces, each described
by a linear inequality. These inequalities are exactly what we call
positivity bounds. Therefore, in this view, our goal is to find the
inequality representation of $\mc C$. Alternatively,  
the {\it extremal} representation follows from the Krein-Milman theorem, which
implies that a {\it salient cone} $\mc C$ is a {convex} hull of its ERs.  The
extremal approach proposed in Ref.~\cite{Zhang:2020jyn} simply follows
this representation, and it determines $\mc C$ by first finding its ERs. 

The two representations of a convex cone are connected by the concept of dual cones:
the bounds of $\mc C$ are the ERs of $\mc C^*$, and vise versa. In other words,
the inequality representation of $\mc C$ is the extremal representation of $\mc
C^*$.  To see this, note that $(\mc C^*)^*=\mc C$ implies $\mc C$ is described
by a set of inequalities,
\begin{flalign}
	\mc C=\left\{M\ |\ M\cdot T\ge0,\ \forall T\in \mc C^*\right\}
	\label{eq:dep}
\end{flalign}
The extremal representation of $\mc C^*$ allows any element $T\in \mc C^*$ to
be written as a positively weighted sum of $E_i\in\mc E$, where $\mc E$ is the
set of ERs in $\mc C^*$, so the above is equivalent to
\begin{flalign}
	\mc C=\left\{M\ |\ M\cdot E\ge0,\ \forall E\in \mc E\right\}
	\label{eq:nondep}
\end{flalign}
This defines $\mc C$ as a set of bounds, each represented by an
ER of $\mc C^*$.  This is exactly the inequality representation of $\mc C$ that
we are looking for. 

Note that although Eqs.~\eqref{eq:dep} and \eqref{eq:nondep} describe
the same cone $\mc C$, Eq.~\eqref{eq:dep} consists of an infinite number
of inequalities, most of which are redundant, as $M\cdot T\ge0$ is guaranteed by
$M\cdot E\ge 0$. In contrast, Eq.~\eqref{eq:nondep} is the complete and independent
set of inequalities that are required to describe $\mc C$.
Thus our goal is to find $\mc E$, the set of all ERs of $\mc
C^*$.  In fact, the idea of the elastic positivity approach is to first construct
explicitly a subset of $\mc C^*$ (see Eq.~\eqref{eq:Q}), and identify its ERs.
This leads to a set of conservative bounds, which we denote by $\mc C^{el}$.
It is conservative because ${\mc C^{el}}^*\subset \mc C^*$ implies $\mc
C^{el}\supset \mc C$.

On the other hand, in the extremal approach, one first finds the ERs of $\mc
C$, $\{e_i\}$.  This can be greatly simplified by taking into account the SM
symmetries.  One can essentially replace the $m^{ij}m^{kl}$ tensor with the projector
operators of the irreps within which the intermediate states are charged, and
rewrite the dispersion relation \cite{Zhang:2020jyn}
\newcommand{\bfr} {{\bf r}}
\newcommand{\ud} {\mathrm{d}}
\begin{flalign}
	M^{ijkl}= \int_{(\epsilon\Lambda)^2}^{\infty}\ud\mu
	{\sum_{X\text{ in }\mathbf{r}}}'
	\frac{|\braket{{X}|\mathcal{M}|\mathbf{r}}
	|^2 }{\pi  \mu^3}
	P_\mathbf{r}^{i(j|k|l)}
	\label{eq:2}
\end{flalign}
where $\bfr$ runs through all irreps of the $SO(2)$ rotation around the forward
scattering axis and the $SU(3)_C\times SU(2)_L\times U(1)_Y$ symmetries; ${ X}$
runs through all intermediate states in the irrep $\mathbf r$;
$P_\bfr^{ijkl}\equiv \sum_\alpha
C^{\bfr,\alpha}_{i,j}(C^{\bfr,\alpha}_{k,l})^*$ are the projective operators of
${\bf r}$; $C^{\bfr,\alpha}_{i,j}$ are the Clebsch-Gordan coefficients for the
direct sum decomposition of $\mathbf{r}_i\otimes\mathbf{r}_j$, with
$\mathbf{r}_i(\mathbf{r}_j)$ the irrep of particle $i(j)$, and $\alpha$ the
label of states in $\mathbf{r}$; $i(j|k|l)$ means that the $j,l$ indices are
symmetrized; see \cite{Zhang:2020jyn} for more details.  With this the cone
$\mc C$ becomes
\begin{flalign}
	\mc C \equiv 
\mbox{cone}\left( \left\{P_\mathbf{r}^{i(j|k|l)} \right\} \right)
	\label{eq:cone2}
\end{flalign}
So the ERs are a subset of $\big\{P_\mathbf{r}^{i(j|k|l)} \big\}$, and
can be easily determined. This gives the extremal
representation of $\mc C$.

In order to extract bounds, the next step is to convert the extremal
representation to the
inequality representation. For polyhedral cones, recall that the ERs (edges) of
$\mc C$ are the bounds (facets) of $\mc C^*$ and vice versa. 
This conversion can be done by a procedure called {\it vertex enumeration}: 
it determines the edges of a polyhedral cone ($\mc C^*$) from its facets, and this
is equivalent to determining the facets of $\mc C$ from its edges or ERs.
There are efficient algorithms to perform this computation, such as the reverse
search algorithm of \cite{Avis,lrs}, which allows for easy transformation
between the two representations.  More generally, for non-polyhedral cones with curved
boundaries, one will have to deal with a continuous version of this problem. We
will discuss this in Section~\ref{sec:extremal}.

The symmetries we have considered so far are not always sufficient to fully determine
the dynamics. In the SMEFT, the physical amplitude $M^{ijkl}$ can be expanded by
the operators, $M^{ijkl}=\sum_\alpha C_\alpha M_\alpha^{ijkl}$, where
$C_\alpha$ are the dim-8 Wilson coefficients, and this defines a subspace $\mc S$
of $\{M^{ijkl}\}$ that is allowed by the SMEFT at the tree level.  Elements in $\mc S$ is represented
by a vector of coefficients: $\vec C=(C_1,C_2,\dots)$, and we also define the
vector of amplitudes from individual operators $O_i$, $\vec M=(M_1,M_2,\dots)$,
with superscript $ijkl$ suppressed.  Now $\mc C$ represents the requirement
from the dispersion relation, while $\mc S$ represents the requirement from
 dynamics, so our goal is to identify their intersection, which
is also a convex cone, $\mc C_S=\mc C\cap \mc S$ (or $\mc C^{el}_S=\mc C^{el}
\cap \mc S$). Specifically, elements of
$\mc C_S$ needs to satisfy
\begin{flalign}
	\vec C\cdot \vec M^{ijkl}T^{ijkl}\ge 0,\ \forall T\in \mc C^*
\end{flalign}
This defines $\mc C_S$ as the dual cone of
\begin{flalign}
\label{CsDef}
	\mc C_S^*=\left\{
        \vec M^{ijkl}T^{ijkl}, T\in \mc C^*
	\right\}
\end{flalign}
(Strictly speaking, the elements of the dual of the r.h.s.~are the vectors $\vec C$,
while $\mc C_S$ is a set of $\vec C\cdot\vec M^{ijkl}$, but we consider them to be
equivalent, as $M^{ijkl}_\alpha$ form a linearly independent basis.)
In the elastic approach, it may be more convenient to work directly with $\mc C_S^{el}$,
because its dual can be directly written down by replacing the $T$ in the
above equation by a subset of $\mc C^*$ (see again Eq.~\eqref{eq:Q}).
On the other hand, it is also possible to work with $\mc C$, as the
bounds of $\mc C$ can be easily converted to the bounds of $\mc C_S$, by
a matching calculation of the physical amplitude.

Before concluding this section, let us mention several notations that we use to
denote various sets of bounds.  In general, we use $\mc C$ with
sub/superscripts for the set of dim-8 Wilson coefficients allowed by some set
of bounds.  Our goal is to identify $\mc C_S=\mc C\cap \mc S$.
Without any approximations, the extremal positivity
gives $\mc C_S$, while elastic positivity relaxes it to $\mc C^{el}_S$.
Due to the complexity of the problem, our analytical and numerical
results in both approaches are sometimes based on approximations, 
and are in general not exactly the same as $\mc C_{S}$ or $\mc C_{S}^{el}$.
We will denote the analytical and numerical results as, say, $\mc C_{A}^{el}$
and $\mc C_{N}^{el}$ respectively.

To compare the constraining power of various sets of bounds, and to assess how
accurately they describe the exact parameter space, one notion that will be
useful in this work is the ``solid angle'' of the parameter space allowed by
some set of bounds, say $\mc C_S$.  In general, to quantify the constraining
power, one can use the volume of the allowed region of the parameter space. The
positivity bounds themselves are however projective, as $\mc C_S$ (as well as
other bounds) is closed under multiplication of a global real and positive
number. Therefore the volume is infinity, but instead,
one can define the ``solid angle'' of the cone $\mc C_S$, $\Omega(\mc C_S)$,
normalized to that of the full parameter space.  

To compute $\Omega(\mc C_S)$,
an efficient way is to sample the Wilson coefficients, say the 10 aQGC coefficients
$F_{T,i}$ with $i = 0, 1, 2, 5, 6, 7, 8, 9, 10, 11$, on a 10-sphere, and count the number of points
that satisfy all bounds.  To sample $F_{T,i}$ uniformly on a $n$-sphere, it is
sufficient to
simply let $F_{T,i}$ sample the standard normal distribution, thanks to the
Muller method and that the bounds are projective.  The Monte Carlo sampling
error can be estimated by the square root of the variance. Suppose one uniformly
samples $N$ points in the 10-sphere and $N_{\mc C_S}$ points satisfy positivity.
Our solid angle is defined by $\Omega(\mc C_S) = N_{\mc C_S}/N$, and its error can be
estimated by the square root of the sampling variance of $\Omega(\mc C_S)$,
which is 
\begin{equation}
\frac{1}{\sqrt{N}}  \sqrt{\frac{[N_{\mc C_S} (1-\Omega(\mc C_S))^2+(N-N_{\mc C_S})(0-\Omega(\mc C_S))^2]}{N-1}}\simeq \sqrt{\frac{(1-\Omega(\mc C_S))\Omega(\mc C_S)}{N}} \simeq \frac{\Omega(\mc C_S)}{\sqrt{\Omega(\mc C_S) N}}
\end{equation}
where in the last $\simeq$ we have assumed $\Omega(\mc C_S)\ll 1$. To achieve a
relative error of $0.1\%$, one needs $N \Omega(\mc C_S) $ to be greater than $10^6$.
The above discussion applies not only to $\mc C_S$ but also to all other bounds
that we will present.

\subsection{The elastic positivity approach}
\label{sec:elasticapproach}

In the elastic positivity approach, bounds are derived from elastic scatterings
of superposed states. Conceptually this is a simple way to obtain bounds.  The
approach however has two drawbacks. The first is that the resulting bounds are
not guaranteed to be complete.  The second is that the extraction of the full set of bounds
quickly becomes difficult as the number of low-energy modes involved increases.
We have briefly discussed these bounds in Section~\ref{sec:dispersion}. 

From the geometric point of view, finding the elastic positivity bounds is equivalent
to constructing a subset of $\mc C^*$:
\begin{flalign}
	\mc Q\equiv \mbox{cone}\left(\left\{u^iv^ju^kv^l\right\}
	\right) \subset \mc C^*
\label{eq:Q}
\end{flalign}
so that
\begin{flalign}
	\mc C^{el} = \mc Q^* \supset \mc C
\end{flalign}
The elements of $\mc Q$ represent the elastic scattering channels between
$u^i\ket{i}$ and $v^i\ket{i}$.
The fact that $\mc Q\subseteq \mc C^*$ follows Eq.~(\ref{eq:QQ}).
$M^{ijkl}\in \mc C^{el}$ is equivalent to $P(u,v)\ge0$. In addition,
$\mc Q$ is often a proper subset of $\mc C^*$, see Ref.~\cite{Zhang:2020jyn},
which means that the elastic approach in general only
gives conservative bounds. The problem of finding $\mc C_S$ is then relaxed to
finding the dual of $\mc Q$ projected on $\mc S$.
\begin{flalign}
	\mc C_S^{el}= \mbox{cone}\left(\left\{ \vec M^{ijkl}u^iv^ju^kv^l \right\}\right)^*
	=\mbox{cone}\left(\vec p(u,v)\right)^*\equiv \mc Q_P^*
	\label{eq:elS}
\end{flalign}
where we define the vector of quartic polynomials, $\vec p(u,v)$, whose
components are $p_\alpha(u,v)\equiv M_\alpha^{ijkl}u^iv^ju^kv^l$.
Our final goal is to identify the set of ERs of $\mc
Q_P=$cone$\left(\vec p(u,v)\right)$, which we denote by $\mc E_{P}$. The
positivity bounds are then given by
\begin{flalign}
	\vec C\cdot \vec p\ge0,\ \forall \vec p \in \mc E_{P}
	\label{eq:EP}
\end{flalign}

How do we find $\mc E_P$?
It is often possible to identify the bounds of $\mathcal{Q}_P$ by inspecting
the structure of $\vec p(u,v)$. Often, $\mathcal{Q}_P$ turns out to be
a polyhedral cone, and in that case a vertex enumeration will immediately give
$\mc E_P$. To illustrate this, consider the four-Higgs operators as an example.
They are the $S$-type aQGC operators, and are defined as \cite{Degrande:2013rea}
\newcommand{\dd}[1]{D_{#1}\Phi}
\newcommand{\du}[1]{D^{#1}\Phi}
\newcommand{\dg}[1]{(#1)^\dagger}
\begin{flalign}
O_{S,0}=[\dg{\dd{\mu}}\dd{\nu}]\times[\dg{\du{\mu}}\du{\nu}],
\\
O_{S,1}=[\dg{\dd{\mu}}\du{\mu}]\times[\dg{\dd{\nu}}\du{\nu}],
\\
O_{S,2}=[\dg{\dd{\mu}}\dd{\nu}]\times[\dg{\du{\nu}}\du{\mu}].
\end{flalign}
Define the real field components
\begin{flalign}
	\Phi=\left(
	\begin{array}{c}
		\phi_1+i\phi_2\\
		\phi_3+i\phi_4
	\end{array}
	\right),
\end{flalign}
and $u=(u_1,u_2,u_3,u_4)$, $v=(v_1,v_2,v_3,v_4)$ to represent the superposition
of the real Higgs components.
Assuming they are real for the moment (shortly we will show $u,v$ being
complex does not give rise to any new bounds), by an explicit calculation of the $HH\to HH$ amplitude,
we find
\begin{flalign}
	&\vec p(u,v)\propto (X+Y+Z,Y,X+Y)
	\\
	&X=\frac{1}{2}\left(-u_4 v_1-u_3 v_2+u_2 v_3+u_1
   v_4\right)^2+\frac{1}{2}\left(-u_3 v_1+u_4 v_2+u_1 v_3-u_2
   v_4\right)^2\ge0
   \\
   &Y= \left(u_1 v_1+u_2 v_2+u_3 v_3+u_4 v_4\right)^2\ge0
   \\
   &Z= \left(u_2 v_1-u_1 v_2+u_4 v_3-u_3 v_4\right)^2\ge0
\end{flalign}
and this defines $\mathcal{Q}_P=$cone$(\{\vec p(u,v)\})$ as a triangular cone, with 3 facets
\begin{flalign}
	&\vec p(u,v)\cdot (0,-1,1)\ge0 \label{eq:fac1}\\
	&\vec p(u,v)\cdot (0,1,0)\ge0\\
	&\vec p(u,v)\cdot (1,0,-1)\ge0 \label{eq:fac2}
\end{flalign}

Now, any $\vec p(u,v)$ could lead to a necessary bound $\sum_ip_i(u,v) F_{S,i}\ge0$, 
but our goal is the complete set of independent bounds, which are given by $\mc E_P$, the
ERs of this triangular cone.  They can be obtained from the 3 facets by a
vertex enumeration: $\mc E_P=\{(1,0,0),(1,0,1),(1,1,1)\}$. These give the
following bounds:
\begin{flalign}
	&F_{S,0}\ge0\,,\\
	&F_{S,0}+F_{S,2}\ge0\,,\\
	&F_{S,0}+F_{S,1}+F_{S,2}\ge0\,.
\end{flalign}
It is easy to see that cone$(\{\vec p(u,v)\})$ fill the entire triangular cone
described by Eqs.~\eqref{eq:fac1}-\eqref{eq:fac2}, and so the three bounds
above give indeed the best and valid positivity bounds for the four-Higgs operators.

The example shown here is a simple one with only 3 ERs in a 3-dimensional space,
so the problem is somewhat trivial and can be easily solved in other ways. However,
formulating the solution with the help of vertex enumeration is convenient, in
particular when the dimension of the parameter space and the number of ERs
become larger. We will demonstrate this in Section~\ref{sec:nonfac}.  More
generally, $\mathcal{Q}_P$ may not be polyhedral, as for example in the
problem considered in this work, where $W$- and $B$-bosons are both taken into
account.  Still, it is possible to identify $\mc E_P$ by inspecting the
boundary of $\mc Q_P$.  We will demonstrate this approach in
Section~\ref{sec:nonfac}.

We have mentioned that complex values of $u,v$ do not need to be considered, as
they lead to no additional bounds. Here we give a proof.
First note that according to Eq.~(\ref{eq:1}), $M^{ijkl}$
has the following
crossing symmetries:
\begin{flalign}
	M^{ijkl}=M^{ilkj}=M^{kjil}=M^{klij}\,.
	\label{eq:crossings}
\end{flalign}
For any $u,v\in \mb C^n$, we write $u=r+is$, $v=p+iq$, with $r,s,p,q\in \mb R^n$.
Let $T^{ijkl}=u^iv^ju^{*k}v^{*l}$. Recall that the elastic positivity
Eq.~(\ref{eq:elasticpositivity}) requires that $P(u,v)=T^{ijkl}M^{ijkl}\ge0$.
Since $M$ is crossing symmetric, we can symmetrize $T$ w.r.t.~the same
symmetries in Eq.~(\ref{eq:crossings}):
\begin{flalign}
	P(u,v)&=T^{ijkl}M^{ijkl}=\frac{1}{4}\left( 
        T^{ijkl}+T^{ilkj}+T^{kjil}+T^{klij}
	\right)M^{ijkl}
	\nonumber\\&=
	\left( r^ip^jr^kp^l+s^ip^js^kp^l+r^iq^jr^kq^l+s^iq^js^kq^l \right)M^{ijkl}
	\nonumber\\&=
	P(r,p)+P(s,p)+P(r,q)+P(s,q)
\end{flalign}
Since $r,s,p,q\in\mb R^n$, $P(r,p)$ being PSD for real $r,p$ is a sufficient
condition for $P(u,v)$ to be PSD for complex $u,v$. Thus restricting $u,v$
to $\mb R^n$ is sufficient.

This condition essentially comes from the crossing symmetry of $M$, which is
present because we work with real modes. For example, if instead of $W^i_{x,y}$
we used the helicity basis, $W^i_{\pm}$, the above would not hold. Fortunately, it
is always possible to work with a basis in which the crossing symmetry is
manifest. This is exactly why we choose the linear polarization basis for the
vector modes.  We should also mention that, in Section~\ref{sec:fac}, we consider
$u,v$ to be factorizable in the polarization and gauge spaces, i.e.
\begin{flalign}
	u^i=x^a\alpha^b, \  v^i=y^a\beta^b,\quad i=(a,b) \label{eq:uv_fac}
\end{flalign}
with $a,b$ the polarization index and the gauge index, respectively,
so that the problem of determining a PSD polynomial can be simplified.
In this case the crossing symmetry for one set of variables is lost, and so
in principle we must consider complex values. In practice, as we will see, with
the factorization assumption, this does not significantly increase the difficulty.

\subsection{The extremal positivity approach}
\label{sec:extremalRep}

The extremal positivity approach
proposed recently in Ref.~\cite{Zhang:2020jyn} directly constructs the allowed
parameter space as the convex hull of a set of ``potential'' ERs
(PERs), which can be identified as the projectors of the irreps of the
symmetry groups, under which the particles $i,j$ are charged. 
This approach has the advantage that, in principle,
it always gives the best bounds as required by the dispersion relation.

To illustrate this idea, consider the $W$-boson-only case which
has been already discussed in Ref.~\cite{Zhang:2020jyn}. Briefly, 
following Eq.~\eqref{eq:cone2}, one needs the projection operators for
$\mathbf{3}\otimes\mathbf{3}=\mathbf{1}\oplus\mathbf{3}\oplus\mathbf{5}$ in
$SU(2)_L$,
\begin{flalign}
	&P^{(1)}_{\alpha\beta\gamma\sigma}=\frac{1}{N}\delta_{\alpha\beta}\delta_{\gamma\sigma},
	\ 
	P^{(2)}_{\alpha\beta\gamma\sigma}=\frac{1}{2}\left( \delta_{\alpha\gamma}\delta_{\beta\sigma}
	-\delta_{\alpha\sigma}\delta_{\beta\gamma}\right) ,
	\nonumber\\
	&P^{(3)}_{\alpha\beta\gamma\sigma}=\frac{1}{2}\left(
	\delta_{\alpha\gamma}\delta_{\beta\sigma}
	+\delta_{\alpha\sigma}\delta_{\beta\gamma}\right)
	-\frac{1}{N}\delta_{\alpha\beta}\delta_{\gamma\sigma} ,
\end{flalign}
where $N=3$, as well as the projectors of the $SO(2)$ rotation around the
forward direction, for
$\mathbf{2}\otimes\mathbf{2}=\mathbf{1}\oplus\mathbf{1}\oplus\mathbf{2}$, which
is similar but with $N=2$. Combining both types of projectors with the
$\beta,\sigma$ indices symmetrized gives 9 PERs, among which 5 are linearly independent
and 8 are extremal.
Thus $\mathcal{C}$ is determined via the extremal representation, as a 8-edge
polyhedral cone $\mathcal{C}$ in a 5-dimensional space.

\begin{figure}[ht]
	\begin{center}
		\includegraphics[width=0.45\linewidth]{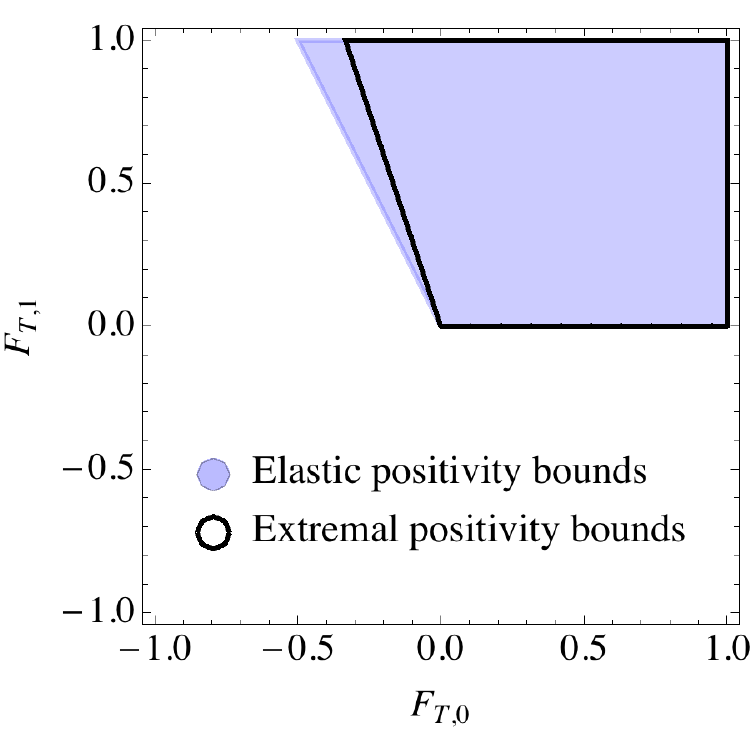}
	\end{center}
	\caption{A comparison of elastic positivity bounds
		and extremal positivity bounds on the $F_{T,0}$-$F_{T,1}$
		plane. All other coefficients are fixed to 0.}
	\label{fig:ww}
\end{figure}

To obtain bounds, one needs to represent the cone in the inequality
representation. Since $\mc C$ is a polyhedral cone, the problem is a simple
vertex enumeration.  Once we have the facets of $\mc C$,
taking an intersection with the physical space $\mc S$ gives
$\mc C_S$ as a cone with 6 facets, each of them representing a
linear bound:
\begin{flalign}
	&F_{T,2}\ge0, \\
	&4F_{T,1}+F_{T,2}\ge0, \\
	& F_{T,2}+8F_{T,10}\ge0,\\
	& 8F_{T,0}+4F_{T,1}+3F_{T,2}\ge0, \\
	&12F_{T,0}+4F_{T,1}+5F_{T,2}+4F_{T,10}\ge0,
	\label{eq:new1} \\
	&4F_{T,0}+4F_{T,1}+3F_{T,2}+12F_{T,10}\ge0.
	\label{eq:new2}
\end{flalign}
The last two bounds, Eqs.~(\ref{eq:new1}) and (\ref{eq:new2}), are actually
tighter than the best linear bounds from the elastic positivity approach. 
For illustration, in Figure~\ref{fig:ww} we show this improvement on $F_{T,0}$ and
$F_{T,1}$, assuming the other two operators vanish.  More discussions about this
improvement can be found in Ref.~\cite{Zhang:2020jyn}.

While this approach is powerful and efficient in the above example,
difficulties might arise if 
degenerate irreps appear. For example, a pair of $W$ and
a pair of $B$-bosons can both form a $SU(2)_L$ singlet, and a projector
can be constructed for any linear combination of them.
The ERs are then continuously parameterized
by a real parameter, $r$, that represents the mixing of the two irreps.
As we will see, when both $W$ and $B$-bosons are considered,
the extremal presentation of $\mc C$ contains a set of discrete ERs, $\vec e_i$, 
plus a set of continuous ones, $\vec e_i(r)$.  One then has to determine the
convex hull of an infinite number of rays.

Numerically, a possible solution
is to sample $\vec e_i(r)$ with a sufficiently large number of discrete $r$
values. This gives a set of numerical PERs, which, after intersecting the physical
subspace, forms an approximation of $\mc C_S$, and we will call it $\mc C_N$.
In principle, one could adopt the same vertex enumeration algorithms to find
its facets.  However, since $\mc C_N$ is inscribed to $\mc C_S$, one has to
choose a sufficiently large $N$, the number of the sampled rays,
to avoid over-constraining. The
number of facets grows quickly with $N$, and as a result, this approach becomes
not very useful in practice.  Instead, we find the following approach more
suitable for large $N$: instead of presenting positivity inequalities and then
checking whether a point $\vec f$ is included in $\mc C_N$, we can directly
check if there exists a set of positive weights $w_i\ge0$ and a subset of 
discretely sampled PERs, $\vec e_{\mathrm{num},i}$, such that $\sum_i
w_ie_{\mathrm{num},i}=\vec f$, i.e.~$\vec f$ is a positively weighted sum of
$\vec e_{\mathrm{num},i}$.  This problem can be recast into a linear
programming problem, and therefore can be solved efficiently with standard
tools.

On the other hand, finding the full set of analytical bounds is often difficult
if there are many PERs that depend on free real parameters.  We will see that the resulting
analytical bounds are not simply linear inequalities, but instead they form a
series of homogeneous polynomial inequalities with increasing degrees, i.e.~linear
inequalities, quadratic inequalities, cubic inequalities, and so on.  In this
work we will show how to derive up to quadratic level bounds for the
transversal aQGC operators.  Due to this ``truncation'' at the quadratic order,
these analytical results are incomplete and describe a slightly larger
cone $\mc C_{A}\supsetneq\mc C_S$. In contrast, the numerical approach
describes $\mc C_N\subsetneq \mc C_S$, which is slightly over-constraining, so
the exact $\mc C_S$ is bounded by the two methods, with $\Omega(\mc
C_N)<\Omega(\mc C_S)<\Omega(\mc C_{A})$.  By comparing the solid angles
$\Omega(\mc C_N)$ and $\Omega(\mc C_A)$, we can get a conservative estimate of
the errors of both approaches. As we will see, their difference in $\Omega$
shows up at the 3rd decimal place, which means that both the numerical bounds
and the truncated analytical bounds are extremely close to the truth.

\section{The elastic positivity bounds}
\label{sec:elastic}
In this section, we extract the positivity bounds using the conventional elastic
positivity approach, i.e.~following Eq.~(\ref{eq:psdpoly}).  We have shown that
$u,v$ only need to be real vectors. Taking this into account, the problem is
still a difficult one, equivalent to the determination of a PSD polynomial with
16 variables. Furthermore, these variables are not connected by a single
symmetry group: there are two sets of quantum numbers, helicity and gauge, and
in addition $W$ and $B$ are not in the same gauge multiplet. 

We will consider three different methods.
In the first one, we restrict the $u,v$ vectors to those with a specific form,
and therefore simplifying the problem, at the cost of a loss in completeness of
the resulting bounds. More explicitly, we assume that $u,v$ can be
factorized as the tensor products of two vectors, one representing the
(complex) superposition in the helicity space, and the other
representing the (complex) superposition in the gauge space (including $W$ and $B$).
We will show that by doing so, the problem is factorized into a vertex
enumeration problem in the helicity space with 3 variables, plus two
quadratic programming problems in the gauge space with 6
variables, both can be solved, analytically and systematically. The
results are, of course, only conservative.

A second method is more general and aims at the full set of bounds.  Using
Eq.~(\ref{eq:EP}) and the definition of an ER, we can directly look for the ERs
of $\mc Q_P$. This approach works well when only the $W$-boson modes are
considered. If the hypercharge boson $B$ is also included, $\mc Q_P$ becomes
not polyhedral, and the identification of its boundary becomes much more difficult.
We are not able to find a systematic approach for this identification. 
Our approach will be based on inspection, but is not always exact.  Resulting
bounds are therefore, strictly speaking, still incomplete, but we will show
that the difference from the complete bounds is tiny, at only the per mille
level.

Finally, we will also consider a fully numerical approach, which directly
checks if a given set of coefficients are allowed by positivity. It proceeds by
minimizing $P(u,v)$ numerically w.r.t.~to $u,v$, and checking that the minimum
is positive.  This gives in principle the best elastic positivity bounds. The
drawback is that no inequalities can be obtained.

The three approaches will be presented in Sections~\ref{sec:fac},
\ref{sec:nonfac}, and \ref{sec:num}. 

\subsection{The factorization assumption}
\label{sec:fac}

In this section we will consider the following factorization for the superposition vectors:
\begin{flalign}
	u^i=x^a\alpha^b, \  v^i=y^a\beta^b,\quad i=(a,b) \label{eq:faca}
\end{flalign}
where $x,y$ represent superpositions of different polarization states, and
$\alpha,\beta$ represent superpositions of different gauge components,
including $W$ and $B$, such that
\begin{flalign}
	&u^i\ket{i}=x^1\alpha^1 \ket{B_R}+\sum_{b=2}^4 x^1\alpha^b \ket{W_R^{b-1}}
	+x^2\alpha^1 \ket{B_L}+\sum_{b=2}^4 x^2\alpha^b \ket{W_L^{b-1}}
	\\
	&v^i\ket{i}=y^1\beta^1 \ket{B_R}+\sum_{b=2}^4 y^1\beta^b \ket{W_R^{b-1}}
	+y^2\beta^1 \ket{B_L}+\sum_{b=2}^4 y^2\beta^b \ket{W_L^{b-1}}
\end{flalign}
where the subscripts $R,L$ indicate 
positive and negative helicity states, respectively.
This parameterization gives a subset of $\mc Q_P$, and so the resulting bounds
are conservative.  We will denote these bounds by $\mc C^{el}_{AF}$.

With this factorized form, $P(u,v)$ can be written as
\begin{equation}
P(u,v)=\sum^4_{i,k,m,r=1} \sum^2_{j,l,n,s=1}\alpha_i\beta_k\alpha^*_m\beta^*_r x_j y_l x^*_n y^*_s M_{ijklmnrs}\geq0\label{eq:fac_amp0}
\end{equation}
and we have divided the amplitude by a positive factor $4\alpha^2\pi^2/\Lambda^4$.
Explicitly, we find that $P(u,v)$ can be written in a bilinear form:
\begin{align}
	&P(u,v)=\sum_{k=1}^3\sum_{l=1}^7C_{kl}(\{F_{T,i}\}) 
	a_k(x_1,x_2,y_1,y_2)
	b_l(\alpha_1,\ldots,\alpha_4,\beta_1,\ldots,\beta_4), \label{eq:fac_amp}
\end{align}
where $C_{kl}$ is a linear function of the coefficients $F_{T,i}$.  We omit its
explicit form here, and refer to Eqs.~\eqref{eq:ckl1}-\eqref{eq:pol4} in
Appendix \ref{sec:app_fac}.  $a_k$, $b_l$ are quartic polynomials of
$x,y,x^*,y^*$, and of $\alpha,\beta,\alpha^*,\beta^*$, respectively.  They can be
written as
\begin{align}
&a_1 = |x_1y^*_2+x_2y^*_1|^2, a_2 = |x_1y_1-x_2y_2|^2, \nonumber\\
&a_3 = |x_1y^*_2-x_2y^*_1|^2 \label{eq:def_a}{ ,}\\
&b_1 = |\langle \vec{u},\vec{v}^*\rangle|^2 ,
b_2 = |\langle \vec{u}, \vec{v} \rangle|^2,
b_3 = |\vec{u}|^2|\vec{v}|^2, \label{eq:b123}\\
&b_4 = \alpha_1\beta_1\langle \vec{u} ,\vec{v}^* \rangle +c.c.,\\
&b_5 = \alpha_1\beta^*_1\langle \vec{u} ,\vec{v}\rangle +c.c.,\\
&b_6 = |\beta_1|^2 |\vec{u}|^2 +|\alpha_1|^2|\vec{v}|^2,\label{eq:b6}\\
&b_7 = |\alpha_1|^2|\beta_1|^2{ ,}\label{eq:b7}
\\
&\vec{u} \equiv (\alpha_2,\alpha_3, \alpha_4)^T, \vec{v} \equiv (\beta_2,\beta_3, \beta_4)^T{ .}
\end{align}
{Here, the inner product is defined by $\langle \vec{u},\vec{v} \rangle \equiv
\vec{u}^\dagger\cdot\vec{v}={\sum^4_{i=2}}\alpha^*_i\beta_i$.}
Note that the polarization parameters, $x,y,x^*,y^*$, are present only in terms of
$a_1$, $a_2$, and $a_3$.  This can be traced back to
the conservations of the total angular momentum and parity.
We give a brief explanation in Appendix~\ref{sec:app_fac2}.
The gauge parameters $\alpha,\beta,\alpha^*,\beta^*$ are present only in terms
of the $b_i$ variables.  $b_{1,2,3}$, $b_{4,5,6}$, and $b_7$ correspond
to the $WW$, $WB$ and $BB$ scattering channels, respectively.

Now the problem is factorized to the determination of $P(u,v)\ge0$
w.r.t.~the $a$ and $b$ variables separately. To see this, note that $a_1,a_2,a_3$
can take any real values that satisfy
\begin{flalign}
	a_1\ge0,\ a_2\ge0,\ 0\le a_3\le a_1+a_2\,,
\end{flalign}
where the last inequality holds because
\begin{flalign}
	a_1+a_2-a_3=|x_1y_1+x_2y_2|^2
\end{flalign}
These inequalities define a 3-dimensional polyhedral cone $\mc A$.
Defining $M_k=C_{kl}b_l$, we see that $P(u,v)\ge0$ is equivalent to
\begin{flalign}
	M_k a_k \ge0,\ \forall (a_1,a_2,a_3)\in \mc A \quad \Rightarrow \quad M_k\in \mc A^*
\end{flalign}
Since we know the bounds of $a_i$, which are the facets of $\mc A$, by vertex
enumeration we get the facets of $\mc A^*$, which are the bounds on $M_k$, as follows:
\begin{flalign}
	M_1\ge0,\ M_2\ge 0,\ M_1+M_3\ge0,\ M_2+M_3\ge0
	\label{eq:facstep1}
\end{flalign}
So the positivity w.r.t.~the helicity superposition is equivalent to a simple vertex enumeration.

To proceed, note that Eq.~(\ref{eq:facstep1}) implies
that 4 linear combinations of $C_{kl}b_l$ must be non-negative, for all possible
$b$'s satisfying Eqs.~(\ref{eq:b123})-(\ref{eq:b7}).
After a change of variables, the latter can be identified by a set of quadratic
inequalities. The problem can hence be turned into a set of
{quadratically-constrained quadratic programming problems,}
which are the minimization of a quadratic polynomial of variables that satisfy a set
of quadratic inequalities. Due to the symmetries of the problem, it turns out that one
needs to solve two such problems with at most 6 variables. These problems can
be solved analytically, and details are given in Appendix~\ref{sec:app_fac}.
In total, we obtain nine linear, three quadratic, and one cubic conditions, for
$P(u,v)\ge0$. These are exactly the bounds that define $\mc C^{el}_{AF}$. They are:
\begin{description}
	\item[Linear]
\begin{flalign}
&2F_{T,0}+2F_{T,1}+F_{T,2}\ge0\label{eq:2210_positive}\\
&F_{T,2} + 4 F_{T,10} \ge0\label{eq:0014_positive}\\
&4F_{T,1}+F_{T,2}\ge0 \label{eq:0410_positive}\\
&F_{T,2} \ge0\label{eq:0010_positive}\\
&2F_{T,0}+F_{T,1}+F_{T,2}+2F_{T,10}\ge0\label{eq:2112_positive}\\
&2F_{T,8}+F_{T,9}\ge0\label{eq:fg_positive}\\
&F_{T,9}\ge0  \label{eq:fg_positive3}\\
&4F_{T,6}+F_{T,7}\ge0\label{eq:0410mix_positive}\\
&F_{T,7}\ge0\label{eq:ct7_positive}
\end{flalign}
\item[Quadratic]
\begin{flalign}
&4\sqrt{[2(F_{T,0} + F_{T,1}) + F_{T,2}](2F_{T,8} + F_{T,9})}
\nonumber\\&\qquad\ge 
{\mathrm{max}\left[-2(2F_{T,5} + 2F_{T,6} + F_{T,7}), 4F_{T,5} + F_{T,7}\right]}\label{eq:quad1}\\
  &{2\sqrt{F_{T,9}(F_{T,2}+4F_{T,10})}\ge \mathrm{max}\left[-(2F_{T,11}+ F_{T,7}),2F_{T,11}\right]} \label{eq:quad2}\\
 &2\sqrt{[4F_{T,10}+4 (F_{T,0} + F_{T,1}) + 3F_{T,2}] (4F_{T,8} + 3F_{T,9})}\ge|2F_{T,11} + 4F_{T.5} + F_{T,7}|\label{eq:quad3}
\end{flalign}
\item[Cubic]
\begin{flalign}
&[4(2F_{T,1}+F_{T,2})(4F_{T,8}+3F_{T,9})-(2F_{T,6}+F_{T,7})|2F_{T,11}+4F_{T,5}+F_{T,7}|] \nonumber\\
&\quad \times(-|2F_{T,11}+4F_{T,5}+F_{T,7}|+2F_{T,6}+F_{T,7})\ge 0\nonumber\\
&\text{or }4F_{T,0}+2F_{T,1}+2F_{T,2}+4F_{T,10}\nonumber\\
&\quad\quad \ge\frac{(2F_{T,1}+F_{T,2})(2F_{T,6}+F_{T,7}-|2F_{T,11}+4F_{T,5}+F_{T,7}|)^2}
{4(2F_{T,1}+F_{T,2})(4F_{T,8}+3F_{T,9})-(2F_{T,6}+F_{T,7})^2}\label{eq:cubic}
\end{flalign}

\end{description}

As a check, we have also obtained the explicit values of $u,v$ vectors for each
bound. This information specifies the elastic channel from which a positivity bound
can be derived.
For the nine linear conditions, the corresponding elastic channels are
\begin{flalign}
\begin{array}{rcl}
	\mbox{Bounds} & \qquad & \mbox{Channels}\ (\ket 1+\ket 2
	\to \ket 1\ +\ket 2)
	\\\hline
	2F_{T,0}+2F_{T,1}+F_{T,2}\ge0,&& { \ket 1 = \ket{W_x^1},\ \ket 2=\ket{W_x^1}}
	\\
	F_{T,2} + 4 F_{T,10} \ge0,&&  { \ket 1 = \ket{W_x^1},\ \ket 2=\ket{W_y^1}}
	\\
	4F_{T,1}+F_{T,2}\ge0,&& { \ket 1 = \ket{W_x^1},\ \ket 2=\ket{W_x^2}}
	\\
	F_{T,2}\ge0,&& { \ket 1 = \ket{W_x^1},\ \ket 2=\ket{W_y^2}}
	\\
	2F_{T,0}+F_{T,1}+F_{T,2}+2F_{T,10}\ge0,&& { \ket 1 = \ket{W_L^-},\ \ket 2=\ket{W_L^+}}
	\\
	2F_{T,8}+F_{T,9}\ge0,&& { \ket 1 = \ket{B_x},\ \ket 2=\ket{B_x}}
	\\
	F_{T,9}\ge0,&&  { \ket 1 =\ket{B_x},\ \ket 2=\ket{B_y}}
	\\
	4F_{T,6}+F_{T,7}\ge0,&& { \ket 1 = \ket{B_x},\ \ket 2=\ket{W^1_x}}
	\\
	F_{T,7}\ge0,&& { \ket 1 = \ket{B_x},\ \ket 2=\ket{W^1_y}}
	\end{array}
	\label{eq:lineartable}
\end{flalign}
Similar results for quadratic and cubic conditions are given in appendix~\ref{sec:app_fac}.

The above results are derived with a factorization assumption on $u,v$, and so
they are incomplete and should be considered as a set of conservative bounds.
The constraining power is reflected by the solid angle of these bounds,
$\Omega(\mc C^{el}_{AF})=0.891\%$.  This is obtained by $10^9$ Monte Carlo
sampling points.  Surprisingly, we see that more than $99\%$ of the naive aQGC
parameter space is already ruled out, even with this conservative approach.

\subsubsection{Comparison with previous results}
The bounds we have obtained in Eqs.~(\ref{eq:2210_positive})-(\ref{eq:quad3}),
although incomplete, are sufficient to cover all previous results on the
transversal aQGC bounds presented in Ref.~\cite{Bi:2019phv} and
Ref.~\cite{Remmen:2019cyz}.  In this section we will make a comparison and
demonstrate this coverage. All other approaches that we will consider in this
work give even better results, and so for the rest of the paper we will make no
further comparisons.

Let us start with Ref.~\cite{Remmen:2019cyz}. In the gauge-boson sector, the
authors have considered superpositions in the helicity space,
but not those in the gauge space.  The resulting bounds are linear, and are
the same as our Eqs.~\eqref{eq:2210_positive}, \eqref{eq:0014_positive},
\eqref{eq:fg_positive}, \eqref{eq:fg_positive3}, \eqref{eq:0410mix_positive},
and \eqref{eq:ct7_positive}.  In particular, among the 5 bounds we have
obtained in the $W$-boson sector, only the first two in
Eq.~\eqref{eq:lineartable} were presented in Ref.~\cite{Remmen:2019cyz}, as the
other three require the two incoming particles to be orthogonal in the gauge
space.  In addition, quadratic and cubic bounds were also not obtained, as
these require superposing both $W$- and $B$-bosons.\footnote{A set of quadratic
	bounds were presented in Eq.~(96) of Ref.~\cite{Remmen:2019cyz},
constraining the potential sizes of CP-violating operators. They arise from the
superposition in the helicity space. In this work we have focused on
CP-conserving operators.}

Now consider the results in Ref.~\cite{Bi:2019phv}. In that work,
superpositions have been considered in the helicities space with complex
coefficients, and with $W^+$ and $W^-$ used as particle basis, instead of $W^1$
and $W^2$.  In addition, $Z$ and $\gamma$ were used instead of $W^3$ and $B$,
Finally, the operators $O_{T,10}$ and $O_{T,11}$ were missed.  The bounds
obtained for the 4$W$ operators are:
\begin{equation}
\begin{array}{ccc}
      W^-_x W^-_y \\
      W^-_R W^-_R \\
      W^-_R W^-_L \\
      W^-_x W^-_x
    \end{array}
 \left(
    \begin{array}{ccc}
      0 & 0 & 1 \\
      0 & 2 & 1 \\
      2 & 1 & 1 \\
      8 & 12 & 5
    \end{array}
  \right)
 \left(
    \begin{array}{c}
     F_{T,0}\\
     F_{T,1} \\
     F_{T,2}
    \end{array}
  \right)\ge0.
  \label{eq:WW_p2}
\end{equation}
The corresponding scattering channel is shown on the left side of each
bound.\footnote{$W^\pm W^\pm$ and $W^\pm W^\mp$ channels give equivalent
bounds. For simplicity we only use the $W^-$ component.} 
For comparison, we set $F_{T,10}=0$.
The first one in Eq.~\eqref{eq:WW_p2} can be identified as our 
Eqs.~\eqref{eq:0014_positive} and \eqref{eq:0010_positive}.
The second one can be obtained by adding our 
Eqs.~\eqref{eq:0410_positive} and \eqref{eq:0010_positive}.
The third one is the same as our
Eqs.~\eqref{eq:2112_positive}.
This correspondence is clear because
\begin{equation}
 \Ket{W^{\pm}_R}=\frac{1}{\sqrt{2}}\left(\Ket{W_R^1}\mp i\Ket{W_R^2}\right).\label{eq:W}
\end{equation}
Finally, the fourth one is obtained from 4 times 
Eqs.~\eqref{eq:2210_positive} plus \eqref{eq:0410_positive}.

Additional results in Ref.~\cite{Bi:2019phv}, obtained from the $ZZ$, $WZ$, $W\gamma$,
$Z\gamma$, and $\gamma\gamma$ channels,
are also covered by positive linear combinations of our constraints 
\eqref{eq:0410_positive}, 
\eqref{eq:0010_positive}, 
\eqref{eq:fg_positive3}, \eqref{eq:0410mix_positive}, \eqref{eq:ct7_positive},
and by relaxing the quadratic constraint, Eq.~\eqref{eq:quad1}
(or Eq.~\eqref{eq:cond_q1} and Eq.~\eqref{eq:cond_q2}). The latter
can be done with the inequality of arithmetic and geometric means (AM-GM
inequality).  For example, the $\gamma \gamma$ scattering with the same polarization gives
\begin{equation}
8F_{T,0}+8F_{T,1}+4F_{T,2}+4F_{T,5}+4F_{T,6}+2F_{T,7}+2F_{T,8}+F_{T,9}\ge0.\label{eq:pre_example}
\end{equation}
This condition is covered by our condition Eq.~\eqref{eq:quad1} (or \eqref{eq:cond_q1}).
To derive this, we use the AM-GM inequality
\begin{align}
&4\sqrt{ab} \le 4a+b\label{eq:relation}, \quad \forall a,b\ge0
\end{align}
Applying this to the l.h.s.~of Eq.~\eqref{eq:quad1} (or \eqref{eq:cond_q1}) with
\begin{align}
&a=2(F_{T,0}+F_{T,1})+F_{T,2},\ \ b=2F_{T,8}+F_{T,9},
\end{align}
relaxes Eq.~\eqref{eq:quad1} (or \eqref{eq:cond_q1}) to
Eq.~\eqref{eq:pre_example}.  In fact, all similar bounds obtained by scattering
channels involving $Z$ or $\gamma$ with definite polarizations can be covered by
our results, because Eq.~\eqref{eq:W} together with
${\Ket{W^3}}=c_W\Ket{Z}+s_W\Ket{\gamma}$ and
$\Ket{B}=-s_W\Ket{Z}+c_W\Ket{\gamma}$
(with $c_W\equiv\cos\theta_W$ and $s_W\equiv\sin\theta_W$, the cosine and sine of the Weinberg angle, respectively)
allows us to represent all these channels by a pair of properly chosen $u,v$
vectors, which are factorizable under our assumption.

\subsection{General bounds}
\label{sec:nonfac}

In this section we solve the problem without any assumptions on $u$ and $v$
vectors. We allow arbitrary superposition between the following 8 states:
\begin{flalign}
	W_x^1,W_y^1,W_x^2,W_y^2,W_x^3,W_y^3,B_x,B_y
\end{flalign}
with $u,v\in\mb R^8$. 
We proceed as follows.  First, in Section~\ref{sec:Wonly} we only include the
$W$-boson components $W^{1,2}_{x,y}$, but neglect $W^{3}_{x,y}$.
We will show that in this case $\mc Q_P$
is a polyhedral cone, and we can determine the elastic positivity bounds by finding its ERs.
Then in Section~\ref{sec:WandB} we include the hypercharge boson modes, $B_{x,y}$.  Finally,
in Section~\ref{sec:W3} we show that adding the $W^3_{x,y}$ components does not lead
to new results.  Note that the last point is not true if the extremal positivity
approach is used, as we will see later.  In Section~\ref{sec:WandB} we will see
that our approach is unfortunately still not complete, so the resulting bounds,
which we dub $\mc C^{el}_A$, is slightly conservative.

\subsubsection{Analytical bounds for $W$-boson only}
\label{sec:Wonly}
The $W$-boson has two set of indices: polarization and the $SU(2)_L$ gauge group index.
We use $a,b,c,d$ for the former and $\alpha,\beta,\gamma,\sigma$ for the
latter. $\alpha,\beta,\gamma,\sigma$ run from 1 to 3, but we consider the first two components
for the moment.  The particle indices $i,j,k,l$ each correspond to a pair of
indices from the two sets, i.e.~$i=(a,\alpha)$, $j=(b,\beta)$ etc.  The
physical space $\mc S$ must be invariant under the rotation around the
beam axis, the global $SU(2)_L$ transformation, and the $j\leftrightarrow l$ exchange.
A basis for the amplitude can be chosen as
\begin{flalign}
	&M_1^{ijkl}=
\frac{1}{2}[\delta^{\alpha\beta}\delta^{ab} \delta^{\gamma\sigma}\delta^{cd}
	+(s\leftrightarrow u)]\,,
	\\
	&M_2^{ijkl}=
\frac{1}{2}[\delta^{\alpha\beta}\epsilon^{ab} \delta^{\gamma\sigma}\epsilon^{cd}
	+(s\leftrightarrow u)]\,,
	\\
	&M_3^{ijkl}=
\frac{1}{2}[\epsilon^{\alpha\beta}\delta^{ab} \epsilon^{\gamma\sigma}\delta^{cd}
	+(s\leftrightarrow u)]\,,
	\\
	&M_4^{ijkl}=
\frac{1}{2}[\epsilon^{\alpha\beta}\epsilon^{ab} \epsilon^{\gamma\sigma}\epsilon^{cd}
	+(s\leftrightarrow u)]\,,
	\\
	&M_5^{ijkl}=-\epsilon^{\alpha\gamma}\epsilon^{ac}\epsilon^{\beta\delta}\epsilon^{bd}\,,
\end{flalign}
where $s\leftrightarrow u$ represents swapping $b$ and $d$, and $\beta$ and
$\sigma$, simultaneously.  An amplitude in this basis, written as
$M^{ijkl}=\sum_{\alpha=1}^5C_\alpha M_\alpha^{ijkl} =\vec C\cdot \vec M$, can
be matched to the SMEFT amplitude, and we can identify the $\vec C$ vector as
combinations of the dim-8 aQGC coefficients and the dim-6 coefficient $a_W$:
\begin{flalign}
\label{eq:C1def}
	C_1&=4(2F_{T,0}+2F_{T,1}+F_{T,2})\,,
	\\
	C_2&=2(F_{T,2}+4F_{T,10})\,,
	\\
	C_3&=4F_{T,1}+F_{T,2}-36 \bar{a}_W^2,
	\label{eq:C3def}
	\\
	C_4&=F_{T,2}\,,
	\\
	C_5&=2(2F_{T,1}+F_{T,2})\,,
	\label{eq:C5def}
\end{flalign}
where a common factor $16\alpha^2\pi^2/s_W^4\Lambda^4$ has been factored out.
The $\bar{a}_W^2$ contribution comes from the dim-6 operator
$O_W$, which we will neglect in this section. We will come back to this contribution
in Section~\ref{sec:dim6}.

We now need to study the cone $\mc Q_P=\mbox{cone}(\vec p(u,v))$ and find its
ERs. First, since $p_\alpha(u,v)=M_\alpha^{ijkl}u^iv^ju^kv^l$, we observe a
number of complete squares that can be constructed with the components of $\vec p(u,v)$:
\begin{flalign}
	&p_1=(u^{a\alpha}v^{a\alpha})^2
	\\
	&p_2=(u^{a\alpha}v^{b\alpha}\epsilon^{ab})^2
	\\
	&p_3=(u^{a\alpha}v^{a\beta}\epsilon^{\alpha\beta})^2
	\\
	&p_4=(u^{a\alpha}v^{b\beta}\epsilon^{ab}\epsilon^{\alpha\beta})^2
	\\
	&p_1+p_2+p_5=(u^{a\alpha}v^{b\alpha}\sigma_3^{ab})^2+
	(u^{a\alpha}v^{b\alpha}\sigma_1^{ab})^2
	\\
	&p_1+p_3+p_5=(u^{a\alpha}v^{a\beta}\sigma_3^{\alpha\beta})^2+
	(u^{a\alpha}v^{a\beta}\sigma_1^{\alpha\beta})^2
	\\
	&p_2+p_4+p_5=(u^{a\alpha}v^{b\beta}\epsilon^{ab}\sigma_3^{\alpha\beta})^2+
	(u^{a\alpha}v^{b\beta}\epsilon^{ab}\sigma_1^{\alpha\beta})^2
	\\
	&p_3+p_4+p_5=(u^{a\alpha}v^{b\beta}\sigma_3^{ab}\epsilon^{\alpha\beta})^2+
	(u^{a\alpha}v^{b\beta}\sigma_1^{ab}\epsilon^{\alpha\beta})^2
\end{flalign}
The fact that these squares or sums of squares are non-negative defines a
polyhedral cone, $\mc Q_R$, in the space of $\vec p$, with 8 facets represented
by the following $\vec k_i$ vectors:
\begin{flalign}
	&\mc Q_R=\left\{\vec p\ | \vec p\cdot \vec k_i\ge0\right\},\ \mbox{with}
	\\
	&
	\begin{aligned}
	\vec k_1=(1,0,0,0,0)
	\\
	\vec k_2=(0,1,0,0,0)
	\\
	\vec k_3=(0,0,1,0,0)
	\\
	\vec k_4=(0,0,0,1,0)
	\end{aligned}
	\quad
	\begin{aligned}
	\vec k_5=(1,1,0,0,1)
	\\
	\vec k_6=(1,0,1,0,1)
	\\
	\vec k_7=(0,1,0,1,1)
	\\
	\vec k_8=(0,0,1,1,1)
	\end{aligned}
\end{flalign}
and we have $\mc Q_P\subseteq \mc Q_R$ since $\vec p(u,v)\in \mc Q_R\ \forall
u,v\in \mb R^n$. Later we will see that $\mc Q_P=\mc Q_R$.

The next step is to find the set of ERs, $\mc E_P$. Since $Q_R$
is polyhedral with known facets, a vertex enumeration immediately leads to 7 edges:
\begin{flalign}
	&\vec E_i=\hat e_i,\ \mbox{for}\ i=1,\ldots,5\,,
	\\
	&\vec E_6=(1,0,0,1,-1),\ \vec E_7=(0,1,1,0,-1)
\end{flalign}
where $\hat e_i$ are the unit vectors along the 5 axes. One can check that
each $\vec E_i$ can be written as $\vec p_i(u,v)$ for some $u,v\in \mb R^n$.
This means that $\mc Q_R=\mbox{conv}(\mc E_P)\subseteq\mbox{cone}(\vec
p(u,v))=\mc Q_P$, and so $\mc Q_P=\mc Q_R$. The ERs of $\mc Q_R$ are
also those of $\mc Q_P$.  Therefore, the full set of the bounds for
$\mc C_S^{el} = \mc Q_P^*$ are then given by $\vec C\cdot \vec E_i\ge0,\
\forall \vec E_i\in\mc E_P$.

Finally, we plug in Eqs.~(\ref{eq:C1def}-\ref{eq:C5def}), which further reduces
the 7 inequalities to 4. They are given by
\begin{flalign}
\begin{array}{rcl}
	\mbox{bounds} & \qquad & \mbox{channel}\ (\ket 1+\ket 2
	\to \ket 1\ +\ket 2)
	\\\hline
	F_{T,2}\ge0,&& \ket 1 = \ket{W_x^1},\ \ket 2=\ket{W_y^2}
	\\
	4F_{T,1}+F_{T,2}\ge0,&& \ket 1 = \ket{W_x^1},\ \ket 2=\ket{W_x^2}
	\\
	F_{T,2}+8F_{T,10}\ge 0,&& \ket 1 = \ket{W_x^1}+\ket{W_y^2},\ \ket 2 = \ket{W_y^1}-\ket{W_x^2}
	\\
	8F_{T,0}+4F_{T,1}+3F_{T,2}\ge 0,&& \ket 1 = \ket{W_x^1}+\ket{W_y^2},\ \ket 2 = \ket{W_x^1}+\ket{W_y^2}
	\end{array}
	\label{eq:elasticW}
\end{flalign}
From the values of $u,v$ that actually generate the $\vec E_i$,
we could also derive the specific superpositions of states whose scattering
amplitudes lead to the above bounds.  This information is also listed above, to
the right of each bound.

One can easily check that these bounds already supersede those presented
in the previous section, based on the factorization assumption,
i.e.~Eqs.~(\ref{eq:2210_positive})-(\ref{eq:2112_positive}).
In fact, Eqs.~(\ref{eq:0010_positive}) and (\ref{eq:0410_positive}) are the same
as the first two bounds shown above; Eqs.~(\ref{eq:0014_positive}) is a sum of the first
and the third bounds; Eq.~(\ref{eq:2210_positive}) is a sum of the second
and the fourth bounds; and finally Eq.~(\ref{eq:2112_positive}) is a sum of the third
and the fourth bounds. On the other hand, these bounds are still weaker than those
obtained from the extremal approach in Section~\ref{sec:extremalRep}: 
while they can all be reproduced by the extremal approach, the latter
also gives Eqs.~\eqref{eq:new1} and \eqref{eq:new2}, which cannot be obtained by the
elastic approach.

\subsubsection{Analytical bounds for $W$-boson and $B$-boson}
\label{sec:WandB}
We have seen that, for the $W$-boson case, the elastic positivity problem
can be nicely solved by finding the ERs of the polyhedral cone $Q_P$. This
is because the symmetries of the problem are sufficient to restrict the
number of ERs to be finite.  Adding the hypercharge boson $B$, the set of ERs
becomes infinite. This significantly increases the difficulty of the problem.

Following a similar approach, we first specify a basis for $\mc S$, whose elements
are invariant under both gauge transformations and rotations along the
beam axis.  Instead of the amplitude basis, we directly give all $\vec p(u,v)$.
We denote
$u^i=(r^{a\alpha},p^{c})$, $v^j=(s^{b\beta},q^{d})$, etc.,
where $a,b$ are the polarization indices of the $W$-boson modes,
$\alpha,\beta$ are the gauge indices for the $W$-boson modes,
and $c,d$ are the polarization indices for the $B$-boson modes.
We found that the following $p_\alpha (u,v)$
\begin{flalign}
\label{eq:jaksdf}
	&p_1(u,v)=(r^{a\alpha}s^{a\alpha})^2
	\\
	&p_2(u,v)=(r^{a\alpha}s^{b\alpha}\epsilon^{ab})^2
	\\
	&p_3(u,v)=(r^{a\alpha}s^{a\beta}\epsilon^{\alpha\beta})^2
	\\
	&p_4(u,v)=(r^{a\alpha}s^{b\beta}\epsilon^{ab}\epsilon^{\alpha\beta})^2
	\\
	&p_5(u,v)=-4
	\left(r^{a\alpha}r^{b\beta}\epsilon^{ab}\epsilon^{\alpha\beta}\right)
	\left(s^{a\alpha}s^{b\beta}\epsilon^{ab}\epsilon^{\alpha\beta}\right)
	\\
	&p_6(u,v)=\left(p^cq^c\right)
	\left(r^{a\alpha}s^{a\alpha}\right)
	\\
	&p_7(u,v)=\left(p^cq^d\epsilon^{cd}\right)
	\left(r^{a\alpha}s^{b\alpha}\epsilon^{ab}\right)
	\\
	&p_8(u,v)=\sum_{\alpha=1,2}\left(
        r^{a\alpha}q^a+s^{a\alpha}p^a
	\right)^2
	\\
	&p_9(u,v)=\sum_{\alpha=1,2}\left(
	r^{a\alpha}q^b\epsilon^{ab}+s^{a\alpha}p^b\epsilon^{ab}
	\right)^2
	\\
	&p_{10}(u,v)=\left(p^aq^b\sigma_1^{ab}\right)
	\left(r^{a\alpha}s^{b\alpha}\sigma_1^{ab}\right)
	+\left(p^aq^b\sigma_3^{ab}\right)
	\left(r^{a\alpha}s^{b\alpha}\sigma_3^{ab}\right)
	\\
	&p_{11}(u,v)=\left(p^aq^a\right)^2
	\\
	&p_{12}(u,v)=\left(p^aq^b\epsilon^{ab}\right)^2
	\label{eq:12ps}
\end{flalign}
can be constructed from a set of basis amplitudes that are
sufficient to describe the SMEFT amplitudes.
Similar to the previous case, the corresponding coefficients $C_1,\cdots
C_{12}$ can be mapped to the Wilson coefficients:
\begin{align}
	\begin{aligned}
&C_1=\frac{16}{s_W^4} \left(2 F_{T,0}+2
   F_{T,1}+F_{T,2}\right) \\
& C_2=\frac{8}{s_W^4} \left(F_{T,2}+4 F_{T,10}\right) \\
& C_3=\frac{4}{s_W^4} \left(4 F_{T,1}+F_{T,2}\right)
   \\
& C_4=\frac{4}{s_W^4} F_{T,2} \\
& C_5=\frac{8}{s_W^4} \left(2 F_{T,1}+F_{T,2}\right)
   \\
& C_6=\frac{2}{c_W^2 s_W^2} \left(8
   F_{T,5}+F_{T,7}\right) \\
	\end{aligned}
	\qquad
	\begin{aligned}
& C_7=\frac{2}{c_W^2 s_W^2} \left(F_{T,7}+4 F_{T,11}\right) \\
& C_8=\frac{1}{c_W^2 s_W^2} \left(4
   F_{T,6}+F_{T,7}\right) \\
& C_9=\frac{1}{c_W^2 s_W^2} F_{T,7} \\
& C_{10}=\frac{2}{c_W^2 s_W^2} F_{T,7} \\
& C_{11}=\frac{4}{c_W^4} \left(2
   F_{T,8}+F_{T,9}\right) \\
& C_{12}=\frac{2}{c_W^4} F_{T,9}
	\end{aligned}
	\label{eq:12Cs}
\end{align}
with a common factor $4\alpha^2\pi^2/\Lambda^4$ divided.

We now need to study the boundary of the set of all possible $\vec p(u,v)$.
To this end, we notice that a number of inequalities need to be satisfied by $\vec p$.
First, the following $p_i$'s are non-negative, as they are themselves complete
squares or sums of complete squares:
\begin{flalign}
	p_i\ge 0,\ \mbox{for}\ i=1,2,3,4,8,9,11,12
	\label{eq:WBneq1}
\end{flalign}
In addition, same as the $W$-boson case, four
combinations of $p_1,\cdots p_5$ are non-negative:
\begin{flalign}
&p_1+p_2+p_5\ge0,\quad p_1+p_3+p_5\ge0,\nonumber\\
&p_2+p_4+p_5\ge0,\quad p_3+p_4+p_5\ge0\,.
\label{eq:WBneq2}
\end{flalign}
Also, a few other complete squares that are linear in $\vec p$ can be formed:
\begin{flalign}
	&p_8+p_9-4p_6=
	\sum_{\alpha=1,2}\left(r^{a1}q^{b}\sigma_1^{ab}-s^{a1}p^{b}\sigma_1^{ab} \right)^2
	+
	\sum_{\alpha=1,2}\left(r^{a1}q^{b}\sigma_3^{ab}-s^{a1}p^{b}\sigma_3^{ab} \right)^2
	\\
	&p_1+2r_1p_6+r_1^2p_{11}=\left[r_1\left(p^cq^c\right)
	+\left(r^{a\alpha}s^{a\alpha}\right)\right]^2
	\\
	&p_2+2r_2p_7+r_2^2p_{12}=\left[r_2\left(p^cq^d\epsilon^{cd}\right)
	+\left(r^{a\alpha}s^{b\alpha}\epsilon^{ab}\right)\right]^2
	\\
	&p_1+p_2+p_5+2r_3p_{10}+r_3^2(p_{11}+p_{12})=
	\sum_{I=1,3}\left[r_3\left(p^cq^d\sigma_I^{cd}\right)
	+\left(r^{a\alpha}s^{b\alpha}\sigma_I^{ab}\right)\right]^2
\end{flalign}
Here, $r_1,r_2,r_3$ are free parameters. These squares lead to the following inequalities
\begin{flalign}
&	-4p_6+p_8+p_9\ge0,\ 
	p_1p_{11}-p_6^2\ge0,\ 
	p_2p_{12}-p_7^2\ge0,\ 
	\nonumber\\
	& (p_1+p_2+p_5)(p_{11}+p_{12})-p_{10}^2\ge0\,.
\label{eq:WBneq3}
\end{flalign}
Note that the last three inequalities are quadratic. They come from
the fact that, for example, $p_1+2r_1p_6+r_1^2p_{11}=0$ cannot have two distinct
real solutions for $r_1$.
Finally, one additional nonlinear inequality can be written down:
\begin{flalign}
	&p_8+p_9-2\sqrt{p_1+p_2+p_3+p_4+p_5}\sqrt{p_{11}+p_{12}}
	\nonumber\\&
	-2\left(r^{a\alpha}s^{a\alpha}\right)\left(p^cq^c\right)
	+2\left(r^{a\alpha}s^{b\alpha}\epsilon^{ab}\right)\left(p^cq^d\epsilon^{cd}\right)
	=\left( \norm{r} \norm{q}-\norm{s}\norm{p} \right)^2
	\nonumber\\
	~~\Rightarrow\quad &
	p_8+p_9-2\sqrt{p_1+p_2+p_3+p_4+p_5}\sqrt{p_{11}+p_{12}}
	+2\left( \sqrt{p_1p_{11}} + \sqrt{p_2p_{12}} \right)\ge0 &
\label{eq:WBneq4}
\end{flalign}

Similar to the previous case, 
all these inequalities together define a region $\mc Q_R$.
A difference however is that $\mc Q_R$ itself is not a convex cone, because
Eq.~(\ref{eq:WBneq4}) does not describe a convex boundary. Therefore, instead of
$\mc Q_R$, we should consider conv($\mc Q_R$). Due to the free parameters $r_i$ and the nonlinear
inequalities, conv($\mc Q_R$) is not polyhedral.  Also, unlike the $W$-boson
case, conv($\mc Q_R$) is not equal to $Q_P$. Its ERs cannot always be written
as $\vec p_i(u,v)$, so they are not always in $Q_P$. This may imply that
additional inequalities about $p_i$ may be derived to further restrict $\mc
Q_R$, but we have not been able to identify them. We will only use the inequalities
in Eqs.~(\ref{eq:WBneq1}),(\ref{eq:WBneq2}),(\ref{eq:WBneq3}),(\ref{eq:WBneq4}),
and for the ERs of conv($\mc Q_R$), we will only keep the ones that are inside $\mc
Q_P$, to avoid over constraining $\mc C_S^{el}$. For this reason, our approach
does not enumerate all the ERs of $\mc Q_P$, and so our $\mc C_A^{el}$ is 
slightly conservative.

To find the ERs of conv($\mc Q_R$), we simply follow the
definition of ERs, and search in the set $\mc Q_R$.
We first notice that $p_8$ and $p_9$
are only bounded from below. Given $p_8\ge0$ and $p_9\ge0$, one can easily
check:
\begin{flalign}
	&\vec E_{0,1}=(0,0,0,0,0,0,0,1,0,0,0,0)\\
	&\vec E_{0,2}=(0,0,0,0,0,0,0,0,1,0,0,0)
\end{flalign}
are extremal. The rest inequalities only involve $p_8+p_9$. Therefore, we shall
first remove $p_8$ and $p_9$ from the problem, find the ERs of the
rest $p_i$'s, and then for each ER check the minimum value of
$p_8+p_9$, allowed by all inequalities.
Let $x=\mbox{min}(p_8+p_9)$. If $x\le0$, setting $p_8=p_9=0$ is
extremal in $\mc Q_R$; if $x>0$, setting $p_8=x,p_9=0$ and $p_8=0,p_9=x$ leads to
two ERs in $\mc Q_R$.

For the rest $p_i$'s, according to the definition of an ER, 
if $\vec p\in \mc Q_R$ is extremal (with $p_{8,9}$ removed), then if we write
\begin{flalign}
	\vec p=\vec p_a+\vec p_b, \ \vec p_{a,b}\in \mc Q_R
\end{flalign}
the only possible solutions are those $\vec p_{a,b}\parallel \vec p$.
One possibility for this to happen is that $\vec p$ saturates 9 linear bounds,
of the form {$\vec p \cdot \vec k_i=0$}, because
\begin{flalign}
	&\vec p_{a,b} \in \mc Q_R\ \Rightarrow\ \vec p_a\cdot \vec k_i\ge0,~ \vec
	p_b\cdot \vec k_i\ge0,\\
	&\vec p\cdot \vec k_i=0 \quad \Rightarrow\quad \vec p_a\cdot \vec k_i+\vec p_b\cdot \vec k_i = 0
	\qquad \Rightarrow\quad \vec p_{a,b}\cdot \vec k_i=0
\end{flalign}
and since the dimension of $\vec p$ is 10 (with $p_{8,9}$ removed), 9 such
constraints will leave only one possible solution for $p_{a,b}$, up to a scalar
multiplication. 

Alternatively, a quadratic
inequality, if saturated, counts two constraints.  To see this, notice that
there are 3 such inequalities in Eqs.~(\ref{eq:WBneq3}), which all have the
form (if saturated):
\newcommand{\SP}[2]{\left(\vec{p}_{#1}\cdot\vec{k}_{#2}\right)}
\newcommand{\Sp}[2]{\vec{p}_{#1}\cdot\vec{k}_{#2}}
\begin{flalign}
 (\vec p\cdot\vec k_1)(\vec p\cdot\vec k_2) = (\vec p\cdot\vec k_3)^2,\quad
 \mbox{with }\SP{}{1}\ge0,\ \SP{}{2}\ge0
\end{flalign}
where $\SP{}{1,2}\ge0$ are due to other linear bounds in $\mc Q_R$.
Splitting $\vec p$ into $\vec p_a+\vec p_b$,
\begin{flalign}
	&\vec p_{a,b} \in \mc Q_R\ \Rightarrow\ 
(\vec p_{a,b}\cdot\vec k_1)(\vec p_{a,b}\cdot\vec k_2) \ge (\vec p_{a,b}\cdot\vec k_3)^2,\ 
\SP{a,b}{1,2}\ge0,\ 
\nonumber\\
&0\le\left[ \sqrt{\SP{1}{1}\SP{2}{2}}-\sqrt{\SP{2}{1}\SP{1}{2}}\right]^2
\nonumber\\=&
\left[\Sp{1}{1}+\Sp{2}{1}\right]\left[\Sp{1}{2}+\Sp{2}{2}\right]
-\left[ \sqrt{\SP{1}{1}\SP{1}{2}}+\sqrt{\SP{2}{1}\SP{2}{2}}\right]^2
\nonumber\\\le& \SP{}{1}\SP{}{2}-\left(
\left|\Sp{1}{3}\right|+\left|\Sp{2}{3}\right| \right)^2
\le \SP{}{1}\SP{}{2}-\SP{}{3}^2=0\,,
\end{flalign}
so all inequalities between the two zeros on both sides are all saturated,
which implies
\begin{flalign}
	\frac{\Sp{1,2}{1}}{\Sp{}{1}}= \frac{\Sp{1,2}{2}}{\Sp{}{2}}
	= \frac{\Sp{1,2}{3}}{\Sp{}{3}}
\end{flalign}
This counts as two constraints, unless $\Sp{}{1}=0$ or $\Sp{}{2}=0$, in which
case only $\Sp{1,2}{3}=0$ is required, i.e.~the quadratic constraint is
trivialized to one linear constraint (the other one requires
$\Sp{1,2}{1}=0$ or $\Sp{1,2}{2}=0$, which will be already
covered by other linear constraints). 

With these in mind, we should consider all possibilities for $\vec p$ to
saturate 9 such constraints.  Also notice that $p_6$, $p_7$ and
$p_{10}$ are each constrained only by one quadratic inequality, which
means that an ER always has to saturate these 3 inequalities in
Eqs.~(\ref{eq:WBneq3}).  Therefore we shall consider several cases, depending
on how many of the 3 quadratic bounds are trivialized (i.e.~either
$\Sp{1,2}{1}=0$ or $\Sp{1,2}{2}=0$).
For convenience, let us denote the
5-dimensional subspace $\{(p_1,p_2,\cdots, p_5)\}$ by $\mc P_W$, and 
the 2-dimensional subspace $\{(p_{11},p_{12})\}$ by $\mc P_B$.
Note that in Eqs.~\eqref{eq:WBneq1} and \eqref{eq:WBneq2}, there
are 8 linear constraints that apply to $\mc P_W$ and 2 that apply to $\mc P_B$.
Now let us consider the following cases.
\begin{enumerate}
	\item All three quadratic bounds are trivially satisfied. Six
		additional linear bounds need to be saturated. We can take at
		most five in $\mc P_W$, or two in $\mc P_B$, otherwise they are not
		independent.  If we take five in $\mc P_W$ with one in $\mc P_B$,
		all components in $\mc P_W$ will vanish, leaving two ERs:
\begin{flalign}
	&\vec E_{0,3}=(0,0,0,0,0,0,0,0,0,0,1,0)\,,\\
	&\vec E_{0,4}=(0,0,0,0,0,0,0,0,0,0,0,1)\,,
\end{flalign}
	while if we take four in $\mc P_W$ with two in $\mc P_B$, all components in $\mc
	P_B$ will vanish, leaving the seven ERs that we already found in the
	$W$-boson case:
\begin{flalign}
	&\vec E_{0,5}=(1,0,0,0,0,0,0,0,0,0,0,0)\,,\\
	&\vec E_{0,6}=(0,1,0,0,0,0,0,0,0,0,0,0)\,,\\
	&\vec E_{0,7}=(0,0,1,0,0,0,0,0,0,0,0,0)\,,\\
	&\vec E_{0,8}=(0,0,0,1,0,0,0,0,0,0,0,0)\,,\\
	&\vec E_{0,9}=(0,0,0,0,1,0,0,0,0,0,0,0)\,,\\
	&\vec E_{0,10}=(1,0,0,1,-1,0,0,0,0,0,0,0)\,,\\
	&\vec E_{0,11}=(0,1,1,0,-1,0,0,0,0,0,0,0)\,.
\end{flalign}
\item One quadratic constraint is non-trivial. Five additional linear constraints are needed.
	We can only take four in $\mc P_W$ and one in $\mc P_B$, otherwise all three quadratic
	constraints will be trivialized. After adding $p_{8,9}$ and discarding
	the ones that do not admit a solution for $u,v$ satisfying $\vec E=\vec p(u,v)$,
	we find that the following rays are extremal, under the condition $r>0$:
\begin{flalign}
 &\vec E_{1,1} =(0,0,0,0,1,0,0,2 r,0,r,r^2,0    ) \\
 &\vec E_{1,2} =(0,0,0,0,1,0,0,0,2 r,-r,r^2,0   ) \\
 &\vec E_{1,3} =(0,0,0,0,1,0,0,2 r,0,r,0,r^2    ) \\
 &\vec E_{1,4} =(0,0,0,0,1,0,0,0,2 r,-r,0,r^2   ) \\
 &\vec E_{1,5} =(0,1,0,0,0,0,0,2 r,0,r,r^2,0    ) \\
 &\vec E_{1,6} =(0,1,0,0,0,0,0,0,2 r,-r,r^2,0   ) \\
 &\vec E_{1,7} =(0,1,1,0,-1,0,r,0,0,0,0,r^2     ) \\
 &\vec E_{1,8} =(0,1,1,0,-1,0,-r,2 r,2 r,0,0,r^2) \\
 &\vec E_{1,9} =(1,0,0,0,0,0,0,2 r,0,r,0,r^2    ) \\
 &\vec E_{1,10}=(1,0,0,0,0,0,0,0,2 r,-r,0,r^2   ) \\
 &\vec E_{1,11}=(1,0,0,1,-1,r,0,2 r,2 r,0,r^2,0 ) \\
 &\vec E_{1,12}=(1,0,0,1,-1,-r,0,0,0,0,r^2,0    )
\end{flalign}

\item Two quadratic constraints are non-trivial, three linear constraints in $\mc P_W$
	and one in $\mc P_B$ are saturated. We find the following ones parameterized
	by $r>0$:
\begin{flalign}
& \vec E_{1,13}=\left(1,0,0,0,0,-r,0,0,0,r,r^2,0\right) \\
& \vec E_{1,14}=\left(1,0,0,0,0,-r,0,0,0,-r,r^2,0\right) \\
& \vec E_{1,15}=\left(1,0,0,0,0,r,0,4 r,0,r,r^2,0\right) \\
& \vec E_{1,16}=\left(1,0,0,0,0,r,0,0,4 r,-r,r^2,0\right) \\
& \vec E_{1,17}=\left(0,1,0,0,0,0,r,0,0,-r,0,r^2\right) \\
& \vec E_{1,18}=\left(0,1,0,0,0,0,r,0,0,r,0,r^2\right)
\end{flalign}
and two additional ones parameterized by $r,s$, with $r\ge |s|$:
\begin{flalign}
&\vec E_{2,1}=\left(r^2,0,0,r^2-s^2,s^2-r^2,-r,0,0,0,s,1,0\right) \\
&\vec E_{2,2}=\left(0,r^2,r^2-s^2,0,s^2-r^2,0,r,0,0,s,0,1\right) 
\end{flalign}

\item Two quadratic constraints are non-trivial, four linear constraints in $\mc P_W$
	and none in $\mc P_B$ are saturated. We find four ERs parameterized by
	$r,s$, with $r\ge |s|$:
	\begin{flalign}
& \vec E_{2,3}=\left(1,0,0,0,0,s,0,2 (r+s),0,r,s^2,r^2-s^2\right) \\
& \vec E_{2,4}=\left(1,0,0,0,0,s,0,0,2 (r+s),-r,s^2,r^2-s^2\right) \\
& \vec E_{2,5}=\left(0,1,0,0,0,0,-s,2 (r+s),0,r,r^2-s^2,s^2\right) \\
& \vec E_{2,6}=\left(0,1,0,0,0,0,-s,0,2 (r+s),-r,r^2-s^2,s^2\right)
	\end{flalign}

\item Three quadratic constraints are non-trivial. We find that in this case
	all ERs that admit a solution of $\vec E=\vec p(u,v)$ reduce to the
	previous cases.
\end{enumerate}

In total, we find 11 discrete ERs $\vec E_{0,i}$, 18 continuous sets of ERs parameterized
by one free parameter, $\vec E_{1,i}(r)$ with $r>0$, and 6 continuous sets of ERs
parameterized by two free parameters, $\vec E_{2,i}(r,s)$ with $r>|s|$.
Now positivity bounds are given by $\vec C \cdot \vec E_{a,i}\ge0$. For $a=0$ the
bounds are straightforward. For $a=1$, $\vec C \cdot \vec E_{1,i}\ge0$ gives a quadratic
polynomial in $r$, because
\begin{flalign}
	Ar^2+Br+C\ge0,\ \forall r\ge0 
	\quad \Rightarrow\quad A\ge0\ \mbox{and}\ C\ge0\ \mbox{and}\ -B\le2\sqrt{AC}
\end{flalign}
So the resulting bound is a quadratic one, $-B\le2\sqrt{AC}$, while $A\ge0$ and
$B\ge0$ are covered by $\vec E_{0,i}$'s.
For $a=2$, $\vec C \cdot \vec E_{2,i}\ge0$ gives a quadratic polynomial in $r$ and $s$
\begin{flalign}
A r^2 + B s^2 +C r + D s + E \ge0,\ \forall r \ge |s|
\end{flalign}
which is equivalent to
\begin{flalign}
	& A \ge 0\ \mbox{and}\ E \ge 0\ \mbox{and}\ -C \le 2\sqrt{AE}\ \mbox{and}\ A+B \ge 0\label{eq:extra_rs}\\
	\mbox{and}\ &|D|-C\le 2\sqrt{(A+B)E}\\
	\mbox{and}\ & (B\le0,\ \mbox{or}\ A|D| + B C \ge 0,\  \mbox{or}\ 4 A B E \ge A D^2 + B C^2).
\end{flalign}
However, constraints in Eq.~\eqref{eq:extra_rs} are already covered
by $a=0$ and $a=1$ cases, so they can be discarded.  Putting everything
together, we obtain around 40 inequalities.  Numerically, by a Monte Carlo
sampling of the coefficient space we find that some of them are redundant, so
in the following we only list the independent
ones.

\begin{description}
	\item[Linear]
\begin{flalign}
&F_{T,2}\geq 0 \label{eq:l1}\\
\label{eq:nfbound1}
&4 F_{T,1}+F_{T,2}\geq 0 \\
&F_{T,2}+8 F_{T,10}\geq 0 \label{eq:l3}\\
&8 F_{T,0}+4 F_{T,1}+3 F_{T,2}\geq 0 \label{eq:l4}\\
&4 F_{T,6}+F_{T,7}\geq 0 \label{eq:l5}\\
&F_{T,7}\geq 0 \label{eq:l6}\\
&2 F_{T,8}+F_{T,9}\geq 0 \label{eq:l7}\\
&F_{T,9}\geq 0 \label{eq:l8}
\end{flalign}

\item[Quadratic]
\begin{flalign}
  &F_{T,9} \left(F_{T,2}+4 F_{T,10}\right)\geq F_{T,11}^2 \\
  &16 \left(2 \left(F_{T,0}+F_{T,1}\right)+F_{T,2}\right) \left(2
   F_{T,8}+F_{T,9}\right)\geq \left(4 F_{T,5}+F_{T,7}\right){}^2
   \\
  &2 \sqrt{2} \sqrt{F_{T,9} \left(F_{T,2}+8 F_{T,10}\right)} 
  +4F_{T,6}+F_{T,7}-4F_{T,11}\geq0\\
  &4 \sqrt{\left(8 F_{T,0}+4 F_{T,1}+3 F_{T,2}\right) \left(2
   F_{T,8}+F_{T,9}\right)}
 + 8 F_{T,5}+4 F_{T,6}+3 F_{T,7} \geq0
\label{eq:nfbound2}
\end{flalign}

\item[Cubic]
\begin{flalign}
&\left(4 F_{T,0}+F_{T,2}\right) F_{T,7}\geq 4 \left(4
   F_{T,1}+F_{T,2}\right) F_{T,5}
   \,,
   \nonumber\\&\mbox{or}\quad
   4 \left(4 F_{T,1}+F_{T,2}\right)
   \left(8 F_{T,0}+4 F_{T,1}+3 F_{T,2}\right) \left(2
   F_{T,8}+F_{T,9}\right)
   \nonumber\\&
   \geq 16 \left(4 F_{T,1}+F_{T,2}\right)
   F_{T,5}^2+4 \left(4 F_{T,1}+F_{T,2}\right) F_{T,7} F_{T,5}+\left[2
   \left(F_{T,0}+F_{T,1}\right)+F_{T,2}\right] F_{T,7}^2 
\\
&
F_{T,2} F_{T,7}+4 F_{T,10} F_{T,7}+2 F_{T,2} F_{T,11}\geq 0
   \,,
   \nonumber\\&\mbox{or}\quad
4
   F_{T,2}^2 F_{T,9}\geq 4 F_{T,10} F_{T,7}^2+F_{T,2} \left[F_{T,7}^2+4
   F_{T,11} F_{T,7}+8 \left(F_{T,11}^2-4 F_{T,9} F_{T,10}\right)\right]
\label{eq:nfbound3}
\end{flalign}
\end{description}

The results derived above define our analytical elastic positivity
bounds, $\mc C^{el}_A$.  All linear and quadratic bounds from the
factorization approach, presented in Section~\ref{sec:fac}, are
superseded by these bounds.
For the linear bounds, Eqs.~\eqref{eq:0410_positive},
\eqref{eq:0010_positive}, \eqref{eq:fg_positive},
\eqref{eq:fg_positive3}, \eqref{eq:0410mix_positive},
 and \eqref{eq:ct7_positive} in the factorization approach are the same as
 Eqs.~\eqref{eq:nfbound1}, \eqref{eq:l1}, \eqref{eq:l7}, \eqref{eq:l8}, \eqref{eq:l5},
 and \eqref{eq:l6}.
Eq.~\eqref{eq:2210_positive} in the factorization approach is obtained by
adding up Eq.~\eqref{eq:nfbound1} and \eqref{eq:l4}.
Similarly, Eq.~\eqref{eq:0014_positive} can be derived from Eq.~\eqref{eq:l1}
and \eqref{eq:l3}, and Eq.~\eqref{eq:2112_positive} can be derived from
Eq.~\eqref{eq:l3} and \eqref{eq:l4}.  Numerically, we can also confirm that the
quadratic conditions~\eqref{eq:quad1}--\eqref{eq:quad3} from the factorization
approach are covered.

The cubic bound in Eq.~\eqref{eq:cubic} in the factorization approach
is however not fully covered. 
This demonstrates the incompleteness of this approach, namely the inexactness
of our determination of $\mc Q_P$.
The difference is tiny: a randomly chosen point has
only $\sim 10^{-8}$ probability to satisfy all above bounds
but violate Eq.~\eqref{eq:cubic}. Such a difference can be safely ignored in any
realistic application of these bounds.

\begin{figure}[htb]
\begin{center}
\includegraphics[width=.5\linewidth]{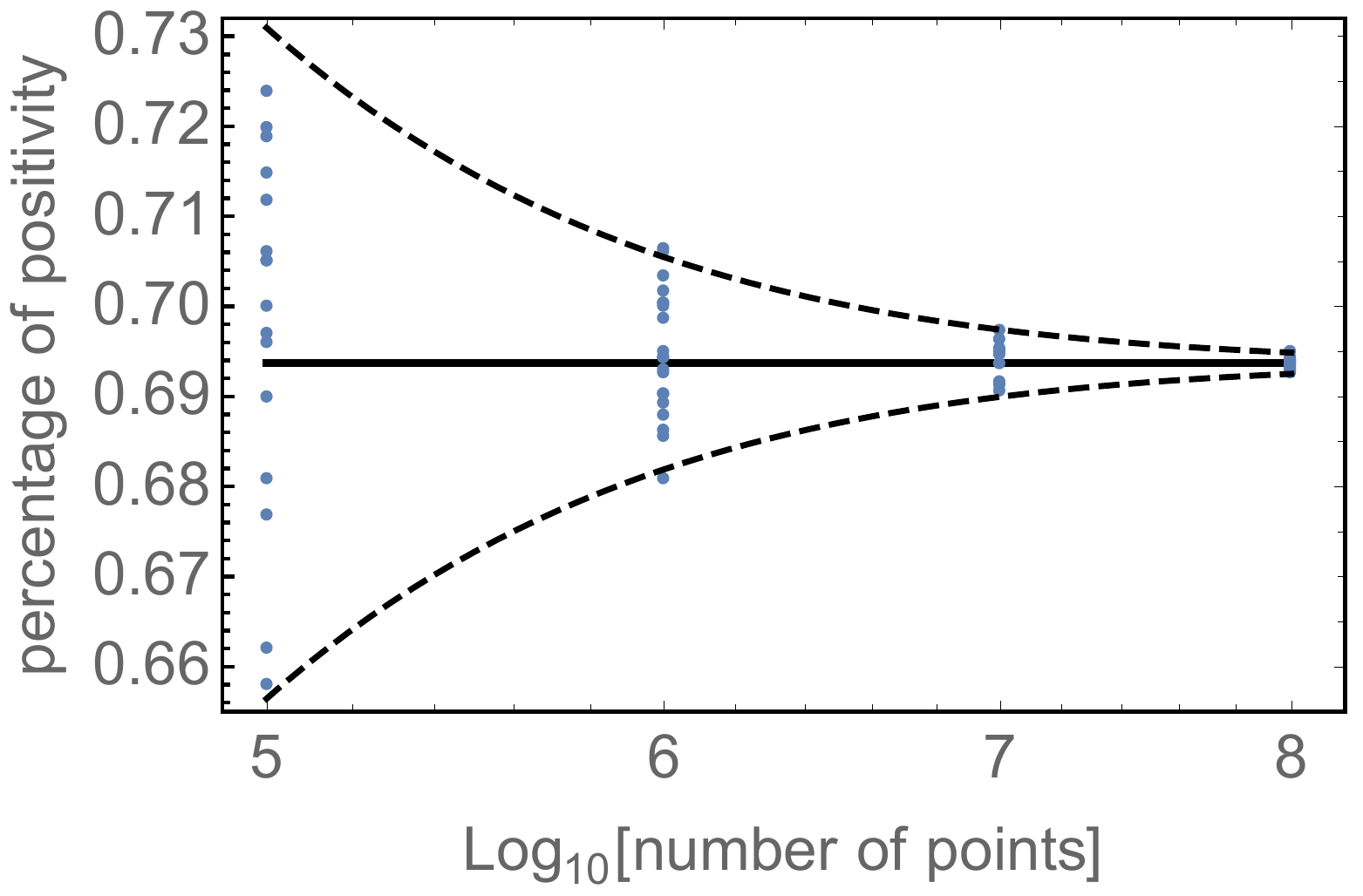}
\end{center}
\caption{Percentages of the parameter space, or solid angle 
	$\Omega(\mc C^{el}_{A})$, of analytical elastic positivity bounds,
	Eqs.~\eqref{eq:l1}-\eqref{eq:nfbound3}, computed from different
	samplings ($10^5, 10^6, 10^7$ or $10^8$ points). Each blue point is a
	sampling. The horizontal axis is the number of sampling points, while
	the vertical axis is the resulting $\Omega(\mc C^{el}_{A})$.  The solid
line (0.6937\%) is the average of 15 largest samplings, each with $10^8$
points, while the dashed lines are 2$\sigma$ errors.}
\label{fig:genAnaMC}
\end{figure}

In terms of constraining power, our analytical elastic positivity bounds correspond to
a solid angle of about $\Omega(\mc C^{el}_{A})=0.694\%$,
which improves the elastic factorization bounds $\Omega(\mc C^{el}_{AF})= 0.891\%$. In Figure
\ref{fig:genAnaMC}, 
we show $\Omega(\mc C^{el}_{A})$ computed from different samplings,
each consisting of up to $10^8$ points.
The average of 15 samplings, with $10^8$ points in each sampling, is
\begin{equation}
\label{Omegaana}
\Omega(\mc C^{el}_{A}) = (0.6937\pm 0.00021)\%
\end{equation}
where the statistical $1 \sigma$ error quoted is the square root of the
sampling variance.

\subsubsection{Adding $W_{x,y}^3$}
\label{sec:W3}
So far we have not used the third component of the $W$-boson in the $SU(2)_L$ triplet
space. The two modes $W_{x}^3$ and $W_{y}^3$ have been ignored. In the following we
show that adding them in the elastic approach does not change the conclusion.

First, we need to extend the basis in Eqs.~(\ref{eq:jaksdf})-(\ref{eq:12ps}) in order to
describe the $P(u,v)$ in this more general case. This can be done by the
following operation:
\begin{itemize}
	\item For $p_i(u,v)$ with $i=1,2,6,7,8,9,10$, simply extend the summation
		of the $\alpha$ index to $\sum_{\alpha=1}^3$.
	\item For $p_i(u,v)$ with $i=3,4,5$, cycle the $\alpha$ and $\beta$
		index and take the sum, e.g.
		\begin{flalign}
			p_i(u,v)\rightarrow
			p_i(u,v)+p_i(u,v)(1\to2\to3\to1)+p_i(u,v)(1\to3\to2\to1)
		\end{flalign}
		where $1\to2\to3\to1$ or  $1\to3\to2\to1$ means that the values
		of the $\alpha,\beta$ indices are cycled.
	\item For $p_i(u,v)$ with $i=11,12$, use the same $p_i(u,v)$.
\end{itemize}
With this new $p_i(u,v)$ and the same $C_i$ values in Eq.~(\ref{eq:12Cs}),
one again has $p_\alpha(u,v)=M_\alpha^{ijkl}u^iv^ju^kv^l$, and $P(u,v)=C_\alpha
p_\alpha(u,v)\ge0$. We need to investigate how $Q_R$ is changed with the above
operation. First, the new components in $u,v$ are $r^{13}$, $r^{23}$,
$s^{13}$, $s^{23}$. Using the $SU(2)_L$ rotations around the first and the second
axes, together with the $SO(2)$ rotation of the polarization space, one can
eliminate 3 of them. Let us keep the $r^{13}$ component. By expanding
the new $p_i(u,v)$'s, we see that only the following ones are changed:
\begin{flalign}
	&p_3(u,v)\to p_3(u,v)+ \left( r^{13} \right)^2
	\left[ \left(s^{11}\right)^2+\left(s^{12}\right)^2 \right]
	\\
	&p_4(u,v)\to p_4(u,v)+ \left( r^{13} \right)^2
	\left[ \left(s^{21}\right)^2+\left(s^{22}\right)^2 \right]
	\\
	&p_8(u,v)\to p_8(u,v)+ \left( r^{13} \right)^2 \left( q^{1} \right)^2
	\\
	&p_9(u,v)\to p_9(u,v)+ \left( r^{13} \right)^2 \left( q^{2} \right)^2
\end{flalign}
These shifts are a positively weighed sum of $\vec E_{0,7}$, $\vec E_{0,8}$,
$\vec E_{0,1}$, and $\vec E_{0,2}$, which themselves are already in
$Q_R$. Therefore these shifts do not create any elements outside cone($Q_R$).
As a result, the ERs of cone($Q_R$) remain unchanged, and so
the resulting bounds are the same.

\subsection{Numerical bounds}
\label{sec:num}

For a given set of Wilson coefficients, it is possible to numerically
verify whether the full set of elastic positivity bounds is satisfied, for
arbitrary polarizations and superpositions of gauge modes.  Sampling the entire
space with Monte Carlo method allows us to numerically determine the region
$\mc C_S^{el}$, and to compute its solid angle $\Omega(\mc C_S^{el})$.  By
comparing the latter with $\Omega(\mc C^{el}_{A})$, we will see that the analytical
bounds obtained in Section~\ref{sec:nonfac} are extremely close to the full
positivity bounds.

To decide whether a given set of Wilson coefficients satisfy the full elastic
positivity bound, we plug the set of Wilson coefficients into the constraint
$P(u,v)\ge0$, which should hold for all $u,v\in \mb R^8$. The naive way forward is to let $u^i$
and $v^j$ randomly sample the 8-sphere for many times, and
check whether $P(u,v)\ge0$ holds. This method however becomes inefficient if we
want to scan the full parameter space or even a continuous sub-region. To speed up
the process, we recast the problem as an autonomous dynamical system 
\cite{Cheung:2016yqr}.

To see how this works, we
combine $u^i$ and $v^j$ into a 16-component vector $x^I=(u^i, v^j)$,
and relax the normalization constraints on $u^i$ and $v^j$. We let
$x^I$ depend on some fictitious ``time'' $t$ and evolve according to an
autonomous dynamical system
\begin{equation}
\frac{d x^I}{d t} = - \frac{d P}{d x^I} ,
\end{equation}
where $P=P(u,v)\ge0$ is the full elastic positivity bound. Then we generate an
initial value for $x^I$, whose components take random values in the interval
$[-1,1]$, and evolve the dynamical system for some time (say $\Delta t=100$).
If the value of $P$ at the final time becomes negative, we know that this set
of Wilson coefficients violates positivity. If the value of $P$ at the final
time is positive, we re-run the autonomous system for more randomly chosen initial
$x^I$'s, to see whether $P$ can become negative. If all these attempts fail,
we declaim that the full elastic positivity bound is satisfied for this set of
Wilson coefficients. 

The reason why this dynamical system approach can guide us
to the choices of $u^i$ and $v^j$ that violate the positivity can be seen from
the fact that $P$ is always decreasing with time
$t$: 
\begin{equation}
\frac{d P}{d t} = \sum_I  \frac{d P}{d x^I} \frac{d x^I}{d t}  = -  \sum_I  \frac{d P}{d x^I} \frac{d P}{d x^I}< 0  .
\end{equation}
away from $x^I=0$ ($d P/d t\neq 0$ because $P$ is a quartic form). It is
prudent that we run the autonomous system for several time as the autonomous
system can have several equilibria in its phase space and also there can be
numerical errors in the evolution. 

\begin{figure}[htb]
\begin{center}
\includegraphics[width=.5\linewidth]{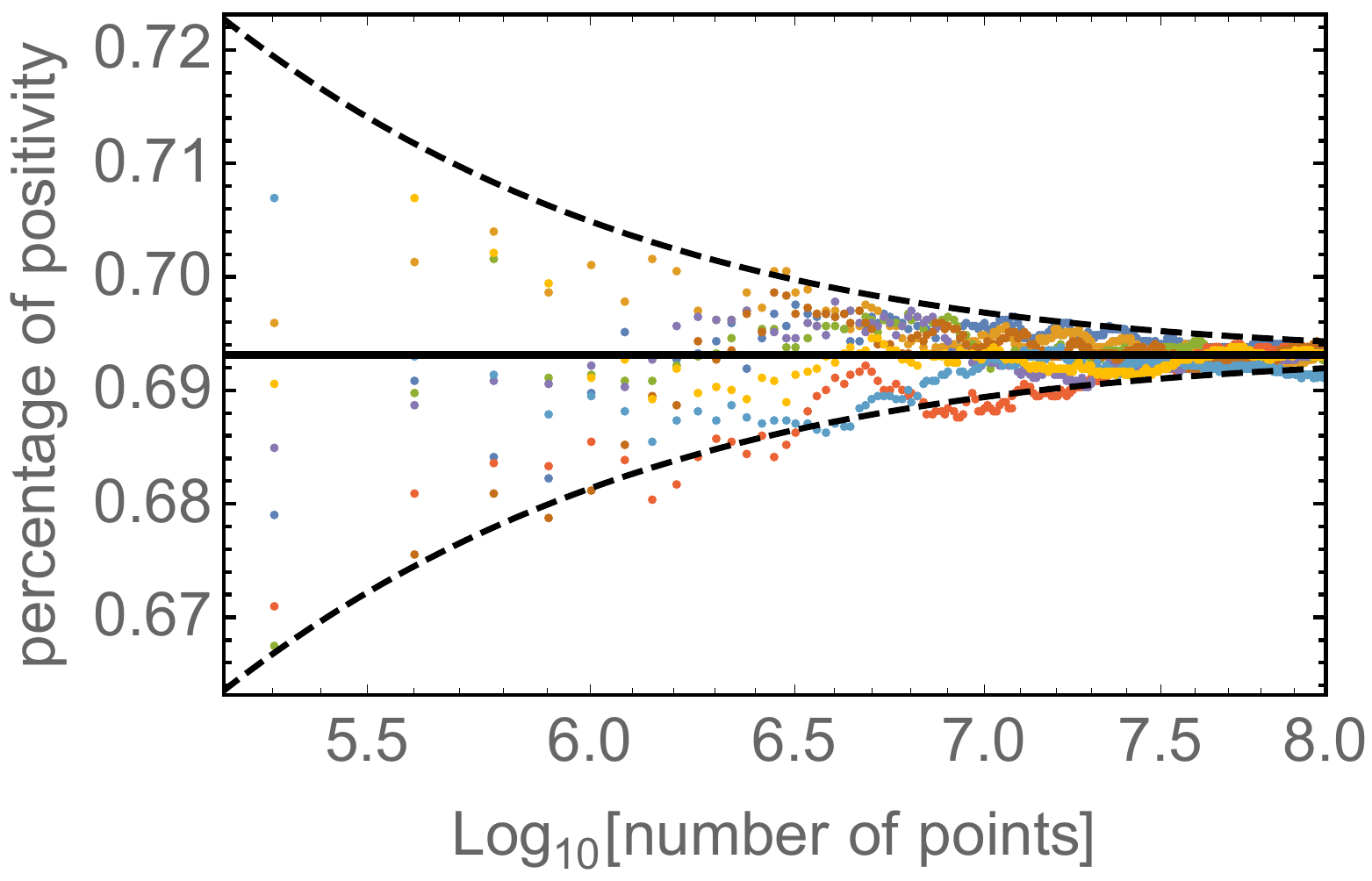}
\end{center}
\caption{Numerical results for the percentages of the parameter space that
satisfy the full elastic positivity bounds. Each point in the plot represents a
sampling, and points with the same color come from the same $10^8$ sampling.
The horizontal axis is the number of points used for the sampling and the
vertical axis is the percentage of these points satisfying the full elastic
positivity bounds. The horizontal solid line (0.6937\%) is the average of ten
$10^8$-points samplings, while the dashed lines are 2$\sigma$ errors (square
root of the sampling variance) for the samplings.}
\label{fig:numMC}
\end{figure}

{The numerical bounds obtained by this approach, denoted by $C^{el}_N$,
can in principle still be conservative if $P$ has some local positive minima
(thus multiple attractors in the phase space of the dynamical system) and all
the initial $x^I$ seeds are accidentally trapped in the attracting basins of
these local minima. We have however checked our results, by varying $\Delta t$
and increasing the number of the initial $x^I$ seeds used for the evolutions,
and found that the resulting $\Omega(\mc C^{el}_N)$ is stable. Therefore, our $C^{el}_N$
should give an accurate description of $C^{el}_S$.}

With this approach, it is straightforward to sample the parameter space
and compute $\Omega(\mc C^{el}_N)$ following the discussion in
Section~\ref{sec:geometry}.
The average of the ten $10^8$ samplings is
\begin{equation}
\label{Omeganum}
	\Omega(\mc C_N^{el}) = (0.6931\pm 0.00026)\%
\end{equation}
where the statistical $1 \sigma$ error quoted is the square root of the
sampling variance. 
With the same method, we have also explicitly verified
that $\mc C^{el}_A\supsetneq \mc C^{el}_N$, which consists of two facts:
1) there are no Wilson coefficients that satisfy the full elastic positivity
but violate the analytical bounds $\mc C^{el}_A$, which means $\mc C^{el}_A$ is
indeed conservative, providing a consistency check for our
results obtained in Section~\ref{sec:nonfac}; and 2)
there are sets of Wilson coefficients that satisfy the analytical bounds $\mc C^{el}_A$
but violate the numerical elastic positivity bounds.
Instances of the latter indeed occur with a probability consistent with the
difference between Eq.~(\ref{Omegaana}) and Eq.~(\ref{Omeganum}), $\sim 0.0006\%$.

To conclude, we have demonstrated that the our analytical results for $\mc C^{el}_A$
are extremely close to the full elastic positivity bounds, with a relative difference
of about $0.1\%$.
We thus conclude that our analytical approach presented in
Section~\ref{sec:nonfac} is, even though incomplete in principle, accurate
enough to capture the full elastic positivity bounds in all practical applications.

\section{The extremal positivity bounds}
\label{sec:extremal}
In this section, we adopt the extremal positivity approach. While
the case for the $W$-boson operators have been discussed already in
Section~\ref{sec:extremalRep}, here we will need to include
the hypercharge boson $B$ as well. We will see that, similar to the elastic positivity
case, this makes the problem more difficult. In particular, the PERs, from
which we construct the convex cone of the physically allowed region $\mc C_S$,
are continuously distributed. The resulting cone is not polyhedral anymore.
We will briefly discuss a general solution to this kind of problems, and 
then present both analytical and numerical results.

Recall that in the $W$-boson case, the PERs are constructed by combining the
projection operators of both the $SU(2)$ and $SO(2)$ group. Adding the $B_{x,y}$ modes
induces more projectors.  Since $B$ lives in the same representation as $W$ in
the $SO(2)$ rotational group, the corresponding projectors of this group remain
the same. We however need to modify the gauge group projectors. Since the
$\mathbf 5$ irrep cannot be formed by two $B$'s or one $W$ and one $B$, we can
keep the corresponding projector $P^{(3)}$. As for $P^{(1)}$ and
$P^{(2)}$, we will ignore their normalization, as this has no
effect on the determination of the parameter space. The $P^{(1)}$ projector can
now be constructed for any linear combination of the singlet in $WW$ and the singlet in $BB$:
\begin{flalign}
	&
	P^{(1)}(r)_{\alpha\beta\gamma\sigma}=\frac{1}{3}d_{\alpha\beta}(r)d_{\gamma\sigma}(r)\\
	&d_{\alpha\beta}(r)=\left\{\begin{array}{ll}
		1\quad & \alpha=\beta=1,2,3 \\
		r\quad & \alpha=\beta=4\\
		0\quad & \mbox{otherwise}
	\end{array}\right.
\end{flalign}
where the indices $\alpha,\beta,\gamma,\sigma$ represent the 3 components
of $W$ and the $B$-boson, for values $1,2,3,4$ respectively. $r$ is a
real number that parameterizes the mixing between the $WW$ and $BB$ singlets.
This means that there is an infinite set of $P^{(1)}$, parameterized by a free
parameter $r$. 

The $P^{(2)}$ can be written in a similar way, for any linear combination
of triplets in $WW$, $WB$ and $BW$:
\begin{flalign}
	& P_S^{(2)}(r_1,r_2)_{\alpha\beta\gamma\sigma}
	=\frac{1}{2}\sum_{i=1}^3f^i_{\{\alpha,\beta\}}(r_1,r_2)f^i_{\{\gamma,\sigma\}}(r_1,r_2)\\
	& P_A^{(2)}(r_1,r_2)_{\alpha\beta\gamma\sigma}
	=\frac{1}{2}\sum_{i=1}^3f^i_{[\alpha,\beta]}(r_1,r_2)f^i_{[\gamma,\sigma]}(r_1,r_2)\\
	&
	f^1_{\alpha\beta}(r_1,r_2)=
	\left(
	\begin{array}{cccc}
		0 & 0 & 0 & r_1 \\
		0 & 0 & 1 & 0 \\
		0 & -1 & 0 & 0 \\
		r_2 & 0 & 0 & 0 
	\end{array}
	\right),\ 
	f^2_{\alpha\beta}(r_1,r_2)=
	\left(
	\begin{array}{cccc}
		0 & 0 & -1 & 0  \\
		0 & 0 & 0 & r_1 \\
		1 & 0 & 0 & 0 \\
		0 & r_2 & 0 & 0 
	\end{array}
	\right),\ 
	f^3_{\alpha\beta}(r_1,r_2)=
	\left(
	\begin{array}{cccc}
		0 & 1 & 0 & 0  \\
		-1 & 0 & 0 & 0 \\
		0 & 0 & 0 & r_1 \\
		0 & 0 & r_2 & 0 
	\end{array}
	\right) 
\end{flalign}
which are freely parameterized by two real numbers, $r_1$ and $r_2$.
The PERs are constructed from the symmetrized projector $P^{(2)}_S$
and the anti-symmetrized $P^{(2)}_A$, which are obtained by further symmetrizing the
$\beta$ and $\delta$ indices. They essentially depend on
at most one free parameter ($r_1+r_2$ or $r_1-r_2$), which we will
simply denote as $r$ and use the notation $P^{(2)}_{S,A}(r)$ instead.

Among all the PERs, there are in total 13 linearly independent terms. We choose
the following ones to form a basis:
\begin{flalign}
	&B_1=(1,3),\ B_2=(3,3),\ B_3=(1,2)_S,\ B_4=(3,2)_S,\ B_5=(1,1)^0,\\
	&B_6=(1,1)^1,\ B_7=(1,1)^2,\ B_8=(3,1)^0,\ B_9=(3,1)^1,\\
	&B_{10}=(3,1)^2,\ B_{11}=(2,2)_S^0,\ B_{12}=(2,2)_S^1,\ B_{13}=(2,2)_S^2. 
\end{flalign}
where $(m,n)$ represents that the projectors $P^{(m)}$ for $SO(2)$ and
$P^{(n)}$ for $SU(2)$ are combined. The subscript $S$ indicates that we take
the symmetric components (i.e.~$P^{(2)}_S$ is used if $m=1,3$, or $P^{(2)}_A$
is used if $m=2$.) The superscript $0,1,2$ means that the free parameter
$r$ is present, and we take the coefficient of the $r^0$, $r^1$, or $r^2$ term,
respectively.  All PERs are at most a quadratic function of $r$.  With this
basis, the full set of PERs can be written as a set of 13-vectors:
\begin{flalign}
&
\vec{e}_1=(1,0,0,0,0,0,0,0,0,0,0,0,0)
\label{eq:PERs1}
\\&
\vec{e}_2=(0,1,0,0,0,0,0,0,0,0,0,0,0)
\\&
\vec{e}_3=(0,0,1,0,0,0,0,0,0,0,0,0,0)
\\&
\vec{e}_4=(0,0,0,1,0,0,0,0,0,0,0,0,0)
\\&
\vec{e}_5=\left(-\frac{1}{6},\frac{1}{6},0,0,-\frac{5}{3},0,0,\frac{5}{3},0,0,\frac{5}{6},0,0\right)
\\&
\vec{e}_6=\left(0,0,-1,1,0,-\frac{3}{4},0,0,\frac{3}{4},0,0,0,1\right)
\\&
\vec{e}_7(r)=\left(0,0,0,0,1,r,r^2,0,0,0,0,0,0\right)
\\&
\vec{e}_8(r)=\left(0,0,0,0,0,0,0,1,r,r^2,0,0,0\right)
\\&
\vec{e}_9(r)=\left(0,0,0,0,0,0,0,0,0,0,1,r,r^2\right)
\\&
\vec{e}_{10}(r)=\left(-\frac{1}{3},\frac{1}{3},-\frac{4 r}{3},\frac{4 r}{3},-\frac{1}{3},0,-r^2,\frac{1}{3},0,r^2,-\frac{1}{3},0,-\frac{4 r}{3}\right)
\\&
\vec{e}_{11}(r)=\left(\frac{1}{2},\frac{1}{2},\frac{r^2}{2},\frac{r^2}{2},-1,-\frac{3 r^2}{8},0,-1,-\frac{3 r^2}{8},0,-\frac{1}{2},r,-\frac{r^2}{2}\right)
\\&
\vec{e}_{12}(r)=\left(1,0,r^2,0,-2,-\frac{3 r^2}{4},0,0,0,0,1,-2 r,r^2\right)
\label{eq:PERs2}
\end{flalign}
Their {conical} hull defines the cone $\mathcal{C}$. 

To find its intersection with the physical subspace, $\mathcal{C}_S$, one can
write the amplitude as $M=\sum_i f_iB_i$, and a comparison with
$M^{ijkl}=\sum_\alpha C_\alpha M_\alpha^{ijkl}$ gives the following relations
between the operator coefficients and $f_i$'s:
\begin{flalign}
	\begin{aligned}
&f_ 1= 4s_W^{-4} \left(8 F_{T,1}+F_{T,2}-8 F_{T,10}\right) \\
&f_ 2= 4s_W^{-4} \left(F_{T,2}+8 F_{T,10}\right) \\
&f_ 3= 4s_W^{-2}c_W^{-2} \left(8 F_{T,6}+F_{T,7}-4 F_{T,11}\right) \\
&f_ 4= 4 s_W^{-2}c_W^{-2} \left(F_{T,7}+4 F_{T,11}\right) \\
&f_ 5= 16 s_W^{-4} \left(6 F_{T,0}+2 F_{T,1}+F_{T,2}-2 F_{T,10}\right) \\
&f_ 6= 3 s_W^{-2}c_W^{-2} \left(8 F_{T,5}+F_{T,7}\right) \\
&f_ 7= 6 c_W^{-4} \left(4 F_{T,8}+F_{T,9}\right) 
\end{aligned}
\qquad
	\begin{aligned}
&f_ 8   = 16 s_W^{-4} \left(F_{T,2}+2 F_{T,10}\right) \\
&f_ 9   = 3 s_W^{-2}c_W^{-2} F_{T,7} \\
&f_ {10}= 6 c_W^{-4} F_{T,9} \\
&f_ {11}= 4 s_W^{-4}\left(F_{T,2}-8 F_{T,10}\right) \\
&f_ {12}= 0 \\
&f_ {13}= 4 s_W^{-2}c_W^{-2} \left(F_{T,7}-4 F_{T,11}\right) \,,
 \\
\label{eq:fFrelation}
\end{aligned}
\end{flalign}
(a common factor $4\alpha^2\pi^2/\Lambda^4$ has been divided.)
These relations can be use to convert the bounds of $\mc C$ to those of $\mc C_S$.
Now we have the cone $\mathcal{C}$ determined through the extremal representation, with all PERs
given in Eq.~(\ref{eq:PERs1})-(\ref{eq:PERs2}), while $\mathcal{C}_S$ is
the set of points in $\mc S$ which, after substituting into the above
relations, are inside $\mathcal{C}$.

Our goal is to find the inequality representation so that the positivity bounds
can be extracted.  While the conversion to this representation can be
easily done in the case of polyhedral cones, the problem is nontrivial
for our non-polyhedral cone $\mc C$,
as obviously the bounds cannot be described by a finite number of linear
inequalities.  Obtaining the full set of analytical bounds is indeed difficult
due to the 6 free parameters in $\vec{e}_7(r)$ to $\vec{e}_{12}(r)$.
Nevertheless, analytical bounds up to the quadratic level is relatively simple
to obtain, and, as an incomplete set of bounds, they give a conservative
description of the allowed physical parameter space.  We will denote these
bounds by $\mc C_{A}$, and will show how they can be derived in
Section~\ref{sec:erana}.  Alternatively, in Section~\ref{sec:ernum}, the numerical
bounds $\mc C_N$ will be obtained, by sampling $\vec
e_i(r)$ with a sufficiently large number of $r$ values.  
In Section~\ref{sec:ercomparison}, we will see that both approaches lead to a
very good approximation of the full set of extremal positivity bounds.

\subsection{Analytical bounds}
\label{sec:erana}
Recall that finding the bounds of $\mathcal{C}$ is
equivalent to finding the ERs of $\mathcal{C}^*$, as $\mc C$ is given by the
collection of points $\vec f$ that satisfy $\vec E_i\cdot \vec f\ge0$, where
$\{\vec E_i\}$ is the ERs of $\mc C^*$.  If $\mc C$ were polyhedral, this would
be a normal vertex enumeration problem.  Unfortunately, due to the $r$ parameters, we
need to solve, in some sense, a continuous version of vertex enumeration. 

For illustration, let us first consider a polyhedral case in which $\mc C$ is a
$d$-dimensional cone constructed from a set of discrete PERs,
$\vec{e}_i$, which are constant vectors. In this case, the ERs of $\mc C^*$ can
be obtained by constructing all possible combinations of $d-1$ $\vec{e}_i$,
i.e.
\begin{flalign}
	E^\alpha = \epsilon^{\alpha_1\alpha_2\dots \alpha_{d-1}\alpha}
	e_{i_1}^{\alpha_1}
	e_{i_2}^{\alpha_2}
	\dots
	e_{i_{d-1}}^{\alpha_{d-1}}
\end{flalign}
that satisfy the following condition:
\begin{flalign}
E^\alpha e_i^{\alpha}=\epsilon^{\alpha_1\alpha_2\dots \alpha_{d-1}\alpha}
	e_{i_1}^{\alpha_1}
	e_{i_2}^{\alpha_2}
	\dots
	e_{i_{d-1}}^{\alpha_{d-1}}
	e_i^{\alpha}
	\equiv\braket{e_{i_1},e_{i_2},\dots,e_{i_{d-1}},e_i}
	\ge0,\quad \forall\ \vec e_i
	\label{eq:523}
\end{flalign}
where $\epsilon$ is the $d$-dimensional Levi-Civita tensor, and $\{\vec
e_{i_1},\vec e_{i_2},\dots,\vec e_{i_{d-1}}\}$ is any combination of $d-1$
$\vec e_i$'s.
This is simply because a facet of $\mc C$ is always in a hyperplane
spanned by a combination of $d-1$ edges, such that the entire cone
is on one side of this hyperplane; the latter condition is guaranteed
by Eq.~\eqref{eq:523}, because any point
$\vec f$ in $\mc C$ can be written as a positive sum of $\vec e_i$.
It's also easy to check that $E^\alpha$ is extremal in $\mc C^*$,
using the fact that $E^\alpha e_{i_1}^\alpha=E^\alpha e_{i_2}^\alpha=\cdots=0$,
i.e.~it saturates $d-1$ bounds of $\mc C^*$.
If a PER which is not really extremal is included in the
combination, it will not give an independent bound.  
Therefore, the positivity bounds can be written as
\begin{flalign}
	\braket{e_{i_1},e_{i_2},\dots,e_{i_{d-1}},f}
	\ge0
\end{flalign}
subject to Eq.~(\ref{eq:523}). This is in fact how vertex enumeration can be solved
by brute force.

If some of the $\vec e_i$'s are functions of a free parameter $r$, the
procedure above still works in principle, but with some notable differences:
\begin{itemize}
	\item If $\vec e_i(r_1)$ is included in the combination, then 
		$\vec e_i{}'(r_1)$, the derivative w.r.t
		$r$ at $r_1$, must also be included, unless
		it is not linearly independent of the rest vectors in the combination. This
		is because $\vec E$ needs to be also orthogonal
		to its neighboring ERs.
	\item The $r$ parameters in all $\vec e_i$'s can be chosen independently.
	\item The condition $\braket{e_{i_1},e_{i_2},\dots,e_{i_{d-1}},e_i} \ge0$
		should be satisfied for any $e_i$ with any choices of $r$.
\end{itemize}
For example, one might construct an ER that depends on 3
		independent parameters:
\begin{flalign}
	E^\alpha(r_1,r_2,r_3) = \epsilon^{\alpha_1\alpha_2\dots \alpha_{d-1}\alpha}
	e_{i_1}^{\alpha_1}(r_1)
	e'_{i_1}{}^{\alpha_2}(r_1)
	e_{i_2}^{\alpha_3}(r_2)
	e'_{i_2}{}^{\alpha_4}(r_2)
	e_{i_3}^{\alpha_5}(r_3)
	e'_{i_3}{}^{\alpha_6}(r_3)
	\dots
\end{flalign}
and this has to satisfy
\begin{flalign}
	E^\alpha(r_1,r_2,r_3)e_i^\alpha(r)\ge0
	\label{eq:ineqexample}
\end{flalign}
for all $\vec e_i$ and all $r\in \mb R$, in order to represent a faithful bound.
If an $\vec e_i$ is a quadratic function of $r$, Eq.~\eqref{eq:ineqexample}
defines a quadratic condition on the components of $E^\alpha(r_1,r_2,r_3)$
(i.e.~$Ar^2+Br+C\ge0\Rightarrow A\ge0,\ C\ge0,\ B^2\le 4AC)$. Therefore, the
resulting bound has the following form: an up to 6-order polynomial (of
$r_1,r_2,r_3$) needs to be PSD, namely $\vec E\cdot \vec f\ge 0$, in a
range constrained by the inequalities \eqref{eq:ineqexample}. 

In our problem, $d=13$ and there are 6 ERs that depend on free
$r$ parameters. From the discussion above, we know that obtaining the full analytical
results is difficult, as it involves the determination of high-degree PSD
polynomials subject to nonlinear constraints. In fact, as soon as $\vec e_i$
depends on a second independent free parameter, the bound involves the
determination of 4th order PSD polynomials, which is already difficult.  To avoid over
complicating this problem, we consider combinations of $\vec e_i$'s that depend, in total,
on at most one common parameter $r_0$, through at most a quadratic function.
In particular, each $\vec e_i(r)$ is only allowed to take either $\vec e_i(0)$,
$\vec e_i(\infty)$, or $\vec e_i(\pm r_0)$, in the construction of $E^\alpha$.
Note that this is different from sampling $\vec e_i(r)$ with only 4 values, not
only because
$\vec e_i{\,}'(0)$,
$\vec e_i{\,}'(\infty)$,
$\vec e_i{\,}'(\pm r_0)$
are also used in the construction, but also because
$E^\alpha(r_0)e_i^\alpha(r)\ge0$ is satisfied for all $r$ values, which are not
sampled but solved analytically. This means that $E^\alpha\in \mc C^*$ 
holds strictly, so the resulting bounds are conservative.

We adopt the above simplified approach and find 10 linear bounds and 59 quadratic
bounds.  Furthermore, the 59 quadratic bounds are very well-approximated by a
subset, constituted by 9 quadratic bounds: a randomly chosen point only has
a chance smaller than $10^{-6}$ to satisfy the subset while violating the
complete set. We will thus replace the 59 quadratic bounds by the subset.
Together, these bounds define our analytical extremal positivity bounds
$\mc C_{A}$. We list our results as follows:

\begin{description}
	\item[Linear]
\begin{flalign}
\label{eq:erbound1}
&F_{T,2}\geq 0 \\
&4 F_{T,1}+F_{T,2}\geq 0\\
&F_{T,2}+8 F_{T,10}\geq 0 \\
&8 F_{T,0}+4 F_{T,1}+3 F_{T,2}\geq 0 \\
&12 F_{T,0}+4 F_{T,1}+5 F_{T,2}+4 F_{T,10}\geq 0 \\
&4 F_{T,0}+4F_{T,1}+3 F_{T,2}+12 F_{T,10}\geq 0 \\
&4 F_{T,6}+F_{T,7}\geq 0 \\
&F_{T,7}\geq 0 \\
&2 F_{T,8}+F_{T,9}\geq 0 \\
&F_{T,9}\geq 0
\end{flalign}

\item[Quadratic]

\begin{flalign}
 & F_{T,9} \left(F_{T,2}+4 F_{T,10}\right)\geq F_{T,11}^2 \\
 & 16 \left(2 \left(F_{T,0}+F_{T,1}\right)+F_{T,2}\right) \left(2
   F_{T,8}+F_{T,9}\right)\geq \left(4 F_{T,5}+F_{T,7}\right){}^2 \\
 & 32 \left(2 F_{T,8}+F_{T,9}\right) \left(3 F_{T,0}+F_{T,1}+2 F_{T,2}+4
   F_{T,10}\right)\geq 3 \left(4 F_{T,5}+F_{T,7}\right){}^2 \\
 & 2 \sqrt{2} \sqrt{F_{T,9} \left(F_{T,2}+8 F_{T,10}\right)}\geq \max
   \left(-F_{T,7}-4 F_{T,11},-4 F_{T,6}-F_{T,7}+4 F_{T,11}\right)\\
 & 4 \sqrt{\left(8 F_{T,0}+4 F_{T,1}+3 F_{T,2}\right) \left(2
   F_{T,8}+F_{T,9}\right)}
   \nonumber\\&\quad 
   \geq \max \left(-8 F_{T,5}-4 F_{T,6}-3 F_{T,7},8
   F_{T,5}+F_{T,7}\right) \\
 & 4 \sqrt{F_{T,9} \left(12 F_{T,0}+4 F_{T,1}+5 F_{T,2}+4 F_{T,10}\right)}
   \nonumber\\&\quad 
 \geq \max
   \left(-F_{T,7}-4 F_{T,11},-4 F_{T,6}-F_{T,7}+4 F_{T,11}\right) \\
 & 4 \sqrt{6} \sqrt{\left(2 F_{T,8}+F_{T,9}\right) \left(12 F_{T,0}+4 F_{T,1}+5
   F_{T,2}+4 F_{T,10}\right)}
   \nonumber\\&\quad 
   \geq \max \left[3 \left(8 F_{T,5}+F_{T,7}\right),-3 \left(8 F_{T,5}+4
   F_{T,6}+3 F_{T,7}\right)\right] \\
 & \sqrt{6} \sqrt{\left(4 F_{T,8}+3 F_{T,9}\right) \left(6 F_{T,0}+2 F_{T,1}+3
   F_{T,2}+6 F_{T,10}\right)}
   \nonumber\\&\quad 
   \geq \max \left(3 \left(2 F_{T,5}+F_{T,11}\right),-3 \left(2
   F_{T,5}+F_{T,7}+F_{T,11}\right)\right) \\
 & 2 \sqrt{\left(12 F_{T,8}+7 F_{T,9}\right) \left(12 F_{T,0}+4 F_{T,1}+5 F_{T,2}+4
   F_{T,10}\right)}
   \nonumber\\&\quad 
   \geq \max \left(-12 F_{T,5}-4 F_{T,6}-5 F_{T,7}-2 F_{T,11},
   12 F_{T,5}+F_{T,7}-2F_{T,11},
   \right.\nonumber\\&\quad \left.
   12 F_{T,5}+F_{T,7}+2 F_{T,11},
   12 F_{T,5}-4 F_{T,6}+F_{T,7}+2 F_{T,11}\right)
\label{eq:erbound2}
\end{flalign}

\end{description}
The first 6 linear bounds are the same as the $W$-boson-only case, which is
expected.  The solid angle $\Omega(\mc C_{A})$ is $0.687\%$, smaller than
the elastic positivity case.

What can we expect if we gradually relax our restriction (that all $\vec e_i$'s
depend on at most a common $r$)?  Since $E^\alpha$
will become higher degree polynomials, it is likely that we will obtain a
series of polynomial inequalities with increasing degrees, i.e.~Eqs.~\eqref{eq:erbound1}-\eqref{eq:erbound2} followed by cubic bounds, quartic
bounds, and so on. In this sense, what we have obtained here is the leading and
the next-to-leading bounds of this series.  However, strictly speaking, we cannot
claim that the full set of linear and quadratic bounds are included in
Eqs.~\eqref{eq:erbound1}-\eqref{eq:erbound2}, as it is always possible to
decompose a degree-$d$ bound to an infinite number of degree-$(d-1)$ bounds.
In any case, one always needs to find a balance between the accuracy of the
results and the complexity of the method. While the above truncation at the
quadratic level is manageable, the resulting bounds are already very
close to the exact extremal positivity bounds, with only $\sim 1\%$ relative difference. 
We will demonstrate this in the next two sections.

\subsection{Numerical bounds}
\label{sec:ernum}
Our numerical
approach to the extremal positivity bounds proceeds by sampling the
continuous PERs, numerically, by a discrete set of rays. Specifically, we 
sample an ER $\vec e_i(r)$ by the following $2N$ values of $r$:
\begin{flalign}
	&r_i=\left\{\begin{array}{ll}
		\tan\frac{i\pi}{2N},&\quad \mbox{for}\ -N+1 \le i\le N-1
		\\
		\infty,&\quad i=N
	\end{array}\right.
	\label{eq:rsample}
\end{flalign}
where for $r_N=\infty$ we define $\vec
e_i(\infty)\equiv\lim_{r\to\infty}r^{-2}\vec e_i(r)$.
Collecting the values of $\vec{e}_7(r) \sim \vec{e}_{12}(r)$
at the above $2N$ points, together with the $\vec{e}_1 \sim \vec{e}_{6}$
that do not depend on $r$, we have in total $12N+6$ numerical PERs, which we
denote by $\vec e_{\mathrm{num},i}$. Their {conical} hull, after taking
an intersection with the physical subspace $\mc S$, defines
the numerical bound $\mathcal{C}_N$.
It is inscribed in and therefore smaller than $\mathcal{C}_S$, but the
difference decreases as we increase $N$, and so the error due to this numerical approximation
can be taken under control. 

\begin{figure}[ht]
	\begin{center}
		\includegraphics[width=.5\linewidth]{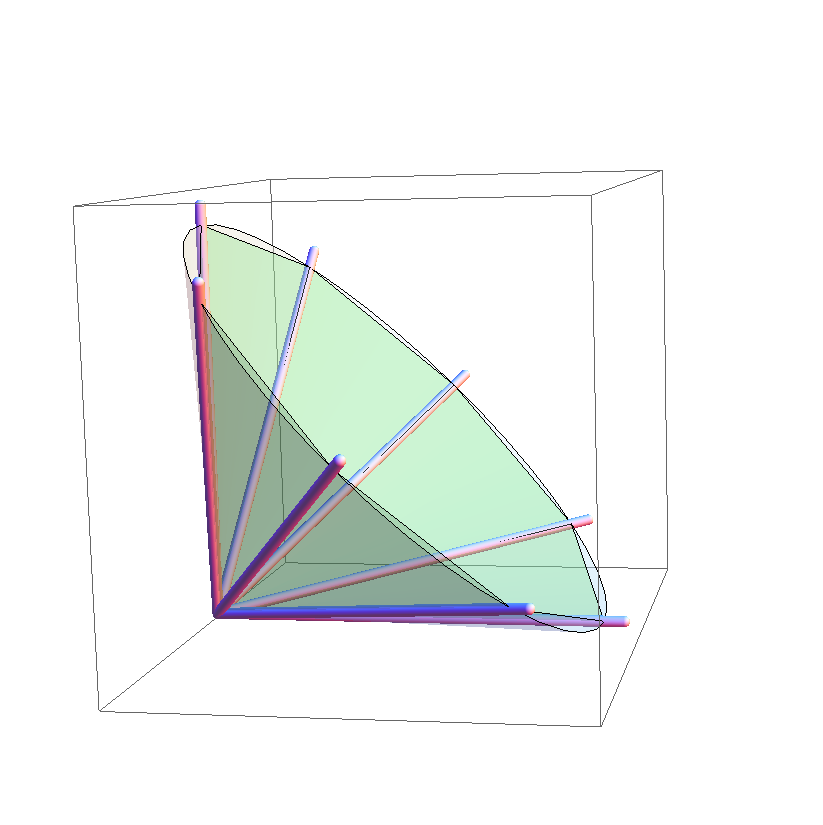}
	\end{center}
	\caption{A circular cone, $(1,\sqrt 2 r,r^2)$, and its inscribed
		octagonal cone, by taking $N=4$ in Eq.~\eqref{eq:rsample}.}
	\label{fig:octagonalcone}
\end{figure}

Let us first estimate how well this approximation is. Note that the conical
hull of a continuous set of ERs of the form $(1,\sqrt2r,r^2)$ represents
a circular cone: by rotating the $x$ and $z$ directions by
$45^\circ$, the rays become
$\big( (1+r^2)/\sqrt{2}$, $\sqrt 2r$, $(1-r^2)/\sqrt{2}
\big)$, which satisfy
\begin{flalign}
	\left( \frac{1+r^2}{\sqrt{2}} \right)^2 =
	\left(\sqrt2r\right)^2+\left( \frac{1-r^2}{\sqrt{2}} \right)^2
\end{flalign}
The $r_i$ values chosen according to Eq.~(\ref{eq:rsample}) generate a set of ERs that
uniformly distribute on this circular cone. In Figure~\ref{fig:octagonalcone} we
show
the case for $N=4$. In this case the eight ER samples form a regular octagonal
cone that is inscribed to the circular cone, $(1,\sqrt 2 r,r^2)$.  To estimate the
difference between the circular cone and the inscribed polyhedral cone with
$2N$ edges, we note that the ratio of the volumes, or the solid angles, between the two
objects, is given by $\frac{n}{2\pi}\sin\left( \frac{2\pi}{n} \right)$, with
$n=2N$.  Since we have 6 continuous sets of ERs, as a rough estimation, we expect
\begin{flalign}
	\frac{\Omega(\mc C_N)}{\Omega(\mc C_S)}\approx
	\left[\frac{n}{2\pi}\sin\left( \frac{2\pi}{n} \right)\right]^6
	=1-\frac{\pi^2}{N^2}+\mathcal{O}(N^{-4})
	\label{eq:volumeratio}
\end{flalign}
This tells us that, to approximate $\mc C_S$ at the per mille level,
taking $N\sim50$ or 100 would be sufficient.

Since $\mc C_N$ is polyhedral, one might think that the most straightforward
way to determine whether a given point $\vec f$ is included in $\mc C_N$, is to
first obtain its facets, as a set of inequalities, and then
check them one by one. For $N\sim \mathcal{O}(100)$, finding the facets of $\mc
C_N$ is, however, not realistic. For $N=5$, with only 66 $\vec e_{\mathrm{num},i}$, we
already have about $400,000$ linear inequalities, thanks to the large dimension
of the problem. We thus take an alternative
approach.  A vector $\vec f=(f_1,f_2,\dots f_{13})$ is included in
$\mathcal{C}_N$, if and only if there exists a set of $12N+6$ real numbers
$w_i\ge0$, such that $\sum_i w_i\vec e_{\mathrm{num,}i}=\vec f$ for the set of
the $12N+6$ numerical PERs, $\vec e_{\mathrm{num,}i}$, 
determined by Eq.~\eqref{eq:rsample}.
In other words, $\vec f$ is a positively weighted sum of $\vec e_{\mathrm{num,}i}$'s.  This can
be written as a linear programming problem:
\begin{flalign}
	\begin{array}{ll}
		\mbox{minimize}\quad
		&0\\
		\mbox{subject to}
		&\sum_i w_i \vec e_{\mathrm{num,}i}=\vec{f},\ w_i\ge0
	\end{array}
	\label{eq:LP}
\end{flalign}
which can be efficiently solved by existing algorithms, such as the simplex method or
the interior point method. Since the objective function is a constant, there is essentially no
minimization, and the algorithm just checks if $\vec f$ is included in
$\mathcal{C}_N$. For example, taking $N=50$, $\mathcal{C}_N$ has 606 ERs.
Using the {\sc LinearProgramming} function in {\sc Mathematica}, checking
the inclusion of $10^5$ points takes less than one minute on a 4-core laptop.
The fraction of points inside $\mathcal{C}_N$ is around $0.68\%$, which is
already a good approximation. 

\subsection{Comparison}
\label{sec:ercomparison}
The analytical bounds defines a cone $\mc C_A$ which contains $\mc C_S$, because
by construction all the $\vec E$ vectors are inside
$\mc C^*$.  On the other hand, the $\mc C_N$ given by the
numerical approach is inscribed in $\mc C_S$. So we have 
\begin{flalign}
	\mc C_N\subset\mc C_S\subset \mc C_A,\quad
	\Omega(\mc C_N)<\Omega(\mc C_S)<\Omega(\mc C_A)
\end{flalign}
and thus a comparison between $\mc C_N$ and $\mc C_A$,
in terms of solid angles, gives us an idea where
exactly $\mc C_S$ lies, and how close $\mc C_N$ and $\mc C_A$ are to the exact
bound. In particular, for $N=250$, we find 
\begin{flalign}
	\Omega(\mc C_A)=0.687\%,\quad \Omega(\mc C_{N})=0.680\%
\end{flalign}
This means that we have
a good estimate of $\mc C_S$, at least at the $1\%$ level, in relative errors.

\begin{figure}[ht]
	\begin{center}
		\includegraphics[width=.49\linewidth]{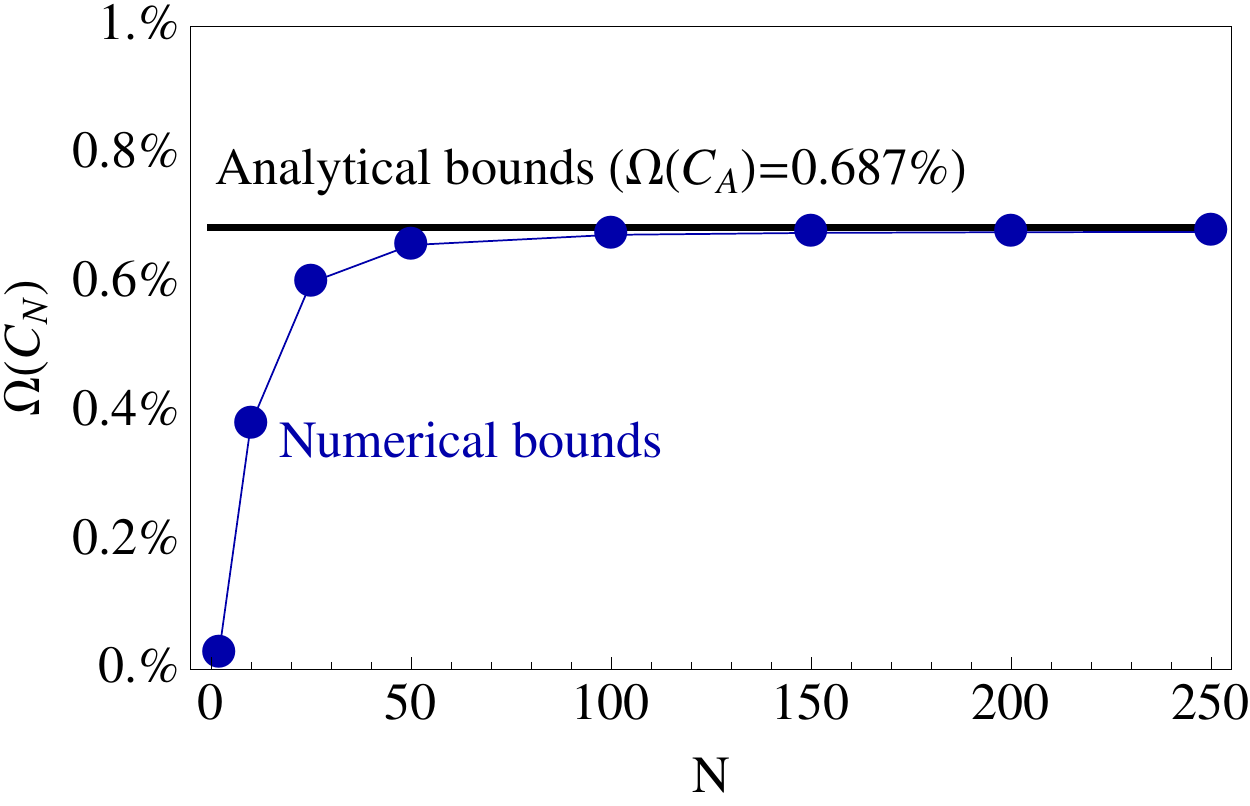}
		\includegraphics[width=.49\linewidth]{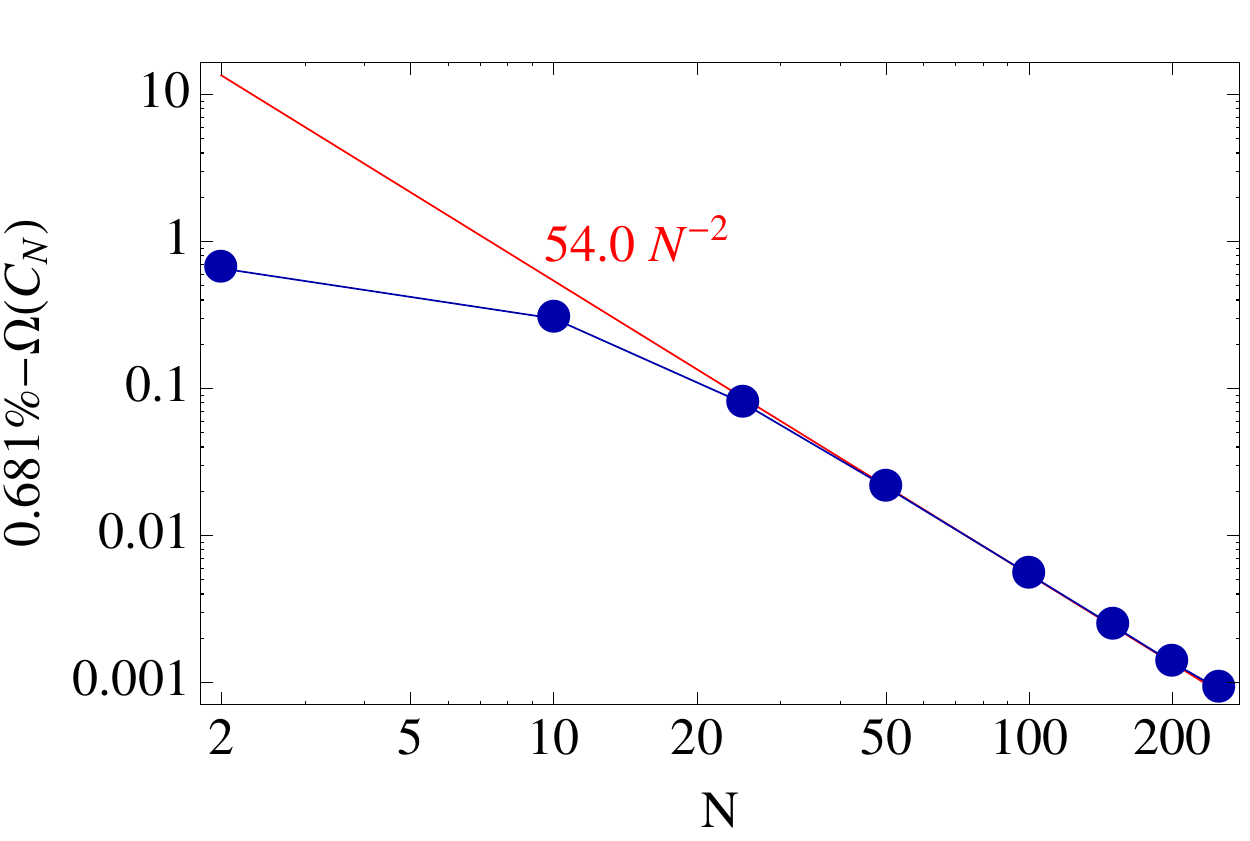}
	\end{center}
	\caption{Solid angle $\Omega(\mc C_N)$ for different $N$ values.}
	\label{fig:Nplot}
\end{figure}

We may also estimate $\Omega(\mc C_S)$ by investigating the behaviour of
$\Omega(\mc C_N)$ as $N$ increases. In the left plot of Figure~\ref{fig:Nplot}, we show how
$\Omega(\mc C_N)$ changes with $N$, and we also compare it with 
$\Omega(\mc C_A)$. The solid angle of $\mc C_S$ is constrained between the black
and the blue lines. 
In addition, for large $N$, Eq.~(\ref{eq:volumeratio}) suggests that
the asymptotic form of $\Omega(\mc C_N)$ is
$\Omega(\mc C_S)(1-aN^{-2})$ for some constant $a$. By fitting $\Omega(\mc
C_N)$ with $N=100,150,200,250$ to this expression, we find that $a\Omega(\mc
C_S)=54.0$ and $\Omega(\mc C_S)=0.681\%$, i.e.
\begin{flalign}
	\Omega(\mc C_N)\approx 0.681\%\left( 1-\frac{79.3}{N^2} \right),\quad
	\mbox{for large $N$}.
\end{flalign}
In the right plot of Figure~\ref{fig:Nplot}, we show the differences between $\Omega(\mc C_N)$
and $0.681\%$, and we see that, indeed, the above expression agrees very well
with our numerical result. This not only indicates that we can confidently
extrapolate $N$ to infinity and obtain
\begin{flalign}
	\Omega(\mc C_S)=0.681\%\,,
\end{flalign}
but also, it illustrates that our numerical approach is reliable and the
fluctuation is well under control.  
In practice, the numerical error due to a finite $N$ is $54.0N^{-2}$, so taking
$N=100$ is already a very accurate approximation, as it leads to a relative
error below $1\%$.

\section{Quadratic dimension-6 contribution}
\label{sec:dim6}

So far we have not considered the quadratic contribution from the dim-6
operator $O_W$. In this section we will briefly discuss how the inclusion of
this operator changes the bounds. Since $O_W$ only enters the $WW$
scattering amplitudes, we will focus on these amplitudes.

First consider the elastic positivity. One can simply restore the $\bar {a}_W^2$
term in Eq.~\eqref{eq:C3def}, which leads to
\begin{flalign}
\begin{array}{rcl}
	\mbox{bounds} & \qquad & \mbox{channel}\ (\ket 1+\ket 2
	\to \ket 1\ +\ket 2)
	\\\hline
	F_{T,2}\ge0,&& \ket 1 = \ket{W_x^1},\ \ket 2=\ket{W_y^2}
	\\
	4F_{T,1}+F_{T,2}\ge 36\bar{a}_W^2,&& \ket 1 = \ket{W_x^1},\ \ket 2=\ket{W_x^2}
	\\
	F_{T,2}+8F_{T,10}\ge 36\bar{a}_W^2,&& \ket 1 = \ket{W_x^1}+\ket{W_y^2},\ \ket 2 = \ket{W_y^1}-\ket{W_x^2}
	\\
	8F_{T,0}+4F_{T,1}+3F_{T,2}\ge00,&& \ket 1 = \ket{W_x^1}+\ket{W_y^2},\ \ket 2 = \ket{W_x^1}+\ket{W_y^2}
	\end{array}
	\label{eq:elasticW2}
\end{flalign}
Comparing with Eq.~\eqref{eq:elasticW}, obviously, the only difference is that
the r.h.s.~of the last two bounds become a non-negative number. Roughly speaking,
the dim-6 coefficients give a positive lower bound on the dim-8 coefficients. On the one hand,
this means that the bounds in Eq.~\eqref{eq:elasticW}, obtained by neglecting $\bar a_W$,
are conservative and valid in general. On the other hand, if the coefficient
of $O_W$ is found to be nonzero in the future, VBS and tri-boson processes will be the next
interesting channels to confirm this, as both $F_{T,2}+8F_{T,10}$ and
$8F_{T,0}+4F_{T,1}+3F_{T,2}$ will have to be non-vanishing. These observations are consistent
with what we found in Refs.~\cite{Zhang:2018shp, Bi:2019phv}.

Now we turn to the extremal approach. While the ERs remain the same, the mapping to
the physical space $\mc S$ will be different, as the latter will now have one more dimension, which
corresponds to the contribution from $\bar a_W^2$. Specifically, the following relations
are modified:
\begin{flalign}
&f_ 1= 4s_W^{-4} \left(8 F_{T,1}+F_{T,2}-8 F_{T,10}-36\bar a_W^2\right) \\
&f_ 2= 4s_W^{-4} \left(F_{T,2}+8 F_{T,10}-36\bar a_W^2\right) \\
&f_ 5= 16 s_W^{-4} \left(6 F_{T,0}+2 F_{T,1}+F_{T,2}-2 F_{T,10}+18\bar a_W^2\right) \\
&f_ 8   = 16 s_W^{-4} \left(F_{T,2}+2 F_{T,10}+18\bar a_W^2\right) \\
&f_ {11}= 4 s_W^{-4}\left(F_{T,2}-8 F_{T,10}+36\bar a_W^2\right)
\label{eq:fFrelation2}
\end{flalign}
To focus on $WW$ scattering, one simply takes the 9 ERs:
$\vec e_1$, $\vec e_2$, $\vec e_5$, $\vec e_7(0)$, $\vec e_8(0)$, $\vec
e_9(0)$, $\vec e_{10}(0)$, $\vec e_{11}(0)$, $\vec e_{12}(0)$,
whose conical hull forms a polyhedral cone with 8 edges. A vertex enumeration
gives the following set of bounds:
\begin{flalign}
	&F_{T,2}\ge0, \\
	&4F_{T,1}+F_{T,2}\ge 36\bar{a}_W^2,
	\label{eq:newnewa} \\
	& F_{T,2}+8F_{T,10}\ge 36\bar{a}_W^2,\\
	& 8F_{T,0}+4F_{T,1}+3F_{T,2}\ge0,
	\label{eq:newnewb} \\
	&12F_{T,0}+4F_{T,1}+5F_{T,2}+4F_{T,10}\ge 0,
	\label{eq:new1new} \\
	&4F_{T,0}+4F_{T,1}+3F_{T,2}+12F_{T,10}\ge 72\bar{a}_W^2.
	\label{eq:new2new}
\end{flalign}

Comparing with the extremal bounds in Section~\ref{sec:extremalRep}, the difference
is that several inequalities become stronger, as the lower bounds become a positive
number ($\bar a_W^2$) rather than 0. The consequence is similar to the case of elastic positivity,
namely that the removal of dim-6 contributions is conservative, and that a nonzero dim-6 coefficient
could also imply non-vanishing dim-8 effects. 
It is likely that this is a general feature of the dim-6 quadratic
contributions in positivity bounds.

\section{Discussion}
\label{sec:discussion}

Recall that the main goals of this work are: 1) to establish the methodology
to obtain complete positivity bounds; 2) to compare the elastic and extremal
positivity approaches; and 3) to get physical results, i.e.~bounds on aQGC
coefficients. Let us discuss these three topics one by one.
\\[5pt]
{\bf \underline{Methodology}} We first summarize the methods that we have adopted in this work.
\begin{itemize}
	\item Elastic positivity:
		\begin{itemize}
			\item The ``factorized'' analytical approach ($\mc
				C^{el}_{AF}$): Assuming that the superposition of
				the helicity states can be factorized from
				those of the additional quantum numbers. This
				reduces the problem
				to a set of quadratically-constrained quadratic
				programming problems. The latter is still
				NP-hard, but if the number of modes is not
				large, they can be solved analytically.
			\item General analytical approach ($\mc C^{el}_{A}$):
				First construct a basis for $M^{ijkl}$ using the
				symmetries of the system, and obtain the $\vec
				p(u,v)$ vectors. Then the boundary of
				$\{\vec p(u,v)\}$ can be identified by inspection
				(e.g.~by finding/constructing complete squares),
				which defines a region $\mc Q_R$. Finding the
				ERs of cone($\mc Q_R$) gives a set of
				analytical bounds.
			\item Numerical approach ($\mc C^{el}_{N}$): For a
				given point in the Wilson coefficient space, minimize
				$P(u,v)$ w.r.t.~$u,v$, by casting the problem
				into a dynamical system and evolve the
				dynamical system.  The sign of the minimum determines whether the point
				satisfies positivity bounds. 
		\end{itemize}
	\item Extremal positivity:
		\begin{itemize}
			\item Analytical approach ($\mc C_{A}$): First
				construct the full set of ERs of $\mc C$ via
				group theoretical considerations. Some of the
				ERs are continuously parameterized by free
				parameters.  To get the linear and quadratic
				bounds, enumerate all ``facets'' determined by
				a combination of ERs, which altogether depend
				on at most one common free parameter.
			\item Numerical approach ($\mc C_{N}$): Sample a
				continuous set of ERs, numerically, with a discrete set of rays.  With
				order $\mc O(10^3)$ sampling rays, use linear
				programming methods to determine the inclusion
				of a given point in the conical hull of these rays.
		\end{itemize}
\end{itemize}

\noindent {\bf \underline{Comparison of elastic and extremal positivity}}
There are several aspects to be compared. The first is the constraining power,
i.e.~how tight these bounds are.  This is quantified by the solid
angle $\Omega$. Below we summarize the solid angles we found for all
approaches, in {\it percentage}.
\begin{flalign}
	\begin{array}{|c|ccc|c|cc|}
		\hline
		\mc C^{el}_S & \mc C^{el}_{AF} & \mc C^{el}_A & \mc C^{el}_N
		& \mc C_S & \mc C_A & \mc C_N
		\\\hline
		0.693 & 0.891 & 0.694 & 0.693 & 0.681 & 0.687 & 0.681-54.0N^{-2}
		\\\hline
	\end{array}
\end{flalign}
where $N$ in the last column is one half of the number of sampling rays for
each continuous set of ERs. Clearly, in all cases, more than $99\%$ of the full
parameter space can be excluded by positivity, because the corresponding SMEFTs
cannot be UV-completed.

In principle, the extremal positivity $\mc C_S$ gives
the tightest bounds that can be set on the SMEFT dim-8 operators by using the
axiomatic QFT principles. The conventional approach based on elasticity, $\mc
C^{el}_S$, is a relaxation of $\mc C_S$, as we have showed in
Section~\ref{sec:elasticapproach}. We see that in our problem the relaxation is
small in terms of the solid angle of the positive region, which amounts to
$\Omega(C^{el}_S)=0.693\%$ compared with $\Omega(\mc C_S) = 0.681\%$.
This is however maybe related with the large dimension of the parameter space, 
which dilutes the differences that mainly arise in the $W$ sector. In the
relevant subspaces, the differences
may be larger, as for example shown in Figure~\ref{fig:ww}.

In practice, due to the complexity of the problem, our analytical approaches
are conservative and do not describe exactly the full bounds. We however see that
$\mc C^{el}_A$ and $\mc C_A$ are accurate at the $0.1\%$ and $1\%$ level, 
compared with $\mc C^{el}_S$ and $\mc C_S$ respectively,
which should be sufficient in many circumstances. The ``factorized'' analytical
elastic bounds, $\mc C^{el}_{AF}$, are more conservative, which is also expected
as the factorization assumption used there is a strong one. This remains an interesting
result thanks to the simplicity of the approach, and may be useful when the exactness
of the bounds is not important. Finally, the numerical approach for the elastic positivity
$\mc C_{N}^{el}$ should describe $\mc C_S^{el}$ exactly, while that for the extremal
positivity $\mc C_{N}$ can be over-constraining, if $N$ is not sufficiently
large. This effect becomes negligible as soon as we take $N\gtrsim 100$.

A second aspect, which may be even more important, is whether an approach can
be systematically algorithmized, i.e.~clearly defined as a set of procedures
that definitely lead to the answer. An approach with this feature is more
valuable as it is applicable to more complicated problems, for example to
derive bounds for more SMEFT operators. Our numerical approaches clearly satisfy this
criteria.  For the analytical approaches, the extremal positivity $\mc C_A$
wins in this aspect, as the enumeration of both the projective operators and the
``facets'' can be algorithmized.
We are not able to algorithmize the elastic positivity $\mc C_A^{el}$,
mainly because of the difficulty in identifying the boundary of $\{\vec p(u,v)\}$. We
are not aware of a systematic way for doing this, and in this work this was done
by inspection: we searched for various kinds of complete squares by trying
different combinations. This is neither systematic nor complete, and is
tedious. While in this work it has worked adequately, as the resulting bound is
only $0.1\%$ weaker than the exact one, it is more likely to be insufficient
for more complicated problems, as for example for the full set of aQGCs including
longitudinal modes, because further increasing the number of modes seems to
significantly increase the difficulty.  On the other hand, the $\mc
C_{AF}^{el}$ approach has a clearly defined set of procedures which turns the
problem into a quadratically-constrained quadratic programming problem.
Solving the latter, however, is analytically realistic only when the number of modes is not
large, so this approach is also not suitable for more complicated cases.

What exactly is the origin of the difficulty in $\mc C^{el}_A$? We have showed
that elastic positivity corresponds to a relaxation of the $\mc C_S$ cone
to $\mc Q_P^*=\mathrm{cone}\big( \big\{ \vec M^{ijkl}u^iv^ju^kv^l \big\} \big)^*$.
Naively, one might think that a relaxation should reduce the difficulty of the problem.
This is true if our goal was to obtain some positivity bounds: one can easily plug in
random values of $u,v$ and get a bound; but this is not true if our goal was to extract
the complete set of bounds: the set $\mc Q_P$ is more difficult to describe than
$\mc C^*$, due to the quartic nature of its elements.
In this work we had to first extract the boundaries by searching for complete
square relations, and then search for the ERs. On the other hand, comparing with the extremal approach,
the boundaries of $\mc C^*$ is known from the beginning, as they are just the ERs of
$\mc C$, which we directly wrote down using group theory as a guideline.
Therefore, this relaxation is not efficient: it actually increases the difficulty of the
problem, and at the same time only gives weaker results. It does have the
advantage of connecting the bounds to a clear physics picture, i.e.~the elastic
scattering.  However, to obtain an exact description of all bounds, this approach
is certainly not suitable.  We emphasize that having the exact set of bounds
is important not only for testing the axiomatic principles of QFT, but also for
inferring or excluding possible UV states using positivity, see Ref.~\cite{eepaper}.

One last aspect to compare is the speed of the two numerical approaches.  Below we show
the average time it takes to check numerically whether a given point in the space of Wilson
coefficients satisfies positivity bounds, on a single core
(obviously the problem can be easily parallelized when checking many points).
For the extremal approach, we use the simplex method to solve the linear
programming, Eq.~\eqref{eq:LP}.
\begin{flalign}
	\begin{array}{|c|c|c|c|c|}
		\hline
		\mbox{Single-core time}	& \mbox{Elastic}
		&\multicolumn{3}{|c|}{\mbox{Extremal}\ \mc C_N}
		\\\cline{3-5}
		\mbox{[second]}& \mc C_N^{el} & N=50 & N=100 & N=200
		\\\hline
		\mbox{No filter} & 0.1 & 0.002 & 0.004 & 0.007
		\\\hline
		\mbox{With filter} & 0.001 & 0.00019 & 0.00021 & 0.00023
		\\\hline
		\mbox{Error on } \Omega(\mc C_N) & \mbox{N/A} & 3\% & 0.8\% & 0.2\%
		\\\hline
	\end{array}
\end{flalign}
The time used for numerical extremal bounds depends on $N$, one half of the
number of sampling rays per a continuous set of ERs.
``Filter'' refers to whether we first test a given point against the analytical
elastic bounds $\mc C^{el}_A$, so that we only run the numerical determination if these
analytical bounds are satisfied. This procedure significantly accelerates the determination
for a randomly chosen point, 
and should always be used in a practical application (or with $\mc C^{el}_A$ replaced
by $\mc C_A$). Clearly, the extremal approach drastically improves the speed. 
For $N=50$, the speed of the extremal approach is almost 2 orders of magnitude faster.
The bounds in that case are about $\approx 3\%$ over-constraining. For $N=200$
this error is reduced to $\approx 0.2\%$, while the speed is still more than one
order of magnitude faster. Furthermore, the difference in speed is still large after
applying the ``filter''.

~\\
\noindent {\bf \underline{Physics result}}
We have obtained positivity bounds on the full set of the
transversal aQGC operators, and these results provide guidance to the ongoing
VBS and tri-boson measurements at the LHC. Our results represent the best that can be derived
from the axiomatic principles of QFT, and supersede those presented in the literature
whenever relevant. For practical application, we have provided bounds in terms of 
analytical inequalities, which are convenient to use. We recommend using the
extremal analytical bounds $\mc C_A$, in Eqs.~\eqref{eq:erbound1}-\eqref{eq:erbound2},
for reasons discussed above. While they are slightly weaker than the exact
extremal bounds, by about $1\%$ (relative) in solid angle, one can also resort to the
numerical bounds $\mc C_N$, if better accuracy is needed, along with a sampling
number $N\gtrsim100$ to ensure the desired accuracy.

Positivity bounds on the full set of aQGC operators, including longitudinal modes,
is likely to be too difficult in the elastic approach. The extremal approach
should be feasible, but we will leave it to a future work.

\section{Summary and outlook}
\label{sec:summary}

There are at least two ways to derive positivity bounds in the forward limit
for SMEFT operators: the
conventional elastic positivity approach uses elastic scattering amplitudes
from a pair of arbitrarily superposed SM fields, while a more recently
proposed approach, which we dub extremal positivity approach, directly
constructs the allowed parameter space using the extremal representation of
convex cones.  We have provided a unified picture to understand and connect both approaches,
based on the geometry of the coefficient space and using several concepts from
convex geometry.  We have shown that the extremal positivity approach always
describes the best bounds available from the axiomatic principles of QFT, while the
elastic positivity approach relaxes these bounds and are thus often weaker.

As a case study, we have applied both approaches to the scattering amplitude of the
transversal electroweak gauge bosons. Aiming at obtaining the complete set of
bounds, we have established the methodology for both approaches, identified the main
difficulties that arise due to the large number of low-energy modes involved in
the problem, and provided both analytical and numerical solutions, which
are generally applicable to the SMEFT positivity problems and beyond.
We have further compared the two approaches in several aspects. Our main conclusions are
\begin{itemize}
	\item Extremal positivity gives better bounds than elastic positivity. 
		The improvement in the considered case is small, but can become
		sizable when focusing on specific sets of operators.
	\item Analytical solutions are available for both elastic and extremal
		positivity bounds. They are based on certain approximations,
		but the final results can reach an accuracy level of $0.1\%$ and $1\%$,
		respectively.
	\item Analytically, the extremal approach can be easily implemented algorithmically and
		thus applicable to more complicated problems (e.g.~bounds involving
		more operators and more fields). The elastic approach does not seem to
		have this feature, and so it is less suitable for a more
		complicated application.
	\item Numerically, to determine if a given point in the Wilson coefficient space
		is excluded by positivity, the extremal approach is significantly faster 
		than the elastic approach, by about one or two orders of magnitude,
		while the error can be controlled at the per mille level.
	
	\item Neglecting dim-6 operators leads to conservative bounds in both approaches.
		Non-zero dim-6 coefficients, on the other hand, could imply
		that certain linear combinations of dim-8 coefficients must be
		non-vanishing.
\end{itemize}
For these reasons, we conclude that the extremal positivity approach is always
the better option, for extracting the full set of forward limit positivity bounds in SMEFT.
These conclusions should also apply to other EFTs with a large number of low energy
modes, connected with some symmetry groups.

Physically, this study gives the best positivity bounds on the transversal aQGC
couplings.  The analytical results of these bounds have been explicitly
displayed in the relevant sections.
Interested readers should be able to directly dive into these sections and find
these results for direct application.
These bounds should be obeyed by any SMEFT that admits a UV completion consistent
with the axiomatic principles of QFT.
By excluding more than $99\%$ of the transversal aQGC parameter space,
these bounds will provide useful guidance to
both experimental and theoretical studies of the VBS and tri-boson processes.

While these results already represent a significant improvement over the existing
literature, we can foresee several further developments, which we leave for
future studies: 
\begin{itemize}
	\item The full set of aQGC couplings including the longitudinal modes.
		The extremal approach discussed in this work should be applicable to this
		problem.
	\item Methodology: there is still room to improve for all approaches we
		have provided in this work. In particular, higher-order bounds
		(beyond the quadratic ones) will further improve the analytical
		results, provided that an efficient method exists to find them.
	\item A comprehensive study of the SMEFT bounds. This study will pave the
		way towards further applications of positivity bounds in collider
		physics. While the most obvious application is to provide guidance
		to experimental searches wherever relevant, 
		there are other more interesting possibilities, such as testing
		the axiomatic QFT principles in colliders, and
		inferring/excluding the existence of UV states by using
		precision measurements. For the last two possibilities, we
		refer to \cite{eepaper} for a recent application in $e^+e^-$
		scattering.
\end{itemize}

\section*{Acknowledgements}
We thank G.~Durieux, G.~Remmen, F.~Riva, and N.~Rodd for helpful discussions
and comments. We also thank X.~Li for numerically checking the extremal bounds
including the dim-6 coefficient. KY is supported by
the Chinese Academy of Sciences (CAS) President's International Fellowship
Initiative under Grant No.~2020PM0018.
CZ is supported by IHEP under Contract No.~Y7515540U1, and
by National Natural Science Foundation of China (NSFC) under grant
No.~12035008 and 12075256. SYZ acknowledges support from the starting grant
from University of Science and Technology of China under grant No.~KY2030000089
and GG2030040375 and is also supported by NSFC under grant No.~11947301 and
12075233.

\appendix
\section{Some details for the analytical elastic positivity with factorization}
\label{sec:app_fac}

Here we present more details about deriving the bounds in Section~\ref{sec:fac}. In Eq.~\eqref{eq:facstep1} we have removed the $a_i$ dependence and
determined 4 linear inequalities. In terms of $C_{kl}$, they are
\begin{flalign}
	&C_{1l}b_l\ge0,\quad C_{2l}b_l\ge0,
	\label{eq:ckl1}
	\\
	&(C_{1l}+C_{3l})b_l\ge0,\quad (C_{2l}+C_{3l})b_l\ge0
\end{flalign}
and we can write them collectively as
\begin{flalign}
	\sum_jM^i_{bj}b_j\ge0,\quad \mbox{for}\ i=1,2,3,4
	\label{eq:facstep2}
\end{flalign}
with {$M^i_{bj}$'s} given by:
\begin{flalign}
&i=1:\nonumber\\
&M^1_{b1}=\frac{16 F_{T,0} + 8 F_{T,1} + 6 F_{T,2}}{\sw^4},
\ \ M^1_{b2}=\frac{16 F_{T,0} + 8 F_{T,1} + 6 F_{T,2}}{\sw^4}, 
\ \ M^1_{b3}=\frac{4 (4 F_{T,1} + F_{T,2})}{\sw^4}, \nonumber\\
&M^1_{b4}=\frac{8F_{T,5} + 4F_{T,6} + 3F_{T,7}}{2 \cw^2 \sw^2},  
\ \ M^1_{b5}=\frac{8F_{T,5} + 4 F_{T,6} + 3 F_{T,7}}{2 \cw^2 \sw^2}, 
\ \ M^1_{b6}=\frac{4F_{T,6} + F_{T,7}}{\cw^2 \sw^2},  \nonumber\\
&M^1_{b7}=\frac{4 (2 F_{T,8} + F_{T,9})}{\cw^4}.
 \label{eq:pol1}
\end{flalign}
\begin{flalign}
&i=2:\nonumber\\
&M^2_{b1}=\frac{4 (4 F_{T,10} + 4 F_{T,0} + F_{T,2})}{\sw^4},
\ \ M^2_{b2}=\frac{4 (2 F_{T,1} + F_{T,2})}{\sw^4},
\ \ M^2_{b3}=\frac{4 (2 F_{T,1} + F_{T,2})}{\sw^4},\nonumber\\
&M^2_{b4}=\frac{2 F_{T,11} + 4 F_{T,5} + F_{T,7}}{\cw^2 \sw^2},
\ \ M^2_{b5}=\frac{2 F_{T,6} + F_{T,7}}{\cw^2  \sw^2}, 
\ \ M^2_{b6}=\frac{2 F_{T,6} + F_{T,7}}{\cw^2  \sw^2}, \nonumber\\
&M^2_{b7}=\frac{4F_{T,8} + 3 F_{T,9}}{\cw^4}.
 \label{eq:pol2}
\end{flalign}
\begin{flalign}
&i=3:\nonumber\\
&M^3_{b1}=\frac{4 (2 F_{T,1} + F_{T,2})}{\sw^4},
\ \ M^3_{b2}=\frac{4 (4 F_{T,10} + 4F_{T,0} + F_{T,2})}{\sw^4}, 
\ \ M^3_{b3}=\frac{4 (2 F_{T,1} + F_{T,2})}{\sw^4},\nonumber\\
&M^3_{b4}= \frac{2 F_{T,6} + F_{T,7}}{\cw^2 \sw^2},
\ \ M^3_{b5}=\frac{2 F_{T,11} + 4 F_{T,5} + F_{T,7}}{\cw^2 \sw^2},
\ \ M^3_{b6}=\frac{2 F_{T,6} + F_{T,7}}{\cw^2 \sw^2}, \nonumber\\
&M^3_{b7}=\frac{4 F_{T,8} + 3 F_{T,9}}{\cw^4}.
\end{flalign}
\begin{flalign}
&i=4:\nonumber\\
&M^4_{b1}=\frac{2 (8 F_{T,10} + F_{T,2})}{\sw^4},
\ \ M^4_{b2}=\frac{2 (8 F_{T,10} + F_{T,2})}{\sw^4},
\ \ M^4_{b3}=\frac{4 F_{T,2}}{\sw^4}, \nonumber\\
&M^4_{b4}=\frac{4 F_{T,11} + F_{T,7}}{2\cw^2 \sw^2},
\ \ M^4_{b5}=\frac{4 F_{T,11} + F_{T,7}}{2\cw^2 \sw^2},
\ \ M^4_{b6}=\frac{F_{T,7}}{\cw^2 \sw^2}, \nonumber\\
&M^4_{b7}=\frac{2 F_{T,9}}{\cw^4}.
\label{eq:pol4}
\end{flalign}

The first observation here is that the $i=2$ case and the $i=3$ case are in fact identical:
they are related by swapping $b_1$ with $b_2$, and $b_4$ with $b_5$. This
simply amounts to taking $\beta_i\to \beta_i^*$. Since $\beta_i$ are allowed to
take any complex 4-vectors, $i=2$ and $i=3$ lead to identical bounds, and thus
we can omit the $i=3$ case.  A second observation is that not all $M^i_{bj}$'s
are independent. In fact, the following equations hold:
\begin{align}
i=1,4:\quad & M^1_{b1}=M^1_{b2},\ \ M^1_{b4}=M^1_{b5}, \label{eq:c_pol1}\\
i=2:\quad& M^2_{b2}=M^2_{b3},\ \ M^2_{b5}=M^2_{b6}, \label{eq:c_pol2}
\end{align}
i.e.~there are only 5 independent $M^i_{bj}$ coefficients.
This means that we need to solve the following two positivity conditions
for all $b_i$'s:
\begin{align}
i=1,4&:\ (b_1+b_2)M_{b1}+ b_3M_{b3}+(b_4+b_5)M_{b4}
+b_6M_{b6}+b_7M_{b7}\ge0 \label{eq:amp_c1}\\
i=2&:\ b_1M_{b1}+ (b_2+b_3)M_{b2}+b_4M_{b4}
+(b_5+b_6)M_{b5}+b_7M_{b7}\ge0 \label{eq:amp_c2}
\end{align}
Here we omit the superscript $i$ for $M_{bj}$.

Since the $b_i$'s are quartic polynomials of $\alpha$ and $\beta$,
they cannot take arbitrary values.
We need to find the range of $b_i$, as we should only require
Eqs.~\eqref{eq:amp_c1} and \eqref{eq:amp_c2} to hold for $b_i$ within this range.
It is easy to identify the following bounds for $b_i$'s:
\begin{align}
&b_7\ge0, 0\le b_1\le b_3, 0\le b_2\le b_3,\label{eq:rangeb}\\
&|b_4| \le 2 \sqrt{b_1 b_7},\ |b_5| \le 2 \sqrt{b_2 b_7},\ 
b_6 \ge 2\sqrt{b_3 b_7}. \label{eq:rangeb456}
\end{align}
$b_i \ge0$ for $i=1,2,3,6,7$ are simply by definitions. 
$b_1\le b_3$ and $b_2\le b_3$ are obtained from
the Cauchy-Schwarz inequality:
$|\Braket{\vec{u_1}, \vec{u_2}}|^2\leq \Braket{\vec{u_1},\vec{u_1}}\Braket{\vec{u_2},\vec{u_2}}$,
where $\vec{u_1}$ and $\vec{u_2}$ are arbitrary vectors in an inner product space
over $\mathbb{C}$. 
Eq.~\eqref{eq:rangeb456} can be derived from:
\begin{align}
& 4b_1b_7-|b_4|^2 = 4\left[\mathrm{Im}(\alpha_1\beta_1\Braket{\vec{u},\vec{v^*}})\right]^2\ge 0,\\
& 4b_2b_7-|b_5|^2= 4\left[\mathrm{Im}(\alpha_1\beta^*_1\Braket{\vec{u},\vec{v}})\right]^2\ge 0.\\
&
b^2_6-4b_3b_7=(|\beta_1|^2{|\vec{u}|^2}-|\alpha_1|^2{|\vec{v}|^2})^2\ge0.
\end{align}

Now the inequalities \eqref{eq:rangeb} and \eqref{eq:rangeb456}
specify the boundary of the possible range for the $b_i$ parameters.
One still needs to show that this boundary is tight, i.e.~for all $b'_i$ parameters that
satisfy these bounds, there exist a set of $\alpha_i$ and $\beta_i$ parameters
such that $b_i(\alpha_1,\cdots,\alpha_4,\beta_1,\cdots,\beta_4)=b'_i$, or in other words,
all points within this range can be achieved by some superposition in the gauge space.
In fact, for all $b'_i$ that satisfy Eqs.~\eqref{eq:rangeb} and
\eqref{eq:rangeb456}, defining the following variables:
\begin{align}
&\theta_1 = \cos^{-1}\left(\frac{b'_4}{2\sqrt{b'_1 b'_7}}\right),\quad
\theta_2 = \cos^{-1}\left(\frac{b'_5}{2\sqrt{b'_2 b'_7}}\right),\\
&\phi_1 = \cos^{-1}\left(\frac{2b'_1-b'_3}{b'_3}\right),\quad
\phi_2 = \cos^{-1}\left(\frac{2b'_2-b'_3}{b'_3}\right),\\
&r^2 = \frac{1}{2}\left(\frac{b'_6}{\sqrt{b'_3 b'_7}}+\sqrt{\frac{
	{b'}^2_6}{b'_3b'_7}-4}\right),
\end{align}
the following $\alpha_i$ and $\beta_i$ should give the desired {$b'_i$'s} :
\begin{align}
	&\alpha_1 = r {b'_7}^{1/4}\exp\left[i\left(\frac{\theta_1+\theta_2}{2}+\frac{\phi_1+\phi_2}{4}\right)\right],\quad 
	\beta_1 = \frac{ {b'_7}^{1/4}}{r}\exp\left[i\left(\frac{\theta_1-\theta_2}{2}+\frac{\phi_1-\phi_2}{4}\right)\right], \nonumber\\
	&\alpha_2 =\frac{ {b'_3}^{1/4}}{\sqrt{2}},\quad
	\beta_2 = \frac{ {b'_3}^{1/4}}{\sqrt{2}},\quad
	\alpha_3 = \frac{ {b'_3}^{1/4}}{\sqrt{2}}\exp\left[i\left(\frac{\phi_1+\phi_2}{2}\right)\right],\quad
	\beta_3 = \frac{ {b'_3}^{1/4}}{\sqrt{2}}\exp\left[i\left(\frac{\phi_1-\phi_2}{2}\right)\right],\nonumber\\
&\alpha_4 = 0,\quad \beta_4=0, \label{eq:ansatz}
\end{align}
which means that all $b_i$'s that satisfy Eqs.~\eqref{eq:rangeb} and
\eqref{eq:rangeb456} correspond to some $\alpha_i$ and $\beta_i$
parameters.  The fact that $(\alpha_4, \beta_4)=(0,0)$ implies that including
$W^3$ states in this ``factorized'' elastic positivity approach does not give
rise to new bounds.

It is now clear that, for $i=1,4$, we should require that Eq.~\eqref{eq:amp_c1}
holds, subject to the inequalities \eqref{eq:rangeb} and \eqref{eq:rangeb456};
and for $i=2$ (and equivalently $i=3$), we require Eq.~\eqref{eq:amp_c2} to hold,
subject to the same set of inequalities. In terms of $\sqrt{b_1},\sqrt{b_2},\sqrt{b_3},
b_4,b_5,b_6,\sqrt{b_7}$, these are two quadratically constrained quadratic programming
(QCQP) problems, i.e.~we essentially need to minimize the l.h.s.~Eqs.~\eqref{eq:amp_c1}
and \eqref{eq:amp_c2}, which are quadratic polynomials, subject to a set of quadratic
constraints. We can further simplify these problems. For $i=1,4$,
for any $b_i$ within the constraints, one can always change the values of $b_1$ and $b_2$
to $(b_1+b_2)/2$, 
and the values of $b_4$ and $b_5$ to $(b_4+b_5)/2$, 
without changing
the l.h.s.~of Eq.~\eqref{eq:amp_c1}. At the same time the constraints in
Eqs.~\eqref{eq:rangeb} and \eqref{eq:rangeb456} are still satisfied, due to
$2(b_1+b_2)\ge(\sqrt{b_1}+\sqrt{b_2})^2$. Therefore the first QCQP problem
can be written as
\begin{flalign}
	&\mbox{QCQP-1}:\nonumber\\
	&\mbox{minimize}\ \quad
	2b_1M_{b1}+ b_3M_{b3}+2b_4M_{b4}+b_6M_{b6}+b_7M_{b7}
	\nonumber\\
	&\mbox{subject to}\quad
b_7\ge0, 0\le b_1\le b_3, |b_4| \le 2 \sqrt{b_1 b_7}, b_6 \ge 2\sqrt{b_3 b_7}
\label{eq:QGQC1}
\end{flalign}
For the $i=2$ case, we can redefine $b_5+b_6$ as $b_6$, and write
the problem as:
\begin{flalign}
	&\mbox{QCQP-2}:\nonumber\\
	&\mbox{minimize}\ \quad
	b_1M_{b1}+(b_2+b_3)M_{b2}+b_4M_{b4}+b_6M_{b5}+b_7M_{b7}
	\nonumber\\
	&\mbox{subject to}\quad
	 b_7\ge0, 0\le b_1\le b_3, 0\le b_2\le b_3, |b_4| \le 2 \sqrt{b_1 b_7},
	\nonumber\\
	&\hspace{2.05cm} b_6 \ge 2\sqrt{b_3 b_7} - 2\sqrt{b_2 b_7}
\label{eq:QGQC2}
\end{flalign}
In the following, we will show how QCQP-1 can be solved analytically.

Our goal is to find the conditions under which the two QCQP problems:
1) admit a minimum value, and 2) the minimum value is positive.
We will see that each condition corresponds to the target function being
positive at a specific point in the $b_i$ space. In the following, we will always list
the conditions, along with the corresponding $b_i$ points shown in a pair of
parentheses. The conditions will become our positivity bounds, while the $b_i$
values correspond the scattering channels, from which these bounds are derived.

First of all, two conditions can be derived by examining the $b_3=0$ case,
for which we also have $b_1=b_4=0$:
\begin{flalign}
	&\text{Condition: } M_{b6} \geq 0 \label{eq:c1Mb6}\\
	& (b_6>0, \ \ \text{other }b_i=0) \nonumber \\
	&\text{Condition: } M_{b7} \geq 0 \label{eq:c1Mb7}\\
	& (b_7>0, \ \ \text{other }b_i=0) \nonumber
\end{flalign}
Now we consider $b_3>0$. Since the problem is homogeneous in $b_i$, we set $b_3=1$ without
loss of generality.
Knowing $M_{b6}\ge0$, and using $|M_{b4}|\ge0$, we further
minimize the amplitude by taking $b_6\to 2\sqrt{b_7}$ 
and the $b_4$ term to $-4\sqrt{b_1b_7}|M_{b4}|$. Now the problem has two variables:
\begin{flalign}
	&\mbox{minimize}\ \quad
	2x^2M_{b1}+ M_{b3}-4xy|M_{b4}|
	+2yM_{b6}+y^2M_{b7}
	\nonumber\\
	&\mbox{subject to}\quad
	y\geq0, 0\leq x\leq 1
\label{eq:QGQC1_b3_non0_3}
\end{flalign}
where we have let $b_1=x^2$ and $b_7=y^2$, as they are both positive.
The rest is simply to find the $(x,y)$ value that minimize the function.
Note that since $x$ is bounded in $[0,1]$, $M_{b1}$ does not have to be positive
for the target function to admit a minimum.  To proceed, consider two cases,
$M_{b1}<0$ and $M_{b1}\ge0$. For the first case:
\begin{flalign}
&\text{QCQP-1-1:}\ \  M_{b1}<0 \label{eq:case11}\\
&\bullet \text{If } M_{b6}\geq2|M_{b4}|: \label{eq:case11cond0}\\ 
&\quad \text{Condition: } 2M_{b1}+M_{b3}\geq0 \label{eq:c1Mb1b3}\\
&\quad ((x,y)=(1,0)) \nonumber\\
&\quad \text{This condition is {required} even when $M_{b1}\geq0$ and/or $M_{b_6}\leq2|M_{b4}|$.}\nonumber\\
&\bullet \text{If } M_{b6}\leq2|M_{b4}|:\label{eq:case11cond1}\\
&\quad \text{Condition: } 2M_{b1}+M_{b3}-\frac{(2|M_{b4}|-M_{b6})^2}{M_{b7}}\geq0 \label{eq:case11cond}\\
&\quad \left((x,y)=\left(1, \frac{2|M_{b4}|-M_{b6}}{M_{b7}}\right)\right) \nonumber\\
&\quad \text{This condition is {required} even when $M_{b1}\geq0$.}\nonumber
\end{flalign}
Here, condition \eqref{eq:c1Mb1b3} needs to be satisfied regardless of conditions
\eqref{eq:case11} and \eqref{eq:case11cond0},
because the target function needs to be positive for $(x,y)=(1,0)$, which
is in the proper range, i.e.~Eq.~\eqref{eq:QGQC1_b3_non0_3},
Eqs.~\eqref{eq:case11} and \eqref{eq:case11cond0} simply give the conditions
for this point to be the true minimum.
In contrast,
Eq.~\eqref{eq:case11cond1} should be kept, as it is required for
$y\geq0$ in Eq.~\eqref{eq:QGQC1_b3_non0_3}. 
For QCQP-1-2 (i.e.~$M_{b1}\geq0$), we can obtain similar results:
\begin{flalign}
&\text{QCQP-1-2:}\ \ M_{b1}\geq0 \label{eq:case12}\\
&\bullet \text{Condition: } M_{b3}\geq0\label{eq:c1Mb3}\\
&\quad ((x,y)=(0,0))\nonumber\\
&\bullet \text{If } M_{b6}\leq2|M_{b4}|:\\
&\quad \text{Condition: } 2M_{b1}+M_{b3}-\frac{(2|M_{b4}|-M_{b6})^2}{M_{b7}}\geq0\\
&\quad \left((x,y)=\left(1, \frac{2|M_{b4}|-M_{b_6}}{M_{b_7}}\right)\right)\nonumber
\end{flalign}

Collecting all conditions, and converting $(x,y)$ to $\{b_i\}$, the final result
for QCQP-1 is
\begin{flalign}
&\bullet \text{Condition: } M_{b6} \geq 0\\
&\quad (b_6> 0,\ \ \text{other }b_i=0)\nonumber\\
&\bullet \text{Condition: } M_{b7} \geq 0\\
&\quad (b_7> 0,\ \ \text{other }b_i=0)\nonumber\\
&\bullet \text{Condition: }M_{b3} \geq 0\\
&\quad (b_3> 0,\ \ \text{other }b_i=0)\nonumber\\
&\bullet \text{Condition: }2M_{b1}+M_{b3} \geq 0\\
&\quad (b_1=b_3> 0,\ \ \text{other }b_i=0)\nonumber\\
&\bullet  \text{Condition: } 
\text{If } -2M_{b4}\geq M_{b6}\text{ then }
M_{b7}(2M_{b1}+M_{b3})-(2M_{b4}+M_{b6})^2\geq0\\
&\quad \left(b_1=b_3=1, b_4=b_6=2\left(\frac{-2M_{b4}-M_{b6}}{M_{b7}}\right), b_7=\frac{b^2_4}{4}\right)\nonumber\\
&\bullet 
\text{Condition: } 
\text{If } 2M_{b_4}\geq M_{b_6} \text{ then }
M_{b7}(2M_{b1}+M_{b3})-(2M_{b4}- M_{b6})^2\geq0\\
&\quad \left(b_1=b_3=1, b_4=-b_6=-2\left(\frac{2M_{b4}-M_{b6}}{M_{b7}}\right), b_7=\frac{b^2_4}{4}\right)\nonumber
\end{flalign}

For QCQP-2, which can be solved similarly, we simply present the results as below
\begin{flalign}
&\bullet \text{Condition: } M_{b2} \geq 0\\
&\quad (b_2=b_3 > 0,\ \ \text{other } b_i=0)\nonumber\\
&\bullet \text{Condition: } M_{b1}+M_{b2} \geq 0\\
&\quad (b_1=b_3 > 0,\ \ \text{other }b_i=0)\nonumber\\
&\bullet \text{Condition: } M_{b7} \geq 0 \\
&\quad (b_7 > 0,\ \ \text{other }{b_i=0})\nonumber \\
&\bullet \text{Condition: } M_{b5} \geq 0\\
&\quad (b_6 > 0,\ \ \text{other }b_i=0)\nonumber\\
&\bullet \text{Condition: }M_{b7}(M_{b1}+2M_{b2})-M^2_{b4} \geq 0 \label{case2condb4}\\
&\quad \left(b_1=b_2=b_3=1, b_4=-2\frac{M_{b4}}{M_{b7}}, b_6=0, b_7=\frac{b^2_4}{4}\right)\nonumber\\
&\bullet \text{Condition: If }
{(-M_{b4}-M_{b5})(M_{b2}M_{b7}+M_{b4}M_{b5})\geq 0}\\
&\quad \text{then }M_{b1}+M_{b2}-\frac{M_{b2}}{M_{b2}M_{b7}-M^2_{b5}}(M_{b4}+M_{b5})^2 \geq 0\\
&\quad \left(b_1=b_3=1, b_2=\left(\frac{M_{b5}(M_{b4}+M_{b5})}{M_{b2}M_{b7}-M^2_{b5}}\right)^2, 
b_4=2\frac{M_{b2}(-M_{b4}-M_{b5})}{M_{b2}M_{b7}-M^2_{b5}},\right.\nonumber\\
&\left.\quad b_6=2\frac{M_{b2}(-M_{b4}-M_{b5})(M_{b2}M_{b7}+M_{b4}M_{b5})}{(M_{b2}M_{b7}-M^2_{b5})^2},
b_7=\frac{b^2_4}{4} \right)\nonumber\\
&\bullet \text{Condition: If }
{(M_{b4}-M_{b5})(M_{b2}M_{b7}-M_{b4}M_{b5})\geq 0}\\
&\quad \text{then }M_{b1}+M_{b2}-\frac{M_{b2}}{M_{b2}M_{b7}-M^2_{b5}}(M_{b4}-M_{b5})^2 \geq 0\label{eq:case2_r_e}\\
&\quad \left(b_1=b_3=1, b_2=\left(\frac{M_{b5}(M_{b4}-M_{b5})}{M_{b2}M_{b7}-M^2_{b5}}\right)^2, 
b_4=-2\frac{M_{b2}(M_{b4}-M_{b5})}{M_{b2}M_{b7}-M^2_{b5}},\right.\nonumber\\
&\left. \quad b_6=2\frac{M_{b2}(M_{b4}-M_{b5})(M_{b2}M_{b7}-M_{b4}M_{b5})}{(M_{b2}M_{b7}-M^2_{b5})^2},
b_7=\frac{b^2_4}{4} \right)\nonumber
\end{flalign}

Once the programming problems are solved, what is left is simply to plug in
Eqs.~\eqref{eq:pol1} and \eqref{eq:pol4}, respectively, into the conditions of
QCQP-1, and similarly, Eq.~\eqref{eq:pol2} into QCQP-2.  This will give the
positivity bounds in terms of the Wilson coefficients $F_{T,i}$.
To identify the actual elastic channel and the polarization for each bound,
since we have kept track of $a_i(i=1,2,3)$ and $b_j(j=1,\cdots,7)$,
the $(x_i,y_i)$ and $(\alpha_i,\beta_i)$ parameters can be solved using
Eqs.~\eqref{eq:def_a}--\eqref{eq:b7}. The solution is not unique, but for illustration
purposes it suffices to show just one solution.
For the nine linear bounds, this information is shown in Eq.~\eqref{eq:lineartable}.
Below, we give the same information for the quadratic and cubic bounds.

The three quadratic bounds in Eqs.~\eqref{eq:quad1}--\eqref{eq:quad3} are
constructed by the following five conditions, with the corresponding 
scatterings channels ($\ket 1+\ket 2 \to \ket 1\ +\ket 2$) and helicities as follows.
Again, subscripts $R$ and $L$ indicate positive and negative helicity states respectively, while
superscripts for the $W$-boson are $SU(2)$ indices.
\begin{enumerate}
   \item Under the conditions Eqs.~\eqref{eq:2210_positive},~\eqref{eq:fg_positive}, and $2F_{T,5}+2F_{T,6}+F_{T,7}\leq0$,
   the following elastic channel:
   \begin{flalign}
	   &\ket{1}=\ket{2} = \frac{\cw}{\sw}
	   \sqrt{\frac{-(2F_{T,5}+2F_{T,6}+F_{T,7})}{2F_{T,8}+F_{T,9}}}(\Ket{B_R}+\Ket{B_L}) + \Ket{W_R^1}+\Ket{W_L^1}
   \end{flalign}
   gives the bound
   \begin{align}
&2\sqrt{\left[2(F_{T,0}+F_{T,1})+F_{T,2}\right](2F_{T,8}+F_{T,9})}\geq-(2F_{T,5}+2F_{T,6}+F_{T,7})\label{eq:cond_q1}
\end{align}
which is also satisfied when $2F_{T,5}+2F_{T,6}+F_{T,7}\geq0$.

   \item Under the conditions Eqs.~\eqref{eq:2210_positive}, \eqref{eq:fg_positive}, 
 and $4F_{T,5}+F_{T,7}\geq0$, the following elastic channel:
 \begin{flalign}
&\ket{1} = \frac{\cw}{\sw}\sqrt{\frac{4F_{T,5}+F_{T,7}}{2(2F_{T,8}+F_{T,9})}}(\Ket{B_R}+\Ket{B_L}) + \Ket{W_R^1}+\Ket{W_L^1}\\
&\ket{2} = -\frac{\cw}{\sw}\sqrt{\frac{4F_{T,5}+F_{T,7}}{2(2F_{T,8}+F_{T,9})}}(\Ket{B_R}+\Ket{B_L}) + \Ket{W_R^1}+\Ket{W_L^1}
 \end{flalign}
   gives the bound
\begin{align}
&4\sqrt{\left[2(F_{T,0}+F_{T,1})+F_{T,2}\right](2F_{T,8}+F_{T,9})}\geq4F_{T,5}+F_{T,7}\label{eq:cond_q2}
\end{align}
which is also satisfied when $4F_{T,5}+F_{T,7}\leq0$.
\end{enumerate}
The combination of \eqref{eq:cond_q1} and \eqref{eq:cond_q2} gives
condition \eqref{eq:quad1}.
\begin{enumerate}
\setcounter{enumi}{2}
   \item    Under the conditions Eqs.~\eqref{eq:0014_positive}, \eqref{eq:fg_positive3},
   and $2F_{T,11}+F_{T,7}\leq0$, the following elastic channel:
      \begin{flalign}
&\ket{1} = \frac{\cw}{\sw}\sqrt{\frac{-(2F_{T,11}+F_{T,7})}{F_{T,9}}}(\Ket{B_R}-\Ket{B_L}) +\Ket{W_R^1}-\Ket{W_L^1}\\
&\ket{2} = \frac{\cw}{\sw}\sqrt{\frac{-(2F_{T,11}+F_{T,7})}{F_{T,9}}}(\Ket{B_R}+\Ket{B_L}) +\Ket{W_R^1}+\Ket{W_L^1}
   \end{flalign}
   gives the bound
   \begin{align}
&2\sqrt{F_{T,9}(F_{T,2}+4F_{T,10})}\geq-(2F_{T,11}+F_{T,7}) \label{eq:cond_q3}
\end{align}
which is also satisfied when $2F_{T,11}+F_{T,7}\geq0$.
   \item 
   Under the conditions Eqs.~\eqref{eq:0014_positive}, \eqref{eq:fg_positive3} and
   $F_{T,11}\geq0$, the following elastic channel:
   \begin{flalign}
&\ket{1} = \frac{\cw}{\sw}\sqrt{\frac{2F_{T,11}}{F_{T,9}}}(\Ket{B_R}-\Ket{B_L}) +\Ket{W_R^1}-\Ket{W_L^1}\\
&\ket{2} = -\frac{\cw}{\sw}\sqrt{\frac{2F_{T,11}}{F_{T,9}}}(\Ket{B_R}+\Ket{B_L}) +\Ket{W_R^1}+\Ket{W_L^1}
   \end{flalign}
   gives the bound
\begin{align}
&\sqrt{F_{T,9}(F_{T,2}+4F_{T,10})}\geq F_{T,11}\label{eq:cond_q4}
\end{align}
which is also satisfied when $F_{T,11}\leq0$.
\end{enumerate}
The combination of \eqref{eq:cond_q3} and \eqref{eq:cond_q4} gives the
condition \eqref{eq:quad2}.
\begin{enumerate}
\setcounter{enumi}{4}
   \item 
Eqs.~\eqref{eq:0410_positive},
   \eqref{eq:0010_positive}, and \eqref{eq:2112_positive} lead to
   $4(F_{T,0}+F_{T,1})+3F_{T,2}+4F_{T,10}\geq0$; and Eqs.~\eqref{eq:fg_positive},
   \eqref{eq:fg_positive3} lead to $4F_{T,8}+3F_{T,9}\geq0$. With these two
   conditions satisfied, the following elastic channel:
	\begin{flalign}
&\ket{1} = \frac{\cw}{\sw}\sqrt{\frac{2F_{T,11}+4F_{T,5}+F_{T,7}}{4F_{T,8}+3F_{T,9}}}\Ket{B_L} +\Ket{W_L^1}\\
&\ket{2} =  \frac{\cw}{\sw}\sqrt{\frac{2F_{T,11}+4F_{T,5}+F_{T,7}}{4F_{T,8}+3F_{T,9}}}\Ket{B_L} -\Ket{W_L^1}
	\end{flalign}
	gives the bound
\begin{align}
&2\sqrt{[4F_{T,10}+4 (F_{T,0} + F_{T,1}) + 3F_{T,2}] (4F_{T,8} + 3F_{T,9})}\ge|2F_{T,11} + 4F_{T.5} + F_{T,7}|
\end{align}
\end{enumerate}

While all quadratic bounds involve some superposition between the $B$ and the $W^1$
modes, the cubic bound in \eqref{eq:cubic} involves a superposition
of three states: $B$, $W^1$, and $W^2$.
Using $2F_{T,1}+F_{T,2}\geq 0$ {and $2F_{T,6}+F_{T,7}\geq 0$}
derived from the conditions Eqs.~\eqref{eq:0410_positive},
~\eqref{eq:0010_positive}, 
{\eqref{eq:0410mix_positive}, and \eqref{eq:ct7_positive},}
the cubic bound is constructed by the following two cubic conditions.
\begin{enumerate}
\item Under the conditions
{\small
\begin{align}
&[4(2F_{T,1}+F_{T,2})(4F_{T,8}+3F_{T,9})+
 (2F_{T,6}+F_{T,7})(4F_{T,5}+F_{T,7}+2F_{T,11})]\nonumber\\
&\quad\quad\quad\times
 (2F_{T,5}+F_{T,6}+F_{T,7}+F_{T,11})\le0\nonumber 
\end{align}
the following elastic channel:
\begin{align}
\Ket{1}=&\Ket{2}^*\nonumber\\
=&i2\sqrt{2}\cw
\sqrt{(2F_{T,1}+F_{T,2})|2F_{T,5}+F_{T,6}+F_{T,7}+F_{T,11}|}\Ket{B_L}\nonumber\\
&+\frac{1}{2}\sw\times\nonumber\\
&\left(
\sqrt{|4(2F_{T,1} + F_{T,2}) (4F_{T,8} + 3F_{T,9})
-(2 F_{T,6} + F_{T,7})( 4(F_{T,5} + F_{T,6})+3F_{T,7}+2F_{T,11})|}\right.\nonumber\\
&\left.+i \sqrt{
| 4(2F_{T,1} + F_{T,2})(4F_{T,8} + 3F_{T,9})
+(2F_{T,6} + F_{T,7})(4 F_{T,5} + F_{T,7}+2 F_{T,11}) | 
}
\right)\Ket{W_L^1}\nonumber\\
&+\frac{1}{2}\sw\times\nonumber\\
&\left(
\sqrt{|4(2F_{T,1} + F_{T,2}) (4F_{T,8} + 3F_{T,9})
-(2 F_{T,6} + F_{T,7})( 4(F_{T,5} + F_{T,6})+3F_{T,7}+2F_{T,11})|
}\right.\nonumber\\
&\left.-i \sqrt{
|4(2F_{T,1} + F_{T,2})(4F_{T,8} + 3F_{T,9}) 
+(2F_{T,6} + F_{T,7})(4 F_{T,5} + F_{T,7}+2 F_{T,11}) | 
}
\right)\Ket{W_L^2}
\end{align}
}
gives the bound
\begin{align}
 &2F_{T,0}+F_{T,1}+F_{T,2}+2F_{T,10}
\ge \frac{2(2F_{T,1}+F_{T,2})(2F_{T,5} + F_{T,6} + F_{T,7} + F_{T,11} )^2}
{4(2F_{T,1}+F_{T,2})(4F_{T,8}+3F_{T,9})-(2F_{T,6}+F_{T,7})^2}
\end{align}
%
\item Under the conditions
{\small
\begin{align}
&[4(2F_{T,1}+F_{T,2})(4F_{T,8}+3F_{T,9})-(2F_{T,6}+F_{T,7})(4F_{T,5}+F_{T,7}+2F_{T,11})]\nonumber\\
&\quad\quad\quad\times(-2F_{T,5}+F_{T,6}-F_{T,11})\le 0\nonumber
\end{align}
the following elastic channel:
\begin{align}
\Ket{1}=&2\sqrt{2}\cw
\sqrt{(2F_{T,1}+F_{T,2})|2F_{T,5}-F_{T,6}+F_{T,11}|}\Ket{B_L}\nonumber\\
&+\frac{1}{2}\sw\times\nonumber\\
&\left(
\sqrt{|4(2F_{T,1} + F_{T,2}) (4F_{T,8} + 3F_{T,9})+
(2 F_{T,6} + F_{T,7})( 4(F_{T,5} - F_{T,6})-F_{T,7}+2F_{T,11})|
}\right.\nonumber\\
&\left.+i \sqrt{
| 4(2F_{T,1} + F_{T,2})(4F_{T,8} + 3F_{T,9})
-(2F_{T,6} + F_{T,7})(4 F_{T,5} + F_{T,7}+2 F_{T,11}) | 
}
\right)\Ket{W_L^1}\nonumber\\
&+\frac{1}{2}\sw\times\nonumber\\
&\left(
\sqrt{|4(2F_{T,1} + F_{T,2}) (4F_{T,8} + 3F_{T,9})
+(2 F_{T,6} + F_{T,7})( 4(F_{T,5} - F_{T,6})-F_{T,7}+2F_{T,11})|
}\right.\nonumber\\
&\left.-i \sqrt{
|4(2F_{T,1} + F_{T,2})(4F_{T,8} + 3F_{T,9})
-(2F_{T,6} + F_{T,7})(4 F_{T,5} + F_{T,7}+2 F_{T,11}) |}
\right)\Ket{W_L^2}\\
\Ket{2}=&-2\sqrt{2}\cw
\sqrt{(2F_{T,1}+F_{T,2})|2F_{T,5}-F_{T,6}+F_{T,11}|}\Ket{B_L}\nonumber\\
&+\frac{1}{2}\sw\times\nonumber\\
&\left(
\sqrt{|4(2F_{T,1} + F_{T,2}) (4F_{T,8} + 3F_{T,9})
+(2 F_{T,6} + F_{T,7})( 4(F_{T,5} - F_{T,6})-F_{T,7}+2F_{T,11})|
}\right.\nonumber\\
&\left.-i \sqrt{
|4(2F_{T,1} + F_{T,2})(4F_{T,8} + 3F_{T,9})
-(2F_{T,6} + F_{T,7})(4 F_{T,5} + F_{T,7}+2 F_{T,11}) |
}
\right)\Ket{W_L^1}\nonumber\\
&+\frac{1}{2}\sw\times\nonumber\\
&\left(
\sqrt{|4(2F_{T,1} + F_{T,2}) (4F_{T,8} + 3F_{T,9})
+(2 F_{T,6} + F_{T,7})(4(F_{T,5} - F_{T,6})-F_{T,7}+2F_{T,11})
|}\right.\nonumber\\
&\left.+i \sqrt{
|4(2F_{T,1} + F_{T,2})(4F_{T,8} + 3F_{T,9})
- (2F_{T,6} + F_{T,7})(4 F_{T,5} + F_{T,7}+2 F_{T,11})|
}
\right)\Ket{W_L^2}
\end{align}
}
gives the bound
\begin{align}
 &2F_{T,0}+F_{T,1}+F_{T,2}+2F_{T,10}
 \ge \frac{2(2F_{T,1}+F_{T,2})(2F_{T,5} - F_{T,6} + F_{T,11} )^2}{4(2F_{T,1}+F_{T,2})(4F_{T,8}+3F_{T,9})-(2F_{T,6}+F_{T,7})^2}
\end{align}
\end{enumerate}

\section{Polarization dependence in VBS amplitudes}
\label{sec:app_fac2}

In Eq.~\eqref{eq:fac_amp0}, we have observed that the amplitude depends on polarization
parameters $x$ and $y$ only through the three combinations, $a_1$, $a_2$ and
$a_3$, as defined in Eq.~\eqref{eq:def_a}.  In this section, we will
demonstrate that this is a consequence of angular momentum conservation and
parity conservation.  For illustration, we consider the case of $BB$
scattering, but the same reasoning applies in general.

We consider the forward elastic scattering $\ket{1}+\ket{2}\to\ket{1}+\ket{2}$
with the superposed states
\begin{flalign}
&\ket{1}=x_1\ket{B_R}+x_2\ket{B_L}\\
&\ket{2}=y_1\ket{B_R}+y_2\ket{B_L}
\end{flalign}
where $R$ and $L$ denote positive and negative helicity states, respectively. The
matrix element can be expanded in the components of $x$ and $y$:
\begin{flalign}
&\braket{1,2|\mathcal{M}|1,2}\nonumber\\=
&
(x^*_1y^*_1\Bra{B_R,B_R}+x^*_1y^*_2\Bra{B_R,B_L}+x^*_2y^*_1\Bra{B_L,B_R}+x^*_2y^*_2\Bra{B_L,B_L})\nonumber \\
&\mathcal{M}(x_1y_1\Ket{B_R,B_R}+x_1y_2\Ket{B_R,B_L}+x_2y_1\Ket{B_L,B_R}+x_2y_2\Ket{B_L,B_L})\label{eq:hel1}\\
=&|x_1|^2|y_1|^2\Braket{B_R,B_R|\mathcal{M}|B_R,B_R}+|x_2|^2|y_2|^2\Braket{B_L,B_L|\mathcal{M}|B_L,B_L}\nonumber\\
&+x_2 y_2 x^*_1 y^*_1 \Braket{B_R,B_R|\mathcal{M}|B_L,B_L}+x_1 y_1 x^*_2 y^*_2 \Braket{B_L,B_L|\mathcal{M}|B_R,B_R}\nonumber\\
&+|x_1|^2 |y_2|^2 \Braket{B_R,B_L|\mathcal{M}|B_R,B_L}+|x_2|^2 |y_1|^2 \Braket{B_L,B_R|\mathcal{M}|B_L,B_R}\label{eq:hel2} \\
=&(|x_1|^2|y_1|^2+|x_2|^2|y_2|^2)\Braket{B_R,B_R|\mathcal{M}|B_R,B_R}\nonumber\\
&+(x_2 y_2 x^*_1 y^*_1+x_1 y_1 x^*_2 y^*_2)\Braket{B_R,B_R|\mathcal{M}|B_L,B_L}\nonumber\\
&+(|x_1|^2 |y_2|^2+|x_2|^2 |y_1|^2) \Braket{B_R,B_L|\mathcal{M}|B_R,B_L}\label{eq:hel3}\\
=&\left(a_2+\frac{a_1-a_3}{2}\right)\Braket{B_R,B_R|\mathcal{M}|B_R,B_R}
+\frac{1}{2}(a_1-a_3)\Braket{B_R,B_R|\mathcal{M}|B_L,B_L}\nonumber\\
&+\frac{1}{2}(a_1+a_3)\Braket{B_R,B_L|\mathcal{M}|B_R,B_L}, \label{eq:hel4}
\end{flalign}
where, to obtain Eq.~\eqref{eq:hel2}, 10 terms that violate angular momentum
conservation in the forward limit have been dropped (such as $\Braket{B_R,B_R|\mathcal{M}|B_R,B_L}$ and
$\Braket{B_L,B_R|\mathcal{M}|B_R,B_L}$, etc.
);
to obtain Eq.~\eqref{eq:hel3}, parity conservation has been use, as
all operators considered in this work conserve parity.

Eq.~\eqref{eq:hel4} shows that the elastic forward amplitude, superposed in the
helicity space, depends on the superposition parameters $x$ and $y$ only
through the three $a_i$ parameters.

\bibliography{bib.bib}

\end{document}